%% file: ms.tex
\pdfoutput=1
\documentclass[iop,apj,revtex4,astrosymb]{emulateapj}
\usepackage[maxfloats=30]{morefloats}
\usepackage{subfigure}
\usepackage[bookmarks=true,unicode,breaklinks,colorlinks,urlcolor=blue,citecolor=blue,linkcolor=blue]{hyperref}
\usepackage{threeparttable}
\usepackage{bm}
\newcommand{\La}{`bc03\_sa'}
\newcommand{\Lb}{`bc03\_ch'}
\newcommand{\Lc}{`ma05\_sa'}
\newcommand{\Ld}{`ma05\_kr'}

\newcommand\hl[1]{#1}
\gdef\zs{z_\mathrm{s}}
\gdef\zp{z_\mathrm{p}}

\shorttitle{BayeSED: A general approach to fitting the SED of galaxies}
\shortauthors{Han \& Han}

\begin{document}
\title{\hl{BayeSED: A general approach to fitting the spectral energy distribution of galaxies}}

\author{
Yunkun Han\altaffilmark{1,2},
Zhanwen Han\altaffilmark{1,2}
}
\altaffiltext{1} {Yunnan Observatories, Chinese Academy of Sciences, Kunming, 650011, China}
\altaffiltext{2}{Key Laboratory for the Structure and Evolution of Celestial Objects, Chinese Academy of Sciences, Kunming, 650011, China}
\email{hanyk@ynao.ac.cn}
\email{zhanwenhan@ynao.ac.cn}

\begin{abstract}
We \hl{present} a newly developed version of BayeSED, a general Bayesian approach to \hl{the} spectral energy distribution (SED) fitting of galaxies.
The new BayeSED code has been systematically tested \hl{on} a mock sample of galaxies.
The comparison between estimated and inputted value of the parameters show that BayeSED can recover the physical \hl{parameters} of galaxies reasonably well.
\hl{We then} applied BayeSED to interpret the SEDs of a large Ks-selected sample of galaxies in the COSMOS/UltraVISTA field with stellar population synthesis models.
With the new BayeSED code, a Bayesian model comparison of stellar population synthesis models has been done for the first time.
We found that \hl{the model by} \cite{Bruzual2003a}, statistically speaking, has larger Bayesian evidence than \hl{the model by} \cite{Maraston2005a} for the Ks-selected sample.
\hl{Besides, while setting the stellar metallicity as a free parameter obviously increases the Bayesian evidence of both models, varying the IMF has a notable effect only on the \cite{Maraston2005a} model.}
Meanwhile, the physical parameters estimated with BayeSED are found to be generally consistent with \hl{those} obtained with the popular grid-based FAST code, while the former \hl{exhibits} more natural distributions.
\hl{Based on the estimated physical parameters of galaxies in the sample, we qualitatively classified the galaxies in the sample into five populations that may represent galaxies at different evolution stages or in different environments.}
We conclude that BayeSED could be a reliable and powerful tool for investigating the formation and evolution of galaxies from the rich multi-wavelength observations currently available.
A binary version of MPI parallelized BayeSED code is publicly available at \url{https://bitbucket.org/hanyk/bayesed}.
\end{abstract}


\section{INTRODUCTION}
\label{s:intro}
In the formation and evolution of galaxies, the physical processes about the formation and evolution of stars, \hl{the interstellar medium (ISM), and those of super-massive black holes, are expected to be tightly interconnected}.
The theoretical understanding of these processes and their interactions by using semi-analytical \hl{modeling} \citep{Cole2000a,Bower2006a,Croton2006a,Baugh2006a,Marulli2008a,Somerville2008a,Bonoli2009a} and numerical simulations \citep{DiMatteo2005a,Springel2005a,Springel2005b,Hopkins2006b,Hopkins2008b,Hopkins2008a,Springel2010a} {has been greatly advanced} in the last few years.
However, due to the huge complexities presented in these physical processes, our understanding of them and their \hl{mutual interactions} are still far from complete. 

Empirical clues from multi-wavelength observations of galaxies \citep{Bell2003a,Shen2003a,Kauffmann2004a,Baldry2004a,Tremonti2004a,Gallazzi2005a} are very helpful \hl{in providing} crucial insights for our further understanding of the formation and evolution of galaxies.
\hl{Also}, the theoretical understanding of the formation and evolution of galaxies \hl{needs} to be tested, constrained and someday confirmed by many observational results.
Fortunately, all of those complex physical processes involved in the formation and evolution of galaxies have \hl{left some} imprints in the integrated multi-wavelength spectral energy distributions (SEDs) of galaxies.
\hl{Therefore, the multi-wavelength SEDs of galaxies are} a very important source of information for our understanding of those complex physical processes.
The advent of new observing facilities and large surveys at wavelengths from $\gamma$-ray to radio \citep{Lonsdale2003a,Jansen2001a,Martin2005a,Giavalisco2004a,Atwood2009a,Abazajian2009a,Scoville2007a,Davis2007a,Driver2009a,Condon1998a} now allow us to obtain the full SEDs of galaxies.

To extract physical \hl{information} of galaxies from their observed multi-wavelength SEDs, we need some \hl{kind} of theoretical model for the SEDs of galaxies.
The SEDs of most galaxies can be thought of as the superposition of the SEDs of a population of stars with different \hl{masses, ages, metallicities, and so on, that constitute the galaxy}.
\hl{Due to this, stellar population synthesis has been the main method of \hl{modeling} the SEDs of galaxies}, starting from the pioneering works of \cite{Tinsley1972a}, \cite{Searle1973a}, and \cite{Larson1978a}.
From then on, numerous efforts by different groups have been made to improve this technique \citep{Bruzual1983a,Buzzoni1989a,Bruzual1993a,Bruzual2003a,Bruzual2007a,Leitherer1995a,Fioc1997a,Maraston1998a,Maraston2005a,Zhang2005a,Li2008a,Conroy2009a}.
However, some \hl{important} issues  still remain.
For example, some short lived but bright phases of stellar evolution \hl{such as the thermally pulsing asymptotic giant branch, horizontal branch \citep{Catelan2009a,Lei2013a} and blue stragglers} \citep{Tian2006a,Han2007a,Chen2009a} are still not well understood, and they potentially have important effects on the resulting SEDs of galaxies.
\hl{Furthermore, there are issues about the universality of the stellar initial mass function \citep{Padoan1997a,Myers2011a,Dutton2013a}, different parameterizations of the  star formation history \citep{Maraston2010a}, the complex effects  of interstellar medium \citep{Calzetti2000a}, the stochastic nature of stellar population modeling \citep{Buzzoni1993b,Cervino2013b}, and any possible contribution from active galactic nuclei \citep{Polletta2007a,Murphy2009a,Han2012a}.}
These issues represent the large uncertainties in the modeling of galaxy SEDs, and have resulted in the diversity of SED models.
These uncertainties should be properly considered when trying to employ the SED models to \hl{derive} the physical properties of galaxies, or search for physical relations among these properties \citep{Conroy2009a,Conroy2010c,Conroy2010d,Conroy2013e}.

The main method of extracting physical \hl{information} from the multi-wavelength SEDs of galaxies is SED fitting.
By SED fitting, we try to derive one or several physical properties of galaxies by using \hl{certain} fitting methods to compare SED models with observed SEDs.
\hl{In other words, we need to solve the inverse problem: how can the physical properties of galaxies (e.g. stellar ages, stellar masses, star-formation histories, dust extinction and so on) be reasonably derived from quantities that are directly observable (e.g. multi-wavelength photometric SEDs)? }
In the last decade, the technique of SED fitting has been significantly improved \citep[see, e.g.][for a recent review]{Walcher2011a}.
Numerous SED fitting methods and corresponding \hl{software} have been presented by many authors \citep{CidFernandes2005a,Ocvirk2006a,Tojeiro2007a,Koleva2008a,Koleva2009a}, and have been used by even more authors to derive valuable physical \hl{information} about the formation and evolution of galaxies.

\hl{The modeling of a galaxy's SED involves the convolutions of the star formation and evolution history, the stellar initial mass function, and the formation and evolution of its dusty interstellar medium, which has nonlinear effects on the resulting SEDs of galaxies, and many other properties that characterize the formation and evolution of the galaxy \citep{Mo2010a}.}
\hl{Furthermore, apart from uncertainties in the observations, there are many uncertainties in the modeling of galaxy SEDs.}
\hl{Given these complexities and uncertainties, the problem of deriving the physical properties of galaxies from their directly observable properties is generally not invertible in a strict mathematical sense.}
\hl{However, from the perspective of statistical inversion theory \citep{Kaipio2004a,Tarantola2005a}, the inverse problems can be solved by means of Bayesian statistical inference.}
\hl{This is very different from the commonly used approach of solving an optimization problem in the least $\chi^2$ sense \citep{Tinsley1976a,BolzonellaM2000a,WalcherC2006a,MarastonC2006a,Koleva2009a,Pforr2012a,Pforr2013a,Mitchell2013a,Li2013a}, where the main purpose is finding the best fitting result.}
\hl{In Bayesian statistical inference, all quantities (e.g. parameters and fluxes) are modeled as random variables.}
\hl{Therefore, instead of finding a specific value of a parameter that best matches the observations, the solution to an inverse problem is the posterior probability distribution of the quantity of interest, which describes the degree of confidence about the quantity given the available observations.}
\hl{The posterior probability distributions represent our full knowledge about the parameters, with the degree of uncertainty and degeneracies between them manifesting themselves as easily noticeable broad or multi-peaked distributions.}
\hl{Besides, in Bayesian inference, additional information about the problem can be incorporated as priors to constrain the solution even further.}
\hl{The Bayesian inference methods have been successfully used in many fields of physics \citep{Gregory2005a} and especially cosmology to derive the cosmology parameters of the Universe \citep{Lewis2002a,Verde2003a,Cole2005a,Dunkley2009a,Hinshaw2013a}.}

In recent years, Bayesian methods have been used in the field of SED fitting of galaxies by more and more authors.
\cite{Benitez2000a} systematically applied Bayesian inference to \hl{estimate the photometric redshift} of galaxies.
\hl{Notably}, they used Bayesian priors as additional information besides the observed photometric SEDs to give better redshift estimates for the first time.
\cite{Kauffmann2003a}  have used the Bayesian technique to estimate the stellar mass-to-light ratios, dust attenuation corrections and burst mass fractions for a sample of $10^5$ galaxies from the Sloan Digital Sky Survey.
They also presented a rigorous mathematical description of the method, which currently \hl{forms} the basis for Bayesian SED fitting.
\cite{daCunha2008a} presented an empirical but physically motivated model to interpret the SEDs of galaxies from UV to  far-IR consistently, \hl{as well as the corresponding MAGPHYS (Multi-wavelength Analysis of Galaxy Physical Properties) package}.
Similarly, \cite{Noll2009a} presented the code CIGALE (Code Investigating GALaxy Emission) for a Bayesian-like analysis of galaxy SEDs from far-UV to far-IR by fitting the attenuated stellar emission and the related dust emission simultaneously.

Recently, the Markov Chain Monte Carlo (MCMC) algorithm \hl{has} been employed by different authors \citep{Acquaviva2011a,Serra2011a,Johnson2013a} to allow a much more efficient and complete sampling of the parameter space of a SED model than the \hl{grid-based} methods like CIGALE and MAGPHYS.
In \cite{Han2012a}, we described our BayeSED code, where the multimodal nested (MultiNest) sampling algorithm \hl{has} been employed, and applied it to a sample of hyperluminous infrared galaxies as a demonstration.
The use of MultiNest instead of MCMC \hl{allows} us to obtain not only the posterior distribution of all model parameters, but also the Bayesian evidence of the model that can be used as a \hl{generalization} of Occam's razor for quantitative model comparison.
Meanwhile, the principal component analysis and artificial neural networks techniques have been employed to significantly speed up the generation of model SEDs, a major bottleneck for efficient sampling of the parameter space of a SED model.

After the first description in \cite{Han2012a}, the BayeSED code has been significantly improved, including but not limited to the \hl{following}.
Firstly, the new MultiNest algorithm, which is improved by importance nested sampling and \hl{allows} a more efficient exploration of higher-dimensional parameter spaces and more accurate calculation of Bayesian evidence, has been employed.
Secondly, besides the artificial neural networks (ANNs) algorithm, the k-nearest neighbors (KNN) algorithm has been added as another method for efficient interpolation of model SED libraries.
Thirdly, the redshift of a galaxy can be set as a free parameter, and the effect of \hl{the} intergalactic medium and Galactic extinction have been considered.
Fourthly, the main body of BayeSED has been completely rewritten in C++ in an object-oriented programming fashion, and  parallelized with MPI to be able to interpret the SEDs of multiple galaxies simultaneously.
In this paper, we systematically test this new version of \hl{the BayeSED code with a} mock sample of galaxies and apply it to interpret the observed SEDs of a Ks-selected sample of galaxies in the COSMOS/UltraVISTA field with evolutionary population synthesis models.

This paper is organized as follows. 
In \S \ref{s:bayesed}, we describe our BayeSED code, and recent improvements to it. 
We begin in \S \ref{ss:bayes} by introducing the basic idea of Bayesian inference, and its application to the problem of SED fitting.
In \S \ref{ss:sampling}, we introduce the implementation of Bayesian inference with the efficient and robust Bayesian inference tool ---MultiNest\footnote{\url{http://ccpforge.cse.rl.ac.uk/gf/project/multinest/}}, which is capable of calculating the Bayesian evidence of a model and exploring its parameter space, \hl{which could be very complex and of a} moderately high dimension.
To use sampling methods like MCMC and MultiNest, we must be able to evaluate a SED model at any point in the allowed parameter space.
So, in \S \ref{ss:interpolation}, we present the methods of interpolating a model SED library, while using the evolutionary population synthesis model as an example.
In \S \ref{ss:bayesed}, we introduce how the MultiNest algorithm and the interpolating algorithm are combined to build up our BayeSED code.
To test the ability of BayeSED to recover the physical parameters of galaxies from their multi-wavelength photometry, we employ the method of using a mock sample of galaxies in \S \ref{s:application_mock}.
In \S \ref{s:application_real}, we systematically apply our BayeSED code to interpret the SEDs of a Ks-selected sample of galaxies in the COSMOS/UltraVISTA field given by \cite{Muzzin2013a}.
Finally, a summary of our BayeSED method and the results obtained with its application are presented in \S \ref{s:summary}.

\section{BayeSED---Bayesian spectral energy distribution fitting of galaxies}
\label{s:bayesed}

\subsection{Bayesian Inference}
\label{ss:bayes}
In BayeSED, we have employed the Bayesian inference methods to interpret the SEDs of galaxies.
Bayesian methods have been widely used in astrophysics and cosmology \citep[see, e.g.,][for a recent review]{Trotta2008a}. 
They provide a more consistent conceptual basis for dealing with problems of inference in the presence of uncertainties than traditional statistical methods.
For a set of experimental or observational data $\bm d$, and a model (or hypothesis) $M$ with some parameters $\bm \theta$ that are employed to explain them, the Bayes' theorem states that 
\begin{equation}
 P(\bm \theta|\bm d, M) = \frac{P(\bm d|\bm \theta, M) P(\bm \theta|M)}{P(\bm d|M)}.
 \label{eq:bayes_Theorem}
\end{equation}
For SED fitting, $\bm d$ represents the observed SED of a galaxy while $\bm \theta$ represents the parameters of a SED model $M$.
In Equation \ref{eq:bayes_Theorem}, $P(\bm \theta|\bm d, M)$ is called the {\em posterior probability} of parameters $\bm \theta$ given the data $\bm d$ and model $M$.
$P(\bm d|\bm \theta,M)$ is the probability of $\bm d$ given the model $M$ and its parameters $\bm \theta$.
It is also known as likelihood $\mathcal{L}(\bm{\theta})$, which describes how the degree of plausibility of the parameter $\bm \theta$ changes when new data $\bm d$ is considered.
$P(\bm \theta|M)$ is the prior, which describes \hl{knowledge} about the parameters irrespective of the data.
Finally, $P(\bm d|M)$ is a normalization constant called marginal likelihood, also known as Bayesian evidence.

Bayesian inference is generally divided into two categories: parameter estimation and model comparison.
In parameter estimation, the Bayesian evidence $P(\bm d|M)$, as a normalizing factor that is independent of the parameters $\bm \theta$, is usually ignored.
The posterior includes all \hl{information} that can be used for the complete Bayesian inference of the parameter values.
It can be marginalized over each parameter to obtain individual parameter constraints.
So, the posterior probability density function (PDF) of a parameter $\theta_i$ could be obtained as:
\begin{equation}
P(\theta_i|\bm d,M) = \int \mathrm{d}\theta_1 \cdots \mathrm{d}\theta_{i-1} \mathrm{d}\theta_{i+1} \cdots 
\mathrm{d}\theta_{N} P(\bm \theta|\bm d,M).
 \label{eq:bayes_par}
\end{equation}
The Bayesian evidence of a model, which is not important for parameter estimation but critical for model comparison, is given by:
\begin{equation}
 P(\bm d | M) \equiv {\int_{\Omega_M} P(\bm d | \bm \theta, M)
 P(\bm \theta| M){\rm d}\bm \theta},
 \label{eq:bayes_evidence}
\end{equation}
where $\Omega_M$ represents the whole $N$-dimensional parameter space of the model $M$.
It is clear that the Bayesian evidence of a model is just the average of the likelihood weighted by the priors.
However, this simple definition automatically implements the principle of Occam's razor: a simpler theory with compact parameter space should be better than a more complicated one, unless the latter is significantly better for the explanation of observational data.
\hl{Generally, the Bayesian evidence is just larger for a model with a better fit to observations, while smaller for a more complicated model with more free parameters or larger parameter space.}
\hl{The comparison between two models $M_2$ and $M_1$ can be formally expressed as the ratio of their respective posterior probabilities given the observational data set $\bm d$:}
\begin{equation}
\frac{{P({M_2}|\bm d)}}{{P({M_1}|\bm d)}} = \frac{{P(\bm d|{M_2})P({M_2})}}{{P(\bm d|{M_1})P({M_1})}},
  \label{eq:bayes_model}
\end{equation}
\hl{$P({M_2})/P({M_1})$ is the prior odds ratio of the two models, which is often set to be $1$ if none of the two is of special interest.}
\hl{If so, the  Bayes factor, which is defined as}
\begin{equation}
  {B_{2,1}} \equiv \frac{{P(\bm d|{M_2})}}{{P(\bm d|{M_1})}},
  \label{eq:bayes_factor}
\end{equation}
\hl{can be directly used for Bayesian model comparison.}
\hl{According to the empirically calibrated Jeffrey's scale \citep{Jeffreys1998a,Trotta2008a}, $\rm ln(B_{2,1}) > 0, 1, 1.5$ and $5$ (corresponding to the odds of about 1:1, 3:1, 12:1 and 150:1), represent inconclusive, weak, moderate and strong evidence in favor of $M_2$, respectively.}

\subsection{Sampling of high-dimensional parameter space with MultiNest}
\label{ss:sampling}
Commonly, for the problem of Bayesian parameter estimation, we need to solve the $N-1$ dimensional integration of Equation \ref{eq:bayes_par}.
However, it is very hard to obtain an accurate analytical solution for this equation in most cases.
\hl{Moreover, for many problems in astrophysics, we cannot even write down the analytical form of this equation, since the mathematical expression of the priors and/or likelihood function may simply not exist.}
In practice, Bayesian parameter estimation could be achieved more conveniently by taking a set of samples from the parameter space that are distributed according to the posterior $P(\bm \theta|\bm d, M)$, where the posterior \hl{might} be unnormalized.
Then, the estimation of parameters could be obtained by some simple statistics of these samples.

The most widely used sampling method for this is the Markov Chain Monte Carlo (MCMC) method.
The MCMC technique, which is often based on the \hl{Metropolis-Hastings algorithm \citep{Metropolis1953a,Hastings1970a}}, provides an efficient way to explore the parameter space of a model and ensures that the number density of samples is asymptotically proportional to the posterior probability density.
However, the commonly used MCMC methods are very computationally intensive when the posterior distribution is multimodal or \hl{has} large degeneracies between parameters, particularly in high dimensions.
On the other hand, the calculation of Bayesian evidence, which is critical for Bayesian model comparison, cannot be obtained easily by most MCMC techniques. 
This is because the evaluation of the multidimensional integral in Equation \ref{eq:bayes_evidence} is a challenging numerical task.

The nested sampling, firstly introduced by \cite{Skilling2004a}, provides an efficient method to calculate the Bayesian evidence, while also produces posterior inferences as a by-product. 
So, by using the nested sampling method, we are allowed to achieve efficient Bayesian parameter estimation and model comparison simultaneously.
This method \hl{has} been improved further by the works of \cite{Mukherjee2006a} and \cite{Shaw2007a} to \hl{increase} the acceptance ratio and the sampling efficiency.
Building on these works and pursuing further the notion of detecting and characterizing multiple modes in the posterior from the distribution of nested samples, \cite{Feroz2008a} introduced the MultiNest algorithm as a viable, general replacement for traditional MCMC sampling techniques.
With some further development of this algorithm, the resulting Bayesian inference tool was announced  to be publicly released in \cite{Feroz2009a}.
From then on, the MultiNest algorithm \hl{is} becoming more and more popular, and has been successfully applied to numerous inference problems in particle physics, cosmology and astrophysics \citep{Trotta2009a,Feroz2009b,Feroz2010a,Martin2011a,Graff2012a,Karpenka2013b,Kavanagh2014a}.
In \cite{Han2012a}, we have employed the MultiNest algorithm to build our BayeSED code for SED fitting of galaxies.
In the current version of BayeSED, we have employed the most recent version of MultiNest as described in \cite{Feroz2013a}.
The newly developed MultiNest algorithm was largely improved by the \hl{technique known as importance nested sampling (INS), which increases} the accuracy of the calculation of Bayesian evidence by up to an order of magnitude.
To achieve the same level of accuracy, the higher evidence accuracy from INS could potentially speed \hl{up MultiNest by a factor of a few, if fewer live points or higher target efficiency are used}.

\subsection{Interpolation of Model SED Library}
\label{ss:interpolation}
For \hl{the} Bayesian inference of the SEDs of galaxies by using \hl{an} extensive sampling method such as MultiNest (\S~\ref{ss:sampling}), the SED model needs to be able to be evaluated at any point of its parameter space.
However, it would be very computationally expensive to employ a detailed SED model, such as an evolutionary population synthesis model, to generate SEDs during the sampling of a high dimensional and complex parameter space.
Besides, for many SED models, only a pre-computed library of model SEDs is available.
So, it is often very necessary to interpolate a model SED library.
In this subsection, we introduce the interpolation method that we have used in the BayeSED code by taking the evolutionary population synthesis model as an example.

\subsubsection{Building of Evolutionary Population Synthesis Model SED Libraries}
\label{sss:eps}
Currently, the evolutionary population synthesis model is the standard method for \hl{modeling} the SEDs of galaxies.
It is based on the theory of star formation and evolution, the empirical or theoretical stellar spectral library, and the chemical evolution theory of galaxies, \hl{and models the SED of a galaxy as the sum of the contribution from individual stars}.
As mentioned in \S~\ref{s:intro}, there are still many uncertainties in these ingredients of an evolutionary population synthesis model.
So, due to different treatments of these issues, there are many competing evolutionary population synthesis models and many possible parameterizations of the model.

In this paper, we have employed the model of \cite{Bruzual2003a}(bc03) and \cite{Maraston2005a}(ma05), two of the most widely used evolutionary population synthesis models.
For \hl{the} bc03 model, the  initial mass function (IMF) of \cite{Chabrier2003a} and \cite{Salpeter1955a} are \hl{used}, while for the ma05 model, the  initial mass function (IMF) of \cite{Kroupa2001a} and \cite{Salpeter1955a} are \hl{used}.
The star formation histories (SFHs) of galaxies are assumed to be exponentially declining in the form of $\rm{SFR}\propto~\rm{exp}(-t/\tau)$, where $t$ is the time since the onset of star formation and $\tau$ is the e-folding star formation timescale.
To consider the effect of dust attenuation, a uniform dust screen geometry with a \cite{Calzetti2000a} dust extinction law is assumed.

To build SED libraries of the two evolutionary population synthesis models with different assumptions, we have employed a modified version of a grid-based SED fitting code---FAST\footnote{\url{http://astro.berkeley.edu/~mariska/FAST.html}}\citep{Kriek2009a} to actually generate the model SEDs.
We have built four SED libraries, named as ~\La, ~\Lb, \Lc~ and ~\Ld, respectively.
The four SED libraries \hl{cover} a parameter space with $\rm{log}(\tau/yr)$ in the range of $[6.5,11]$ and in steps of $0.10$, $\rm{log}(age/yr)$ in the range of $[7.0,10.1]$ and in steps of $0.05$, and visual attenuation $A_{\rm v}$ in the range of $[0,4]$ and in steps of $0.2$.
\hl{For the bc03 model, the metallicity $Z$ could be $0.004,0.008,0.02$, or $0.05$, while for the ma05 model, $Z$ could be $0.001,0.01,0.02$, or $0.04$.}
In total, there are 243434 model SEDs for \hl{each of the four libraries}.
A summary of these libraries is presented in Table \ref{tab:libs}.
\begin{table*}
  \caption{Summary of model SED libraries}
  \input{libs.tab}
  \tablecomments
  {
  $^a$ In FAST code, `pr' indicate a photometric resolution SED, and the number of wavelengths for a SED is 460.\\
  $^b$ The parameter is in the range of $[6.5,11]$ and in steps of 0.1.
  }
  \label{tab:libs}
\end{table*}

\subsubsection{Dimension Reduction of Model SED with Principal Component Analysis}
\label{sss:pca}
A SED model can be considered as a mapping from parameter $\bm{X}(x_1,x_2,\cdots,x_k)$ to corresponding SED $\bm{S}(f_1,f_2,\cdots,f_n)$, where $x_i$ represents a parameter and $f_i$ represents a flux at a wavelength.
Depending on the resolution of the SED, $n$ could be equal to hundreds or even thousands.
So, for a SED model with many free parameters, the size of \hl{the required} library could be very huge.
For the four SED libraries that we have built in \S \ref{sss:eps}, $n$ is just equal to $460$, and this results in a size of $1.3\rm{Gbyte}$ for each library.
However, due to the continuity of a SED, when the flux at a given wavelength is changed, the fluxes at nearby wavelengths will be changed in a very similar way.
This means that the fluxes at different wavelengths are not completely independent with each other, and the actual \hl{number of dimensions} of the SED could be much less than $n$.
So, it is possible to apply some kind of dimensionality reduction technique to efficiently compress a SED.

One such technique is called principal component analysis (PCA), also known as the Karhunen-Loeve transform.
It is mathematically defined as an orthogonal linear transformation that transforms the original data to a new coordinate system.
The goal of PCA for a SED library is to find an $m$-dimensional ($m \leqslant n$)  linear model of the $n$-dimensional SED library that represents the original SEDs as accurately as possible in a least-squares sense.
In the current version of our BayeSED code, we have employed the PCA algorithm in the SHARK machine learning library\footnote{\url{http://image.diku.dk/shark/sphinx_pages/build/html/index.html}}.

We have applied PCA to the four SED libraries that we have built in \S \ref {sss:eps}.
It should be noted that we take the logarithm of SEDs  before \hl{applying} PCA.
The PCA algorithm in Shark provides two linear models as its outputs.
The first linear model, called `encoder', is a linear transformation from an $n$-dimensional SED $\bm{S}$ to an $m$-dimensional vector $\bm{A}(A_1,A_2,\cdots,A_m)$, where $A_i$ is the amplitude of the $i$-th principal component.
The second linear model, called `decoder', is the inverse transformation of `encoder'.
So, it is a linear transformation from an $m$-dimensional vector $\bm{A}$ to an $n$-dimensional SED $\bm{S}$.
The 'encoder' is used to compress the SEDs of a library, while the `decoder' is used to reconstruct the SEDs.

\begin{figure}
  \centering 
  \includegraphics[scale=0.7]{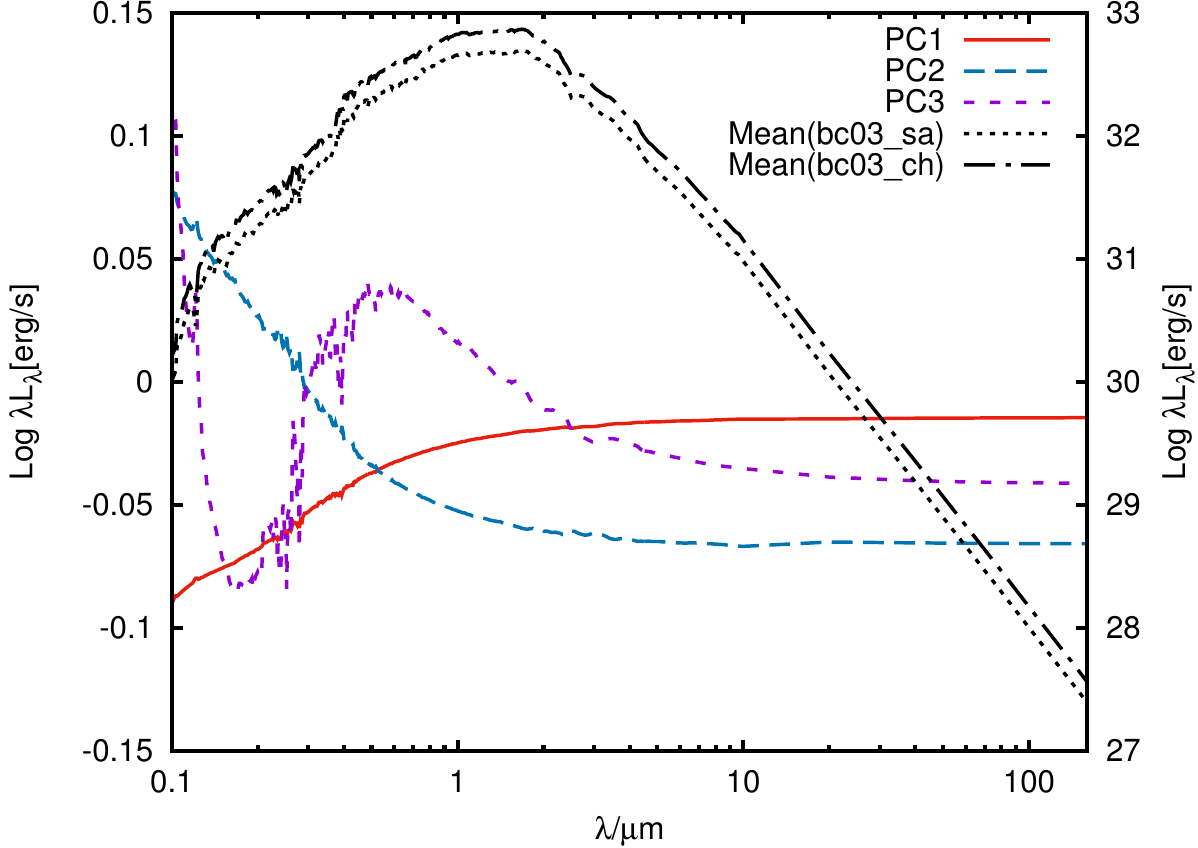}
  \caption
  {
  The first 3 principal components (left axis) and mean spectrum (right axis) of the ~\La~ and ~\Lb~ model SED libraries.
  \hl{  The first 3 principal components contribute $0.82$, $0.16$, and $0.01$ of the overall variance, respectively.}
  }
  \label{fig:pc_bc03}
\end{figure}

\begin{figure}
  \centering 
  \includegraphics[scale=0.7]{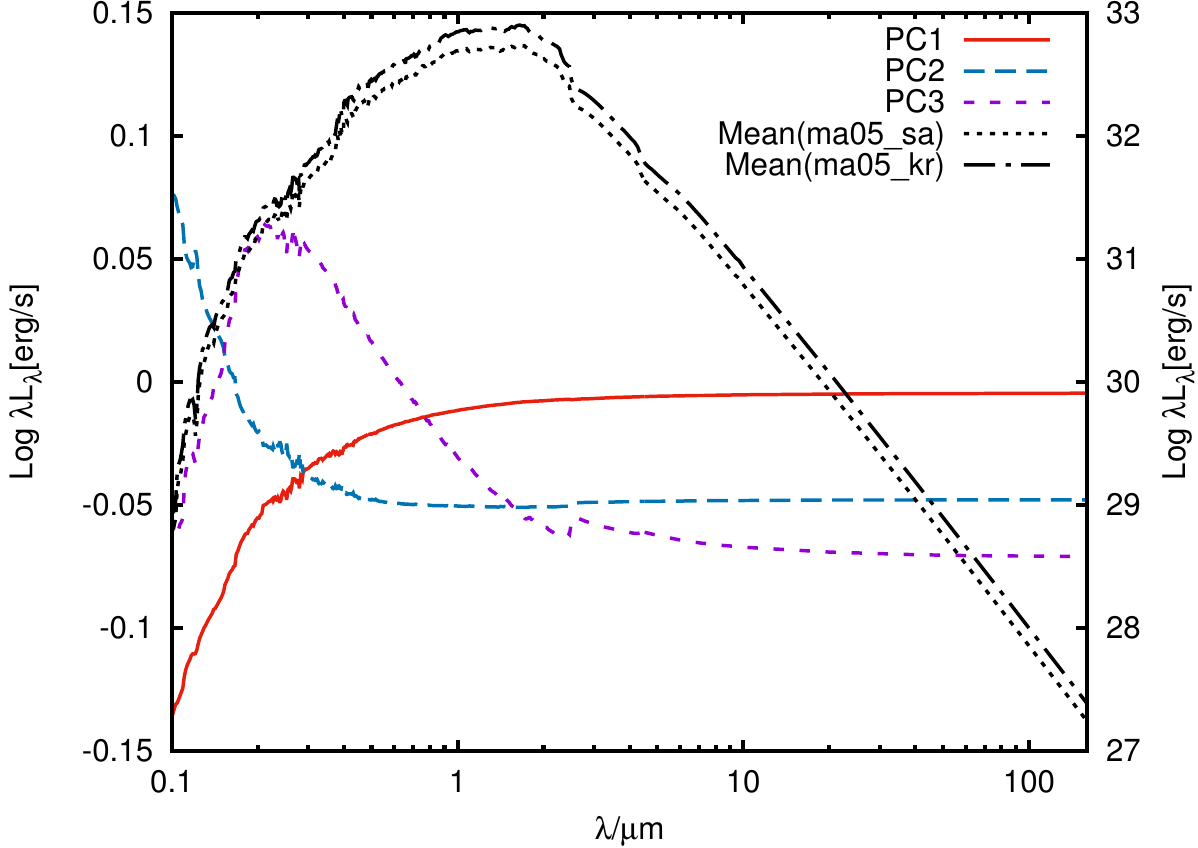}
  \caption
  {
  The first 3 principal components (left axis) and mean spectrum (right axis) of the ~\Lc~ and ~\Ld~ model SED libraries.
  \hl{  The first 3 principal components contribute $0.80$, $0.17$, and $0.02$ of the overall variance, respectively.}
  }
  \label{fig:pc_ma05}
\end{figure}

\begin{figure*}
  \centering 
  \includegraphics[scale=0.75]{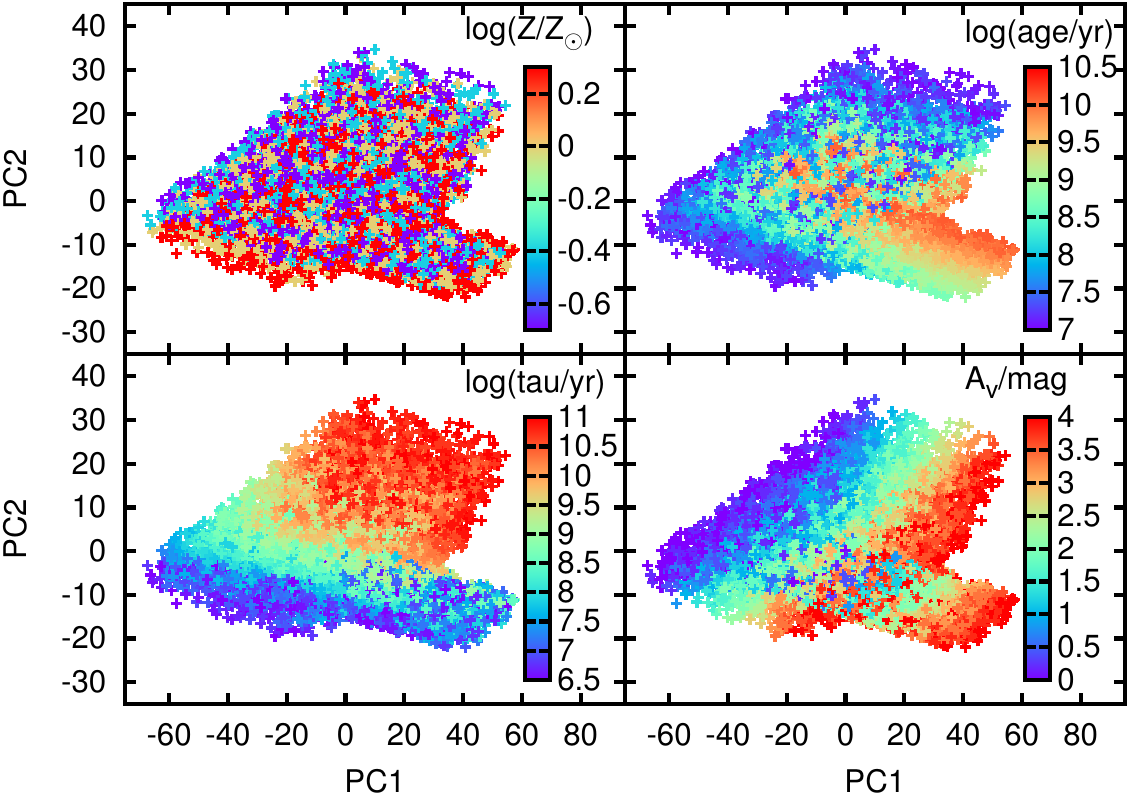}
  \includegraphics[scale=0.75]{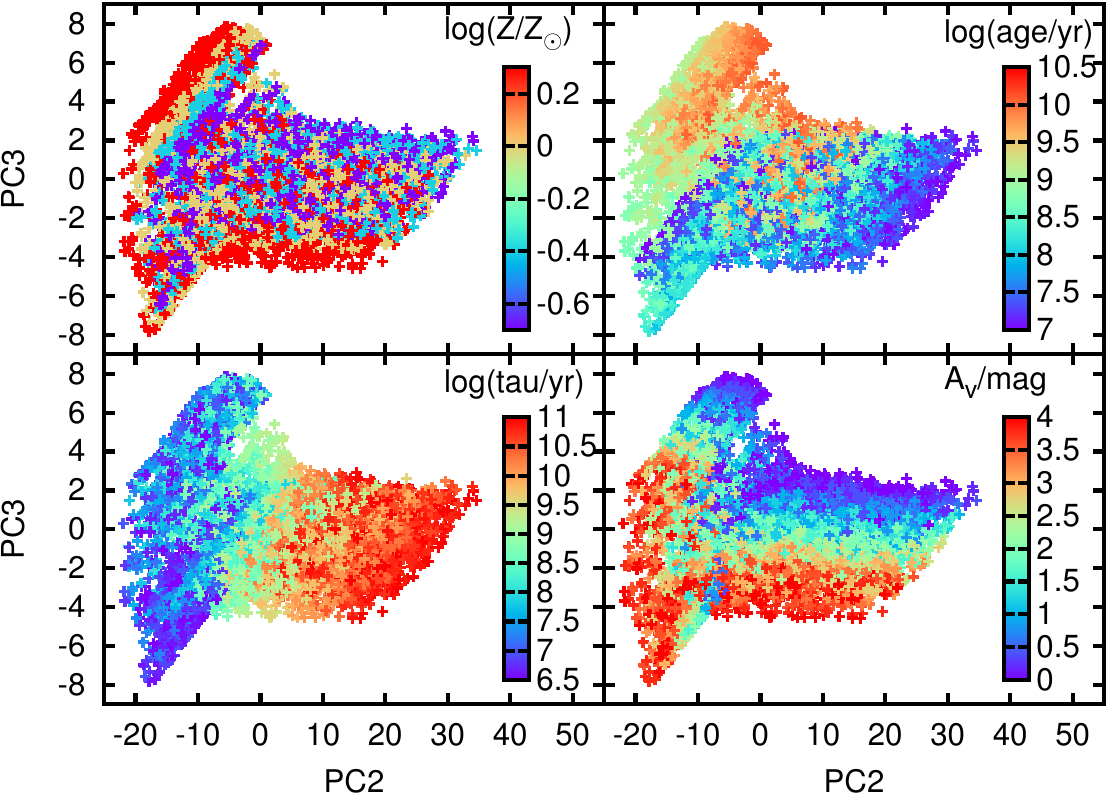}
  \caption
  {
  The distribution of model SEDs for \La~in the space with PC1-PC2 (left) and PC2-PC3 (right) as basis vectors.
  The corresponding physical parameters (stellar metallicity, age, e-folding time, and dust extinction) of SEDs are \hl{represented} by different colors.
  }
  \label{fig:pcs_pars_bc03}
\end{figure*}

\begin{figure*}
  \centering 
  \includegraphics[scale=0.75]{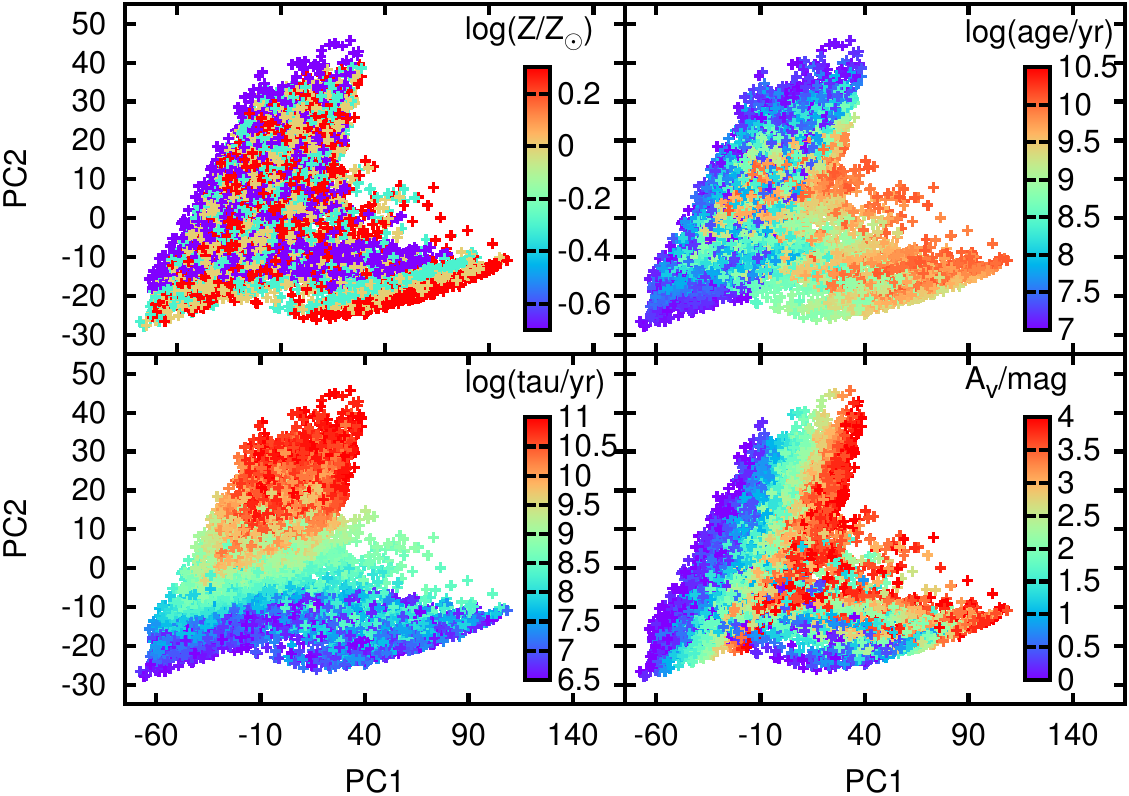}
  \includegraphics[scale=0.75]{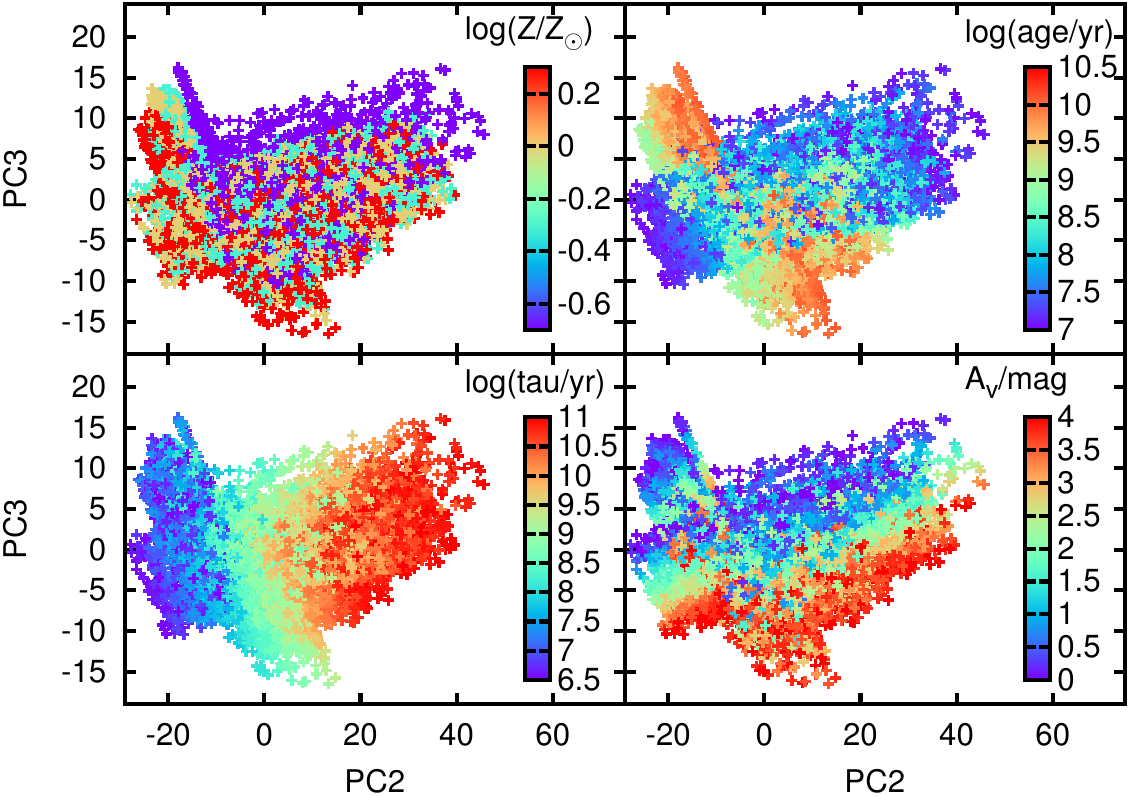}
  \caption
  {
  Similar to Figure \ref{fig:pcs_pars_bc03}, but for \Lc.
  The distributions of SEDs for \Lc~are more complex and different from those for~\La, especially in the PC2-PC3 space.
  }
  \label{fig:pcs_pars_ma05}
\end{figure*}

\begin{figure*}
  \centering 
  \includegraphics[scale=0.75]{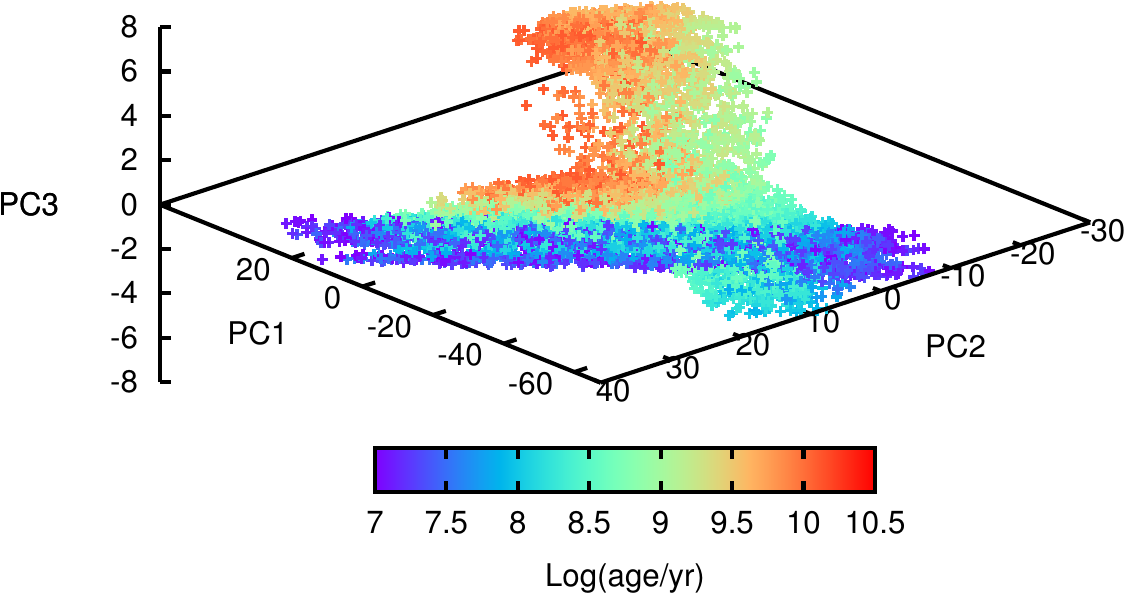}
  \includegraphics[scale=0.75]{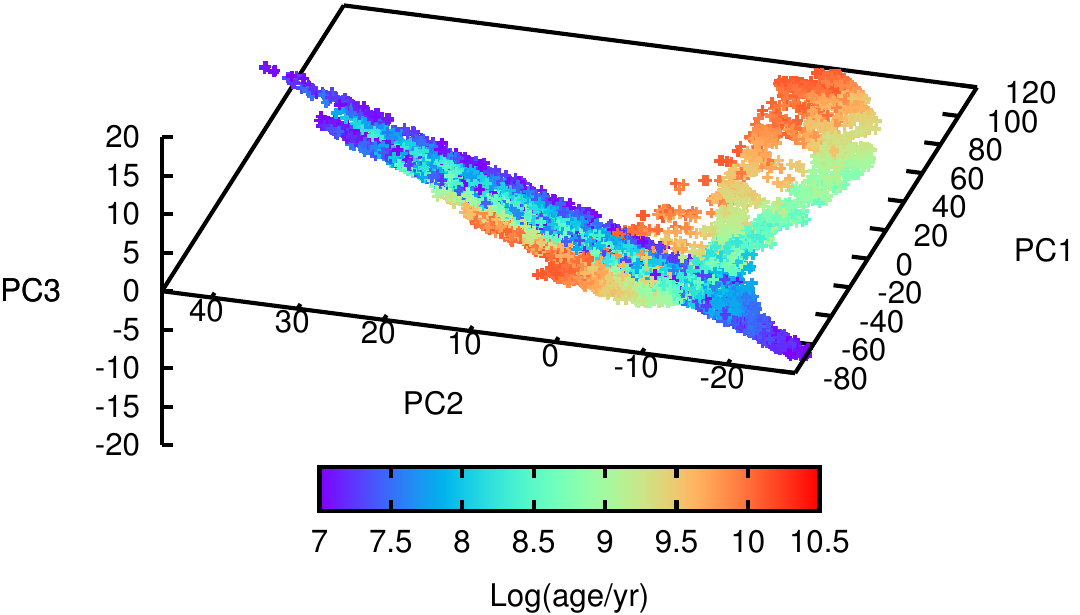}
  \caption
  {
  The distribution of model SEDs for \La~(left) and \Lc~(right) in the space with the first principal components as basis vectors.
  The corresponding age of SEDs are \hl{represented} by different colors.
  We have selected a point of view that best represents the general shape of the distribution of model SEDs in 3D space.
  }
  \label{fig:pcs_age}
\end{figure*}

The first 3 principal components and mean spectrum of the four SED libraries are shown in Figure \ref{fig:pc_bc03} and \ref{fig:pc_ma05}.
As shown in the figures, the low-order principal components, which determine the general shape of a SED, are more smooth than the high-order principal components, which have more detailed features.
For both bc03 and ma05 model, the SED libraries that only differ in the IMF have almost identical principal components, but a slightly different mean spectrum.
However, it is \hl{worthwhile to notice that the SEDs from bc03 and ma05 models follow} different distributions in the space of principal components as shown in Figures \ref{fig:pcs_pars_bc03} and \ref{fig:pcs_pars_ma05}.
While the SEDs from the two models \hl{conform to} somewhat similar distributions in the PC1-PC2 space, they have very different distributions in the PC2-PC3 space.
Meanwhile, the SEDs from the ma05 model show more complex distributions than that from the bc03 model.
Furthermore, the relationship between the amplitudes of principal components and physical parameters are different for the two models.
Generally, the SEDs from each evolutionary population synthesis model shows a unique distribution in \hl{principal component space, a trend which is more obvious in 3D space}, as shown in Figure \ref{fig:pcs_age}.
These differences \hl{may reflect the consequence of different methodologies and treatments of the TP-AGB phase in the two models}.

\begin{figure}
  \centering 
  \includegraphics[scale=0.65]{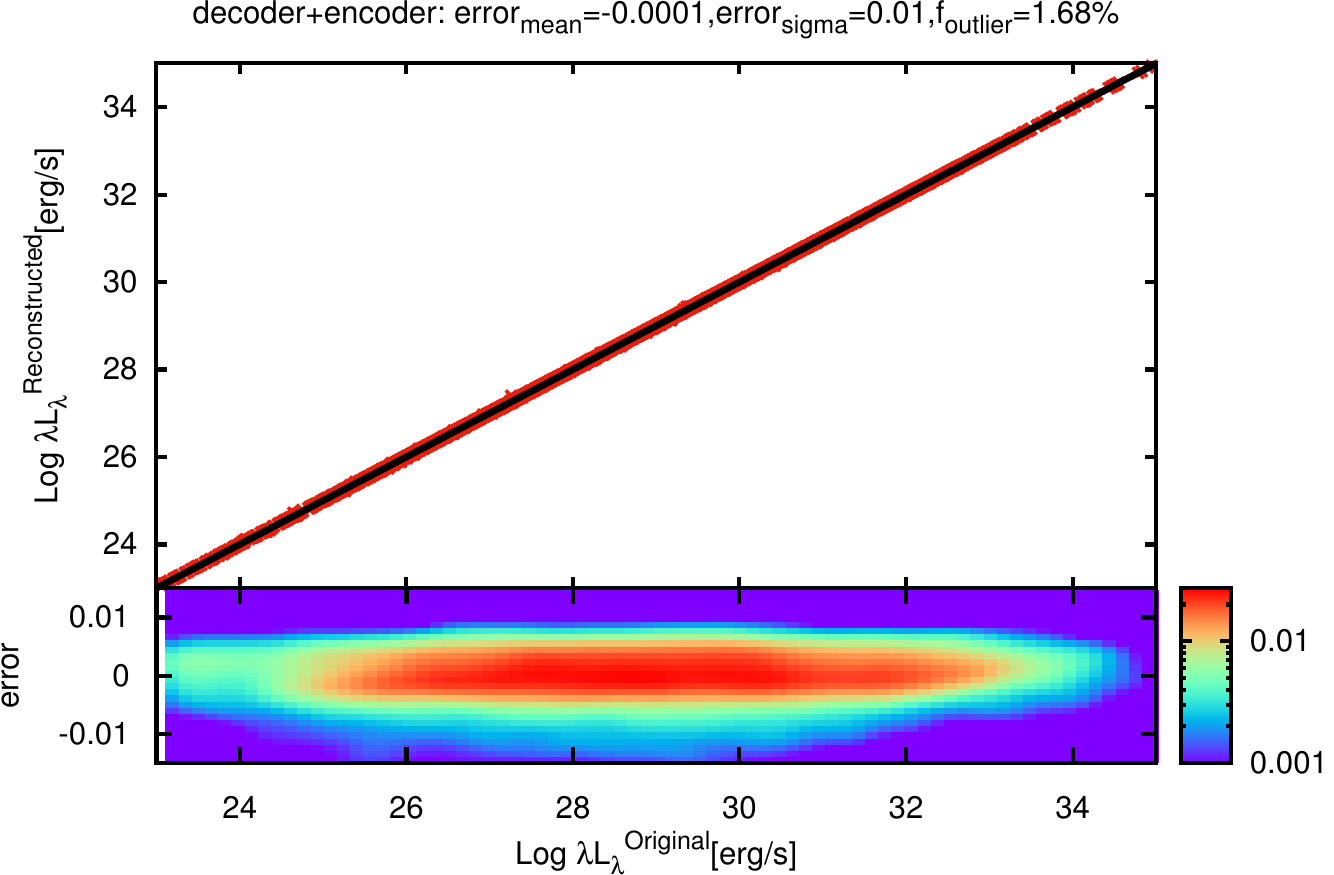}
  \caption{
  \hl{  A test of the PCA method by comparing the original SEDs in the ~\La~ SED library with those reconstructed from the first 23 principal components.}
  \hl{  The probability density distribution (in logarithmic sacle) of errors is shown in the lower panel.}
  \hl{  The mean, standard deviation, and the percentage of outliers (error$>3\sigma$) for the distribution are shown on the top of this figure.}
  }
  \label{fig:test_pca}
\end{figure}

The total number of principal components $m$ is equal to the dimension of the SED $n$, which is $460$ in our case.
However, to compress the original SED libraries, we need to ignore those principal components with much less contribution to the overall variance of the SEDs in the library.
In this work, we choose to ignore those principal components with a contribution $\leqslant 10^{-6}$.
This results in $23$ principal components for the two SED libraries of bc03 model and $26$ principal components for the two SED libraries of ma05 model.
To check the reliability of the PCA method, we have compared the original SEDs in the ~\La~ SED library with those reconstructed from the first 23 principal component.
As shown clearly in Figure \ref{fig:test_pca}, the reconstructed SEDs are almost identical to the original SEDs in most cases.
However, with the application of PCA method, the size of the SED library is reduced to only $23/460 = 5\%$ of the original library.

\subsubsection{Interpolation with Artificial Neural Networks }
\label{sss:ann}
With the application of PCA method, the size of a model SED  library is significantly reduced, since a SED can now be represented by the amplitude of only a few principal components instead of the luminosity at much more wavelengths.
On the other hand, the mapping from parameters $\bm{X}$ to the corresponding SED $\bm{S}$ can now be divided into the mapping from $\bm{X}$ to the amplitudes of principal component $\bm{A}$, and the mapping from $\bm{A}$ to the final $\bm{S}$.
The \hl{latter} mapping is actually the `decoder' obtained in \S \ref{sss:pca}, which is a linear transformation from  $\bm{A}$ to $\bm{S}$.
So, if we can map $\bm{X}$ to $\bm{A}$ very quickly, then we would be able to \hl{evaluate} the original SED model at any point in its parameter space very efficiently.
However, as shown in Figures \ref{fig:pcs_pars_bc03}, \ref{fig:pcs_pars_ma05}, and \ref{fig:pcs_age}, the relationship between $\bm{X}$ and $\bm{A}$ could be very complex.

One method of achieving such an efficient mapping from $\bm{X}$ to $\bm{A}$ is the artificial neural network (ANN) algorithm.
ANNs are mathematical constructs originally designed to simulate some intellectual behaviors of the human brain. 
Just like a human brain, an ANN tries to understand the underling relationship between two set of things (e.g. $\bm{X}$ and $\bm{A}$), from the learning of some instances which obey this relationship.
When the procedure of learning is successfully finished, the ANN could be used to predict the corresponding $\bm{A}$ from any instance of $\bm{X}$, including those have not been learned before.
In the last decade, ANN methods have been successfully used in many problems in astrophysics \citep{Firth2003a,Collister2004a,Carballo2008a,Yeche2010a,Almeida2010a,Silva2011a}.

Currently, there are many kinds of ANNs and the implements of them using different programing languages.
In \cite{Han2012a}, we have modified the widely used ANNz code \citep{Collister2004a}, which is originally built for estimating photometric redshifts, to be suit for interpolation of SED models.
However, to be able to control every components of an ANN algorithm more freely, we have employed a more general and configurable neural network library---Fast Artificial Neural Network Library\footnote{\url{http://leenissen.dk/fann/wp/}}---in the current version of BayeSED.
As ANNz, the ANN algorithm implemented in FANN is the most widely used multi-layer perceptron (MLP) feed-forward network.
A  MLP  network consists of a number of layers of neurons.
Basically, there are tree types layers, which are called input layer, hidden layer, and output layer, respectively.
In a feed-forward network, information propagate through the input layer, hidden layers, and output layer sequentially, without any internal feedback.
Commonly, the network architecture of such an ANN is denoted as $N_{\mathrm{in}}$:$N_1$:$N_2$: $\ldots$ :$N_{\mathrm{out}}$, where $N_{\mathrm{in}}$ is the number of neurons in the input layer, $N_i$ is the number of neurons in $i$th hidden layer, and $N_{\mathrm{out}}$ is the number of neurons in the output layer.

\begin{figure}
  \centering 
  \includegraphics[scale=0.5]{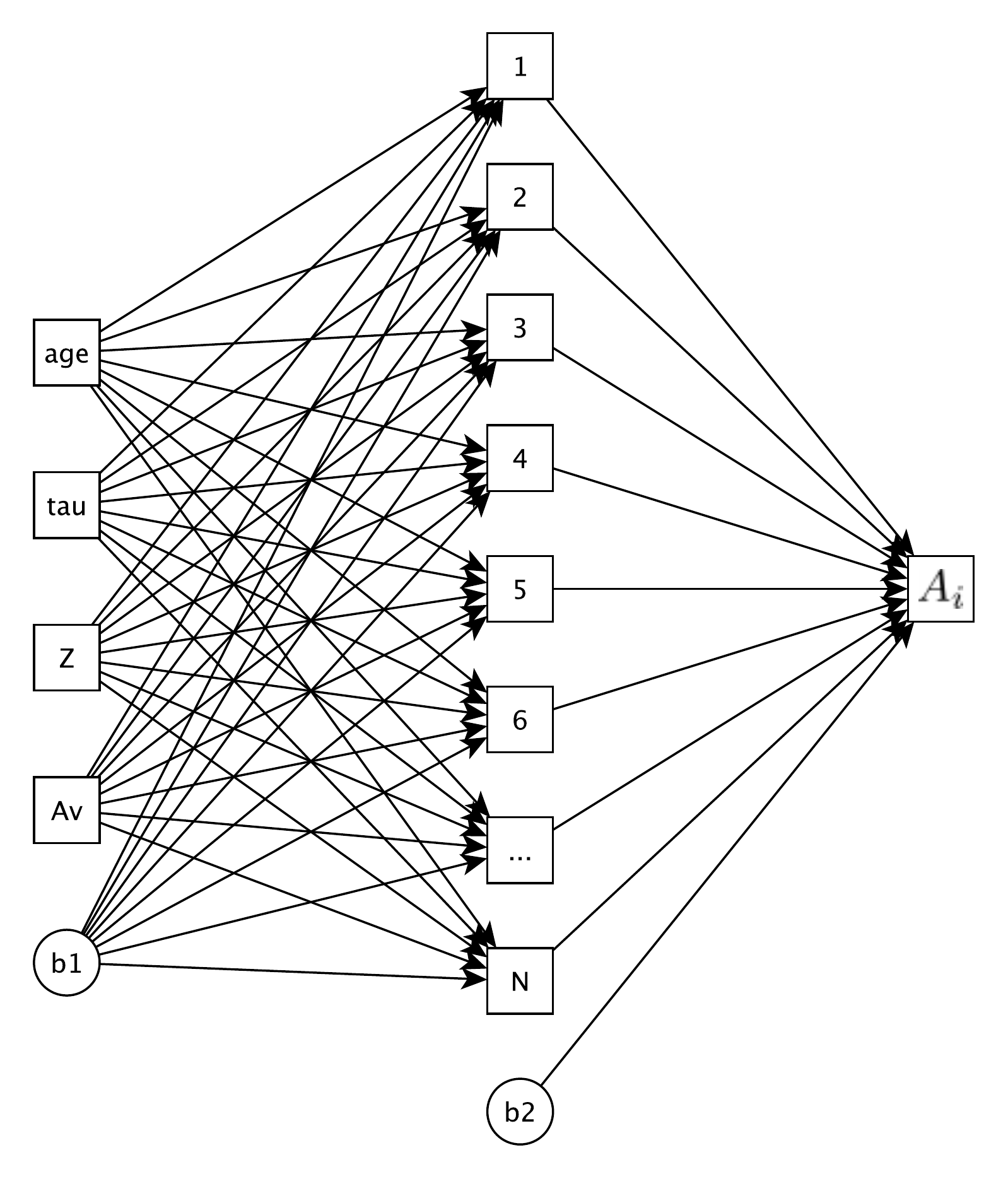}
  \caption
  {
  The network architecture of ANN used for the interpolation of the evolutionary population synthesis model of bc03 and ma05.
  The inputs of this ANN are the 4 parameters of the evolutionary population synthesis model, and the output is the amplitude of a principal component.
  Here, b1 and b2 are two bias neurons, playing as offsets.
  The number of neurons in the hidden layer, N, is set to be 30 for both the bc03 and ma05 models.
  So, the network architecture of ANN used for both bc03 and ma05 models is denoted as 4:30:1.
  }
  \label{fig:ann}
\end{figure}

In Figure \ref{fig:ann}, we show the network architecture of an ANN used for the interpolation of the evolutionary population synthesis model of bc03 and ma05.
The input layer of this ANN have $4$ neurons, corresponding to the $4$ parameters of the evolutionary population synthesis model.
These neurons emit the value of corresponding parameters to the neurons in the next layer.
It is worth to mention that all parameters have been \hl{normalized} to have a zero mean and a standard deviation of $1$ for a better performance of the ANN.
An additional neuron, which is called $b1$, is a bias neuron.
The bias neuron play as an offset and always emits 1.
The capability of an ANN is mainly determined by the structure of its hidden layers.
According to the universal approximation theorem \citep{Cybenko1989a,Kurt1991a,Haykin1999a}, a multilayer feed-forward network with only one hidden layer can approximate any continuous function to arbitrary precision.
However, the neurons in the hidden layer must have a continuous, bounded and nonconstant activation function.
Meanwhile, the number of neurons in the hidden layer needs to be chosen carefully according to the complexity of the problem under consideration.
As the input layer, an additional bias neuron, $b2$, is also needed.
Finally, the last layer gives the output of an ANN.
In our case, there is only one output, which is the amplitude of a principal component.

In principle, more neurons in the hidden layer can give better result, but with the expense of much more training time.
We practically found that a hidden layer with 30 neurons is good enough for the libraries of bc03 and ma05 model.
The choice of activation function for the neurons in the hidden layer and output layer is also crucial.
The FANN library give us many possible choices for the activation function.
For the neurons in the hidden layer, an activation function defined by \cite{Elliott1993a} was chosen.
This activation  function is similar to the commonly used sigmoid activation function, but easier to compute and so faster.
For the neuron in the output layer, a linear activation function was chosen to make sure the output could be scaled to any value.
Since one ANN gives only the amplitude of one principal component, $23$ and $26$ of ANNs with the structure as shown in Figure \ref{fig:ann} need to be trained for the libraries of bc03 and model, respectively.

The training of an ANN is an optimization problem, where we adjust the weights in the ANN to minimize the difference between the outputs of the ANN and that given by the instances in the training data.
The universal approximation theorem of an MLP network states the existence of the solution of such an optimization problem.
However, it told us nothing about how to actually find the solution.
So, the effectiveness of an ANN method largely depends on the algorithm used for training.
The most widely used algorithm for the training of an ANN is the backpropagation algorithm \citep{Rumelhart1986a}.
As the name suggests, the error obtained by propagating an input through the network is then propagated back through the network while the weights are adjusted in such a way that the error becomes smaller.
The original backpropagation algorithm has been further improved by more advanced algorithms like quickprop \citep{Fahlman1988a} and RPROP \citep{Martin1993a,Igel2000a}.
All of these training algorithms have been implemented in the FANN library.
As for our cases, we found that the RPROP training algorithm has the best performance.

Besides the RPROP algorithm, another strategy has been used for the training of ANNs.
The set of input and output data, which are parameters of stellar population synthesis model and corresponding amplitudes of the principal components of SEDs in our case, are sorted randomly and then splitted into three groups.
The first group of data, called `training data', is used by the RPROP algorithm to adjust the weights in the ANN.
During the training of an ANN, the RPROP algorithm tries to minimize the difference between the output of ANN and the results in the `training data'.
However, if too much training is applied to these data, the ANN will eventually over-fit them.
Over-fitting means the ANN will be able to fit the `training data' very precisely, but lose the generalization for other data that have not been used by the RPROP algorithm during the training.
So, another group of data, called `validation data', is used to avoid over-fitting.
During the training, we trace the difference between the output of ANN and the results in the `validation data', which is called error.
The RPROP algorithm will be stopped when the normalized error begin to increase.
Finally, the third group of data, called `testing data' will be used for an ultimate test of the ANN when the training is finished.

We have trained $23$ and $26$ ANNs with the structure as shown in Figure \ref{fig:ann} for the libraries of bc03 and ma05 model, respectively.
Here, we take the training of $23$ ANNs for the ~\La~ SED library as an example.
The SED library has 243432 SEDs in total.
We firstly apply the `decoder', which is obtained by using the PCA method as presented in \S \ref{sss:pca}, to the SEDs in this library to obtain the amplitudes of the first $23$ principal components of them.
Then, the parameters and corresponding amplitudes of principal components are splitted into `training data' ($50\%$), `validation data' ($20\%$) and `testing data' ($30\%$).
These data are used as instances for the training of ANNs.
In practice, we actually train the $23$ ANNs for the ~\La~ SED library simultaneously.
Since the code used for ANN training have been parallelized using OpenMP, the whole  process of training 23 ANNs for this SED library can be finished in about $40$ minutes with 23 Intel 2.20 GHz CPU cores.
Finally, all of these ANNs and other \hl{information} about this SED library are saved to one file, which is only $566$ Kbytes in size.
This single file will be used to replace the original ~\La~ SED library, which is about 1.3 Gbytes in size, when using BayeSED to interpret the observed photometric SEDs of galaxies.

\begin{figure}
  \centering 
  \includegraphics[scale=0.7]{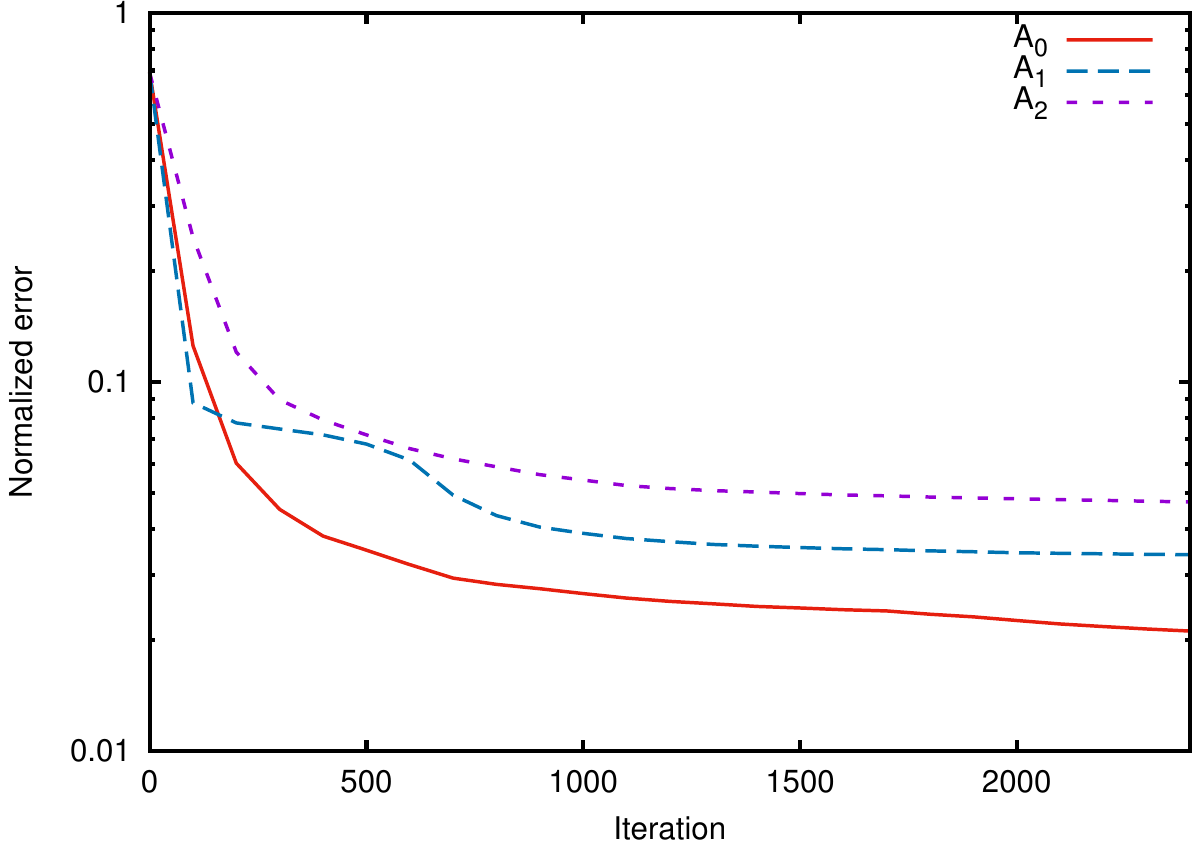}
  \caption
  {
  The normalized error on the `validation data' (from ~\La~ SED library) as a function of the number of iterations for the training of three ANNs with the structure as shown in Figure \ref{fig:ann}.
  }
  \label{fig:ann_train}
\end{figure}

\begin{figure}
  \centering 
  \includegraphics[scale=0.7]{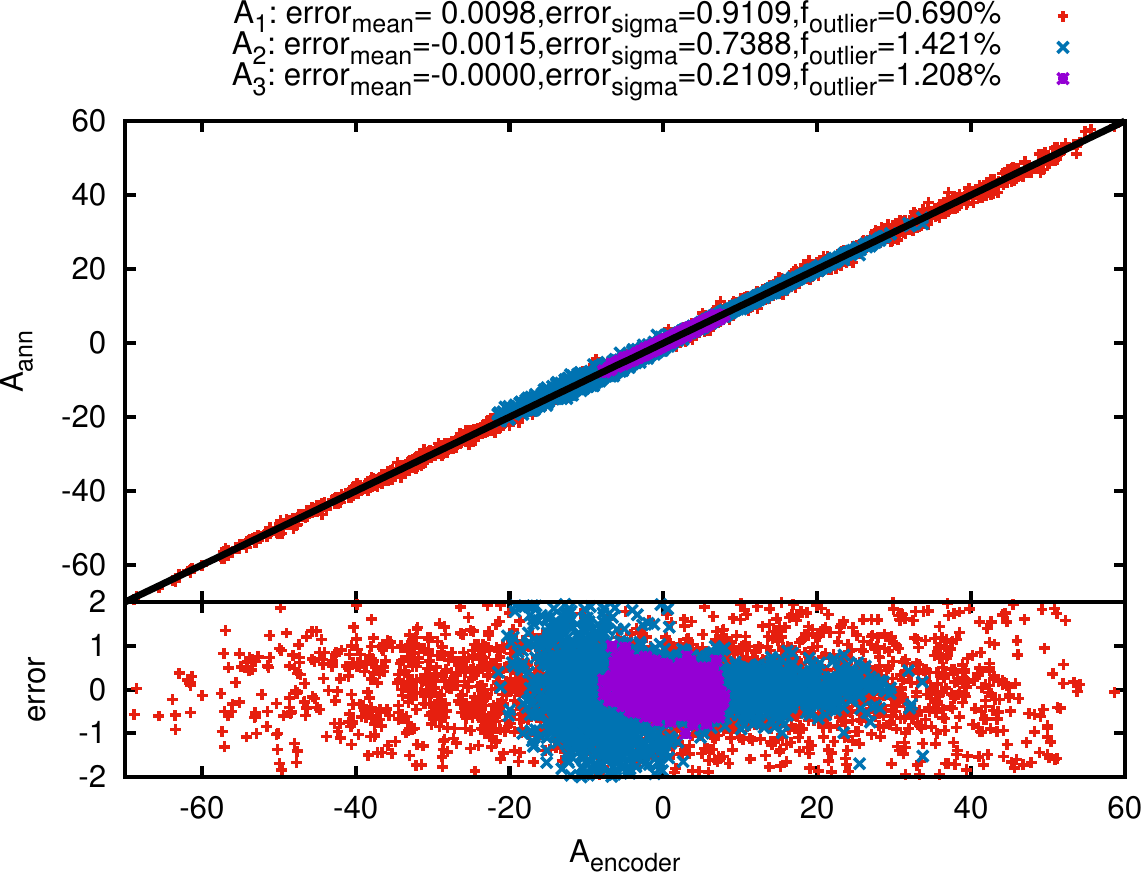}
  \caption
  {
  A test of the ANN interpolation method by comparing the amplitudes of principal components of the original SEDs in the ~\La~ SED library, which is obtained with `encoder', with that generated with ANN.
  The results are for the `testing data' set, which have not been used \hl{in any way} during the training of the ANNs.
  }
  \label{fig:test_ann_a}
\end{figure}

In Figure \ref{fig:ann_train}, we give an example of tracing the normalized error of three ANNs on the `validation data' as a function of the number of iteration during the training.
As shown in the figure, the normalized error of these ANNs have decreased dramatically during the first few hundreds of iterations.
This demonstrated the power of the RPROP algorithm for ANN training.
After about 1000 iterations, the decreasing of error becomes very slow, which means the training tend to be converged.
As mentioned before, the training will be stopped when the error start to increase.
It is worth to notice that the final error for the amplitudes of different principal components are slightly different.
It seems that lower order principal component can be `learned' better by the ANN.
This is good for us, since the lower order principal component is more important for the reconstruction of SED.
We can test the effectiveness of ANN method by comparing the amplitudes of principal components of the original SEDs in the ~\La~ SED library, which is obtained with `encoder', with that generated with ANNs.
This is shown in Figure \ref{fig:test_ann_a} for the `testing data' set.
As mentioned before, these data have not been used by all means during the training of these ANNs.
So, this kind of test should be very rigorous.
As shown in the figure, it is clear that the amplitudes of principal components of the original SEDs can be generated  pretty well by the ANNs.

\begin{figure}
  \centering 
  \includegraphics[scale=0.65]{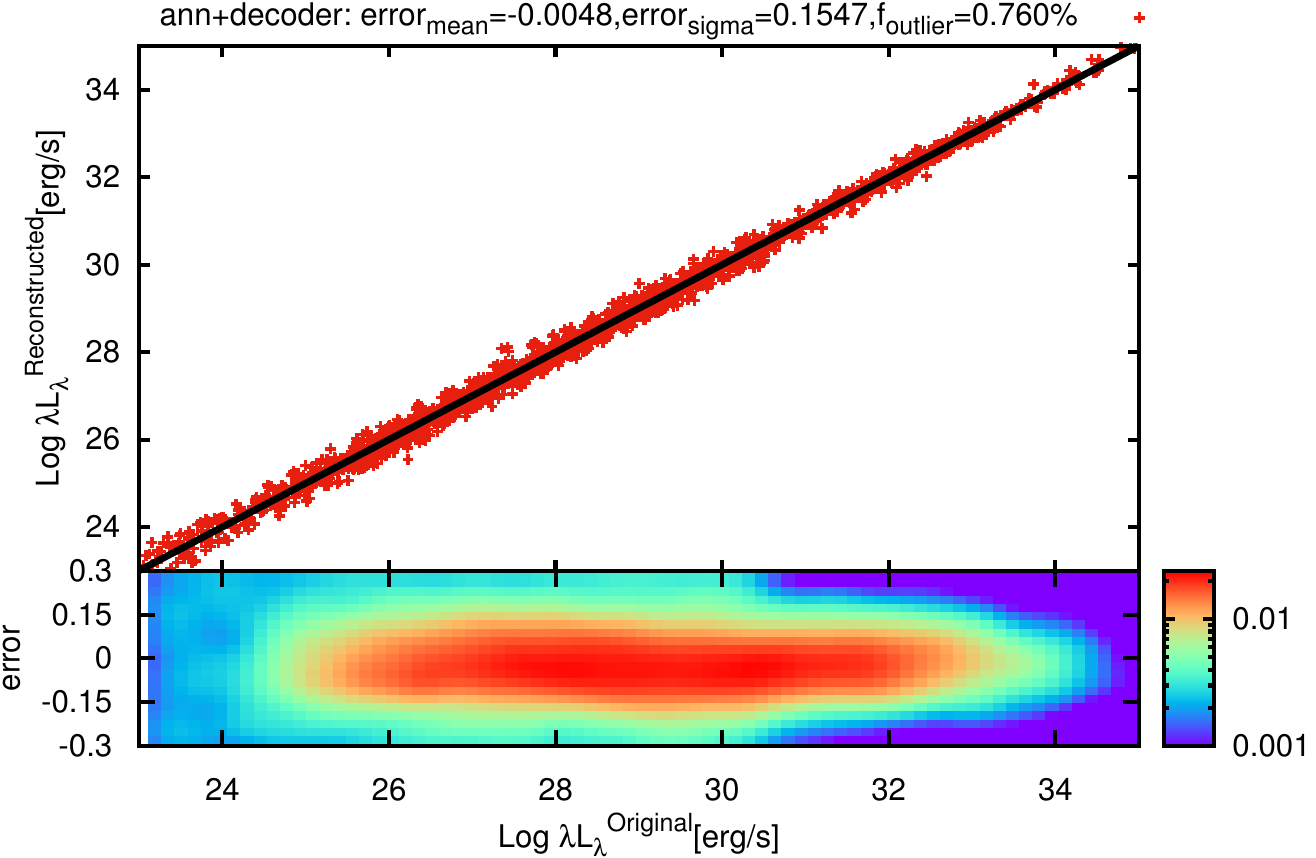}
  \caption
  {
  A test of the ANN interpolation method by comparing the original SEDs in the ~\La~ SED library with that reconstructed by `decoder' from the amplitudes of principal components generated with ANNs.
  \hl{  The probability density distribution of errors is shown in the lower panel.}
  \hl{  The mean, standard deviation, and the percentage of outliers (error$>3\sigma$) for the distribution are shown on the top of this figure.}
  }
  \label{fig:test_ann}
\end{figure}

\begin{figure}
  \centering 
  \includegraphics[scale=0.7]{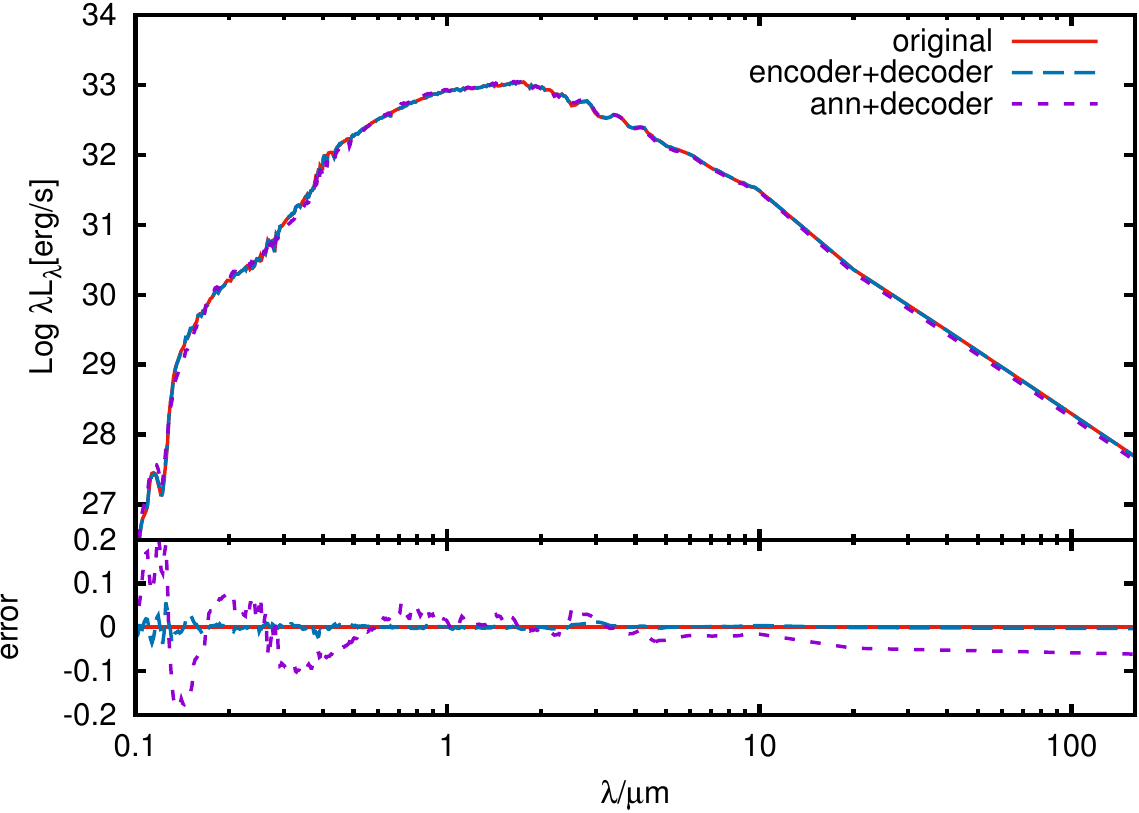}
  \caption{
  \hl{  An example of a SED that is reconstructed with the amplitudes of principal components generated with ANNs (ann+decoder), compared with that using the PCA method only (decoder+encoder) and the original one in the ~\La~ SED library.}
  \hl{  The error induced by PCA is negligible, while that induced by ANNs is larger and seems wavelength dependent.}
  \hl{  However, in most cases, the error is within $0.1$ dex.}
  }
  
  \label{fig:sed_ann}
\end{figure}

Finally, the amplitudes of principal components generated by ANNs can be used to reconstruct the SEDs.
This is the ultimate goal of using the ANN interpolation method.
In Figure \ref{fig:test_ann}, we test the effectiveness of this method by comparing the original SEDs in the ~\La~ SED library with that reconstructed by `decoder' from the amplitudes of principal components generated with ANNs.
As shown in the figure,  the errors for SEDs reconstructed by employing ANN method are slightly larger than that obtained with only the PCA method (Figure \ref{fig:test_pca}).
This is because the ANNs \hl{cannot} predict the amplitudes of principal components without any error.
However, the errors are still very small.
In Figure \ref{fig:sed_ann},  we give an example of SED reconstructed by employing ANN method, and compare it with that obtained using only PCA method and the original one in the ~\La~ SED library.
It is clear that the original SED can be reconstructed pretty well by the amplitudes of principal components generated with ANNs.
So, the method of ANN interpolation of SED library is very successful.

\subsubsection{Interpolation with K-Nearest Neighbors}
\label{sss:knn}
While the ANN method has been proved to be successful for the interpolation of SED library as presented in \S \ref{sss:ann}, there are some reasons for considering other methods for the interpolation of SED library.
Firstly, the network structure of ANN, including the number of hidden layers and the number of neurons in every hidden layer, need to be determined \hl{specially} for the SED library under consideration.
Although an ANN with not too many neurons in one hidden layer has been found to be enough for most problems, there are no simple and general rules for this determination.
Secondly, the training of ANNs may need too much time if the relationship between parameters and corresponding SEDs is too complex (highly nonlinear and/or uncontinuous).
Thirdly, although the trained ANNs may have a good general performance even for those instances that have not been used during the train, they cannot exactly reproduce those instances that do have been used for the training.
One of the methods that can overcome these shortcomings of the ANN method is the interpolation method with K-Nearest Neighbors (KNN).
We will use the KNN method as a complement to the ANN method.

The basic idea of the KNN method is very simple.
To \hl{evaluate} at an arbitrary point of the parameter space of an SED model, where only a limited number of results  have been given in a SED library, we only need to find the first $K$ nearest neighbors of that point and take the average of the results at these points as the result for that point.
The effectiveness of the KNN interpolation method largely relies on how we define the distance between two points, and how we find the first $K$ nearest neighbors.
Commonly, the distance between two points is defined as Euclidean distance:
\begin{equation}
  {D(X,Y)} = \sqrt {\sum\limits_{i = 1}^n {{{({x_i} - {y_i})}^2}} }
  \label{eq:distance}
\end{equation}
To efficiently find the first $K$ nearest neighbors of any point, the knowns points need to be preprocessed into a data structure.
Then, a look-up algorithm is applied to this structure to found the nearest neighbors.
For this, we have employed a modified version of the Nearest-Neighbor-Regression algorithm in the Shark machine learning library.
The library provides two algorithms for the look-up of the $K$ nearest neighbors in a possibly high-dimensional parameter space.
They are called 'SimpleNearestNeighbors' and `TreeNearestNeighbors', respectively.

The `SimpleNearestNeighbors' algorithm is a brute force algorithm, which just evaluate the distance between pair of points one by one.
So, the organization of known points is not important for this algorithm.
The `TreeNearestNeighbors' algorithm, nevertheless, need the known points to be organized as some kind of tree structure in advance.
Generally, the `TreeNearestNeighbors' algorithm is much faster than the `SimpleNearestNeighbors' algorithm if only the dimension of the data points is not too high.
The Shark machine learning library provides tree choices for the tree structure, named as KDTree, KHCTree, and LCTree, respectively.
They all belong to the binary space-partitioning tree.
The KDTree, standing for k-dimensional tree \citep{Friedman1977a}, is the most widely used algorithm for nearest-neighbor search.
It works well in low-dimensional data, but quickly loses its effectiveness as dimensionality increases.
The KHCTree and LCTree, standing for Kernel Hierarchical Clustering tree and Linear Cut tree, respectively, are more advanced tree structures (see the document of Shark library \footnote{\url{http://image.diku.dk/shark/doxygen_pages/html/classshark_1_1_k_h_c_tree.html\#details}}\textsuperscript{,}\footnote{\url{http://image.diku.dk/shark/doxygen_pages/html/classshark_1_1_l_c_tree.html\#details}} for more details about the two tree structures).
We practically found that the LCTree structure has the best performance for our case.

\begin{figure}
  \centering 
  \includegraphics[scale=0.7]{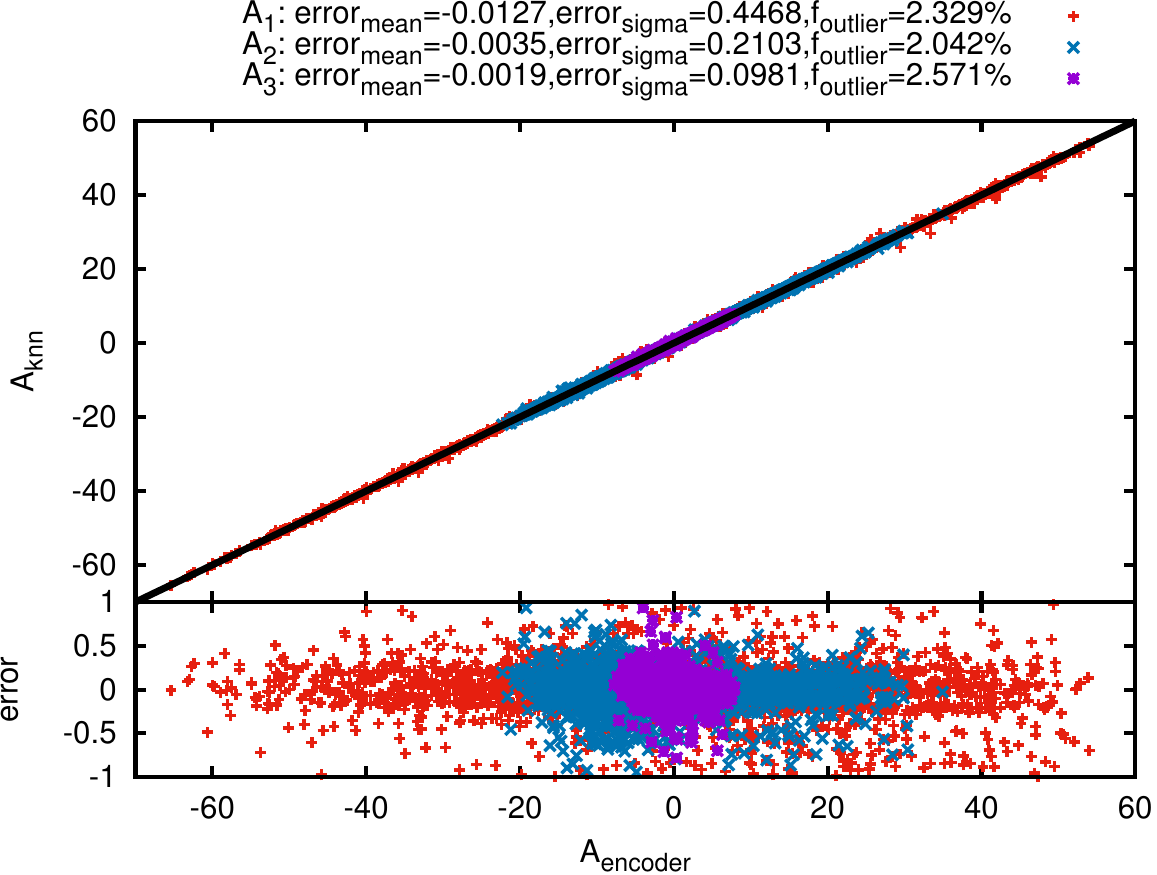}
  \caption
  {
  A test of KNN interpolation by comparing the amplitudes of principal components of the original SEDs in the ~\La~ SED library, which is obtained with `encoder', with that generated with KNN.
  }
  \label{fig:test_knn_a}
\end{figure}

\begin{figure}
  \centering 
  \includegraphics[scale=0.65]{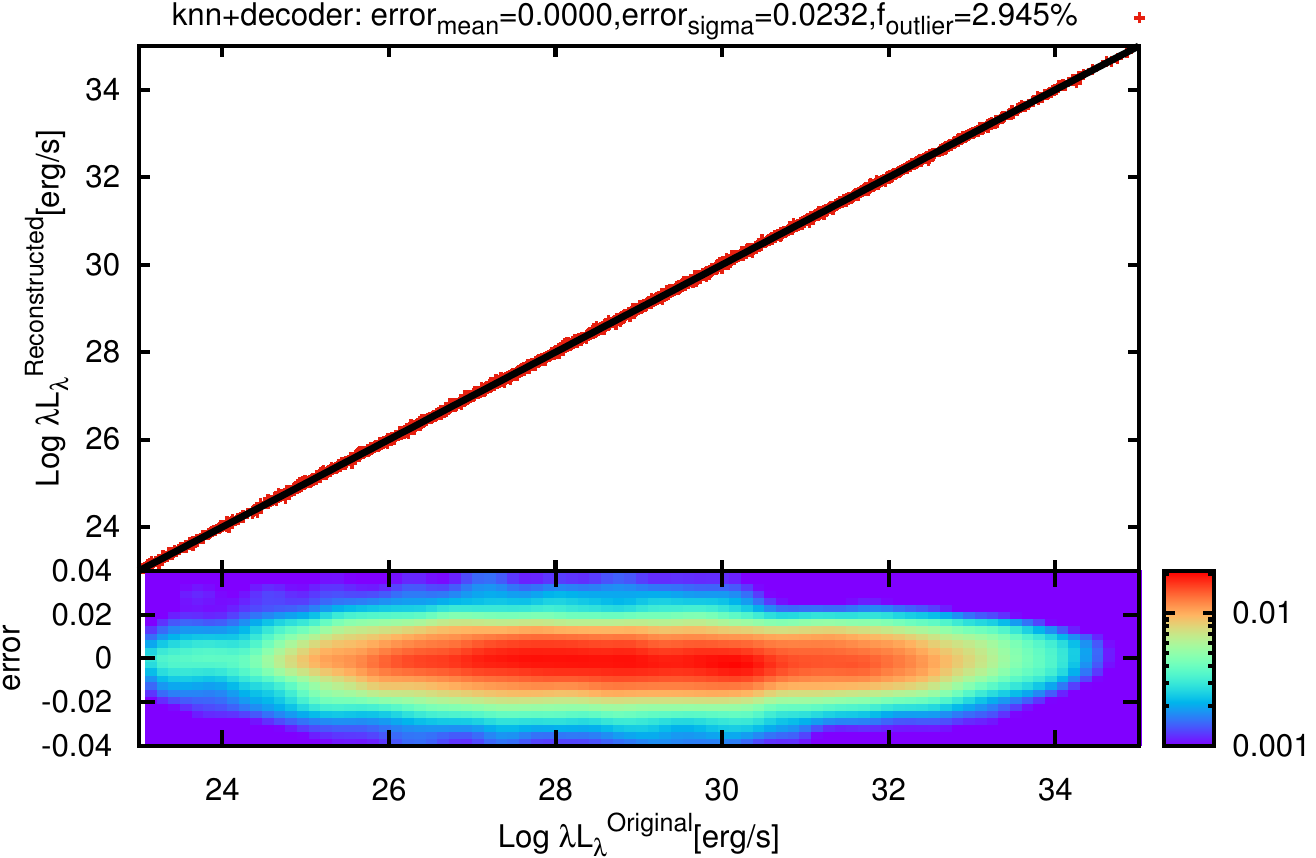}
  \caption{
  \hl{  A test of KNN interpolation by comparing the original SEDs in the ~\La~ SED library with that reconstructed by `decoder' from the amplitudes of principal components obtained with KNN.}
  \hl{  The error induced by KNN interpolation is much smaller than that induced by ANN interpolation (Figure \ref{fig:test_ann}), and only slightly larger than that induced by PCA alone (Figure \ref{fig:test_pca}).}
  }
  \label{fig:test_knn}
\end{figure}

As the ANN interpolation method in \S \ref{sss:ann}, we have applied the KNN interpolation method to predict the amplitudes of principal components.
It is possible for us to predict the corresponding SED directly from the given value of input parameters.
However, for KNN interpolation, we need to save all instances of a SED library to a file in the disk, and reload it into the memory of computer during the sampling of parameter space.
It is clear that the applicability of this method is largely limited by the size of the SED library.
As shown in \S \ref{sss:pca}, the size of a SED library can be reduced to only $5\%$ of the original by using PCA method.
So, by combined with the PCA method, the KNN interpolation method could be more useful in practice.
In Figures \ref{fig:test_knn_a} and \ref{fig:test_knn}, we show a test of KNN interpolation method, which is similar to that for ANN method in Figures \ref{fig:test_ann_a} and \ref{fig:test_ann}.
We should mention that the instances used for this test has not been used to build the LCTree that are use for interpolation.
So, as the case for the test of ANN method using `testing data', this should be a rigorous test.
It is clear that for both the amplitudes of principal components and the reconstructed SEDs, the KNN interpolation method could be even better than the ANN interpolation method.

There is no doubt that the KNN interpolation method has some advantages over ANN interpolation method.
For KNN interpolation, an intensive training process, which is crucial for ANN, is unnecessary.
We only need to store the known instances of a model properly in a data structure that is convenient for searching.
However, this does not mean that KNN method is better than ANN method in all aspects.
For example, the ANN interpolation is much faster than KNN interpolation during the sampling of parameter space.
Besides, the size of data that need to be store for KNN interpolation (e.g., $46$ Mbyte for the ~\La~ SED library) is much larger than that for ANN ($566$ Kbytes).
\hl{Furthermore}, the KNN method is much more sensitive to outliers and local structure of the data than ANN method.
So, in practice we use KNN method as a complement to the ANN method.
Then, it would be worth to check whether the results could be different by using the two methods.

\subsection{The building-up of BayeSED code}
\label{ss:bayesed}
As the last part of this section, we introduce how the MultiNest (\S \ref{ss:sampling}) and ANN (\S \ref{sss:ann}) or KNN (\S \ref{sss:knn}) algorithm are combined to build up our BayeSED code for interpreting the multi-wavelength SEDs of galaxies.

\begin{figure*}
  \centering 
  \includegraphics[scale=0.4]{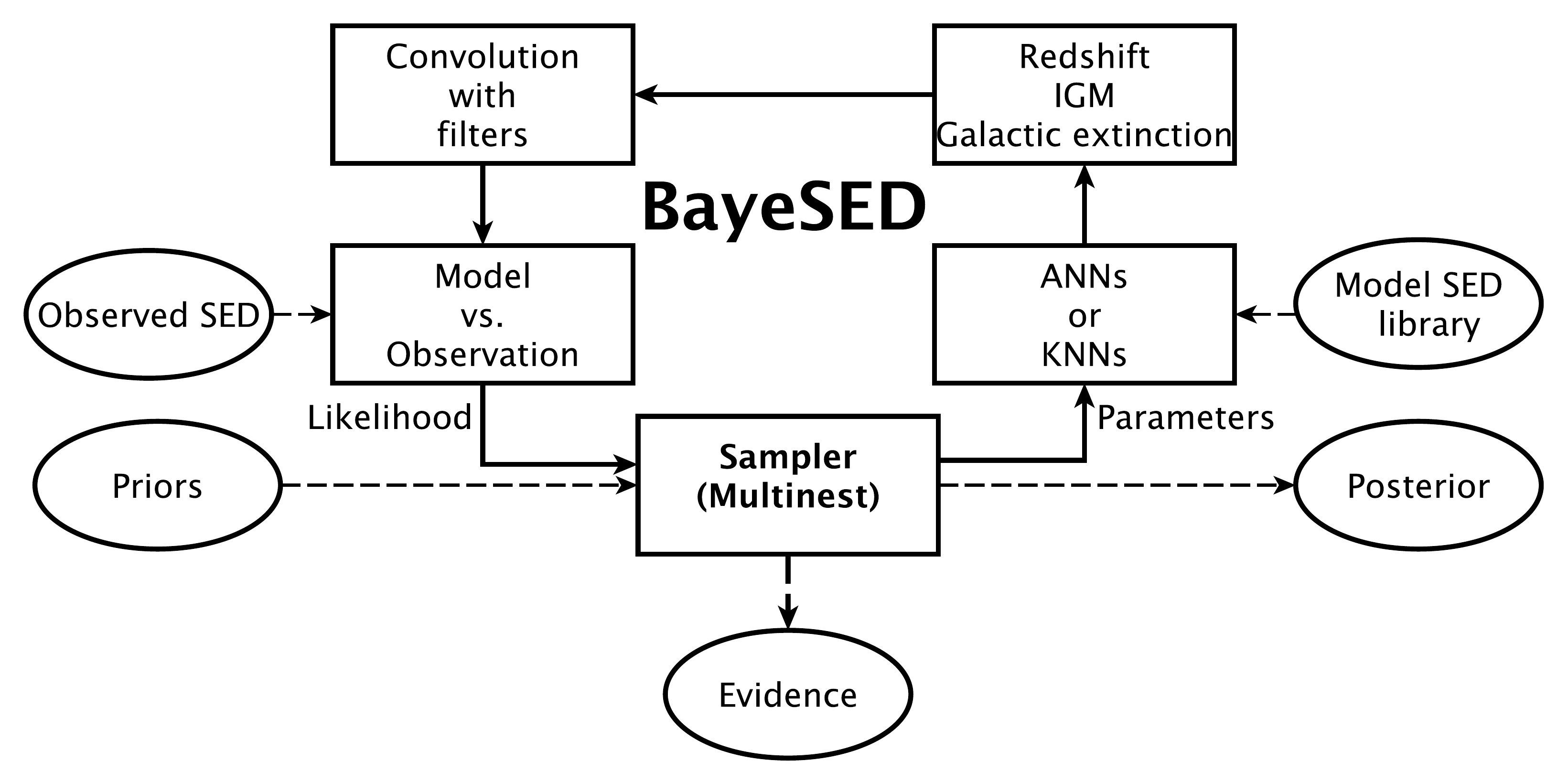} 
  \caption
  {
  The flowchart for interpreting the SED of galaxies with BayeSED. 
  Since BayeSED has been fully parallelized with MPI, this kind of analysis could be done for multiple galaxies, simultaneously.
  }
  \label{fig:bayesed}
\end{figure*}

In Figure \ref{fig:bayesed}, the flowchart of BayeSED code is shown.
The main inputs of BayeSED are the observed multi-wavelength photometric SED, including measurement errors, of a galaxy that needs to be interpreted.
On the other hand, the priors for the SED model that is used to explain the observations are considered as additional inputs.
This includes the allowed ranges for all free parameters of the model, and corresponding distributions of them.
Currently, the distribution of a free parameter is only allowed to be uniform.
So, for parameters with a large dynamic range, the logarithm of them should be used as free parameters instead.
In the future, we plan to provide more choices for the distribution of priors, including those that are physically more informative.

The sampling with MultiNest of the parameter space of a SED model that is used to explain the observations lies at the heart of BayeSED.
During the sampling, the MultiNest sampler continuously request that the likelihood function at a specific point of the parameter space should be computed, \hl{until} the resulting posterior and Bayesian evidence are thought to be converged.
The computation of likelihood at a give point involves the computation of the model SED at that point.
This is achieved by using the ANN algorithm (\S \ref{sss:ann}) or KNN algorithm (\S \ref{sss:knn}) to interpolate a pre-computed SED library.
As mentioned before, the computation of model SED is a major bottleneck for an efficient sampling of the possibly high-dimensional and complex parameter space of a SED model.
Thanks to the ANN and KNN algorithm, this can be done very quickly in our BayeSED code.
The huge SED libraries are only used to build the ANNs or KNNs once and for all.
So, they are no longer needed during the sampling of parameter space, while a small file including all necessary \hl{information} is used instead.

When a model SED is generated with ANNs or KNNs, the effects of cosmological redshift and intergalactic medium (IGM) on the SED are further considered.
In the current version of BayeSED, we provide two options for considering the effects of IGM.
They are based on the prescription of \cite{Madau1996a} and \cite{Meiksin2006a}, respectively.
Optionally, the effect of Galactic dust reddening and extinction \hl{could} be considered by setting the value of $E(B-V)$ at the position of the object, where the R-dependent Galactic extinction curve of \cite{Fitzpatrick1999a} with the ratio of total to selective extinction $R(V)=A(V)/E(B-V)=3.1$ is used.
The redshift of galaxy is considered as an optional free parameter for the fitting of SED.
Then, the redshift and other physical parameters of a galaxy can be obtained simultaneously with our BayeSED code.
However, BayeSED is not optimized for the determination of redshift, while many publicly available codes have been designed for that purpose.
So, the redshift of galaxies could be determined by using other codes and then used in BayeSED.
We will test the reliability of using BayeSED to determine the redshift of galaxies in \S \ref{s:application_real}.
Finally, the model SED is convolved with the transmission function of filters to obtain model fluxes that are directly comparable with multi-wavelength observations.

The value of likelihood at a specific point of the parameter space as requested by the MultiNest sampler is obtained by the comparison between the model fluxes and the corresponding multi-wavelength observations.
Commonly, the distribution of observational errors are assumed to be Gaussian. 
Then, the normalized likelihood function is defined as:
\begin{equation}
  \mathcal{L}(\bm \theta ) = \prod\limits_{i = 1}^{i = n} {\frac{1}{{\sqrt {2\pi } {\sigma _{o,i}}}}\exp ( - \frac{1}{2}\frac{{{{({F_{o,i}} - {F_{m(\bm \theta ),i}})}^2}}}{{\sigma _{o,i}^2}})},
 \label{eq:likelihood1}
\end{equation}
where $\sigma _{o,i}$ is the observational error in the i-th band, $F_{o,i}$ and $F_{m(\bm \theta ),i}$ are the observed flux and model flux in the i-th band, respectively.
In practice, the term ${1/\sqrt {2\pi } {\sigma _{o,i}}}$ is usually omitted, since it is independent of the shape of the likelihood function.
So, in most works of Bayesian SED fitting, the definition of likelihood function is simplified as:
\begin{equation}
  \mathcal{L}(\bm \theta ) = \prod\limits_{i = 1}^{i = n} {\exp ( - \frac{1}{2}\frac{{{{({F_{o,i}} - {F_{m(\bm \theta ),i}})}^2}}}{{\sigma _{o,i}^2}})}.
 \label{eq:likelihood2}
\end{equation}

\hl{In the above definition of likelihood, the observations at different wavelength bands are assumed to be independent and only observational error have been considered.}
\hl{The possible systematic error of the SED model, which could be important, especially for population synthesis models \citep[][]{Conroy2009a,Cervino2013b}, have not been considered yet.}
\hl{The systematic error of a SED model is likely wavelength and model dependent, and so not easy to be considered properly.}
\hl{In the EAZY code \citep{Brammer2008a}, this has been considered as a template error function.}
\hl{However, it is not clear how universal this kind of error function could be.}

\hl{On the other hand, thanks to the application of PCA, an SED can be described by the amplitudes of principal components.}
\hl{Then, the likelihood function could be defined as:}
\begin{equation}
  \mathcal{L}(\bm \theta ) = \prod\limits_{i = 1}^{i = N} {\exp ( - \frac{1}{2}\frac{{{{({A_{o,i}} - {A_{m(\bm \theta ),i}})}^2}}}{{\sigma _{o,i}^2}})},
 \label{eq:likelihood3}
\end{equation}
\hl{where N is number of principal components.}
\hl{In this approach, the model SEDs need not to be reconstructed from the amplitudes of principal component over and over again, while the observed SED needs to be projected to the principal components only once.}
\hl{This is especially useful for the analysis of spectroscopic data \citep{Chen2012a}, where the dimension of the data is much larger than the number of necessary principal components.}
\hl{However, this is less helpful for the analysis of photometric data, where the dimension of the data is comparable to the number of necessary principal components.}
\hl{Besides, it is not so straightforward to project the sparsely sampled photometric SED to the principal components \citep[see also][]{Wild2014a}.}

By using the BayeSED code to interpret the multi-wavelength SED of a galaxy, many outputs could be obtained.
Firstly, we can obtain the Bayesian evidence of the model which is used to explain the observed SED.
Secondly, we can obtain the estimation of all parameters of the model.
\hl{As mentioned in \S \ref{ss:bayes}, the posterior PDF including all information about the parameters.}
However, in practice, it is not possible to report the results of Bayesian parameter estimation with full PDF, especially for a large sample of galaxies.
So, it is very necessary to use some summary statistics instead.
In BayeSED, we provide many summary statistics about a parameter, including mean, median, maximum-a-likelihood (MAL, or best-fit), and maximum-a-posteriori (MAP).
The corresponding error of a parameter is estimated with standard deviation or percentiles of the PDFs.
Besides, the best-fit model SED and corresponding amplitudes of principal components, rest-frame absolute magnitudes, observed-frame apparent magnitudes could be optionally outputted.
Finally, it is worth to mention that BayeSED has been fully parallelized with MPI\footnote{The MPI implemented in the MultiNest algorithm itself is switched off, since we found that it is not efficient for multiple SED-fitting.}.
So, it is possible to interpret the multi-wavelength SEDs of a large sample of galaxies simultaneously, and all results are saved into a single file.

\section{Application to a mock sample of galaxies}
\label{s:application_mock}
As mentioned in \S \ref{s:intro}, by interpreting the SEDs  of galaxies, we try to solve the inverse problem of SED \hl{modeling}: to derive the physical parameters of galaxies from their observed multi-wavelength photometric SEDs.
The ability of a SED-fitting code to solve this problem can be properly tested by using mock samples of galaxies.
So, before applying the BayeSED code to interpret the SEDs of galaxies in a real sample, we will test the reliability of it in this section.

\subsection{Building of Mock Sample of Galaxies}
\label{ss:mock_build}
The starting point for the building of a mock sample of galaxies are a set of model SEDs of galaxies, which are obtained with a model of galaxy SED such as the evolutionary population synthesis model.
\hl{Then, these model SEDs are transformed according to that would be experienced of real galaxies to obtain mock fluxes at multiple wavelength bands.}
\hl{Since the true value of all parameters of the mock observations are known in advance, it is easy to check if they can be recovered properly by a SED-fitting code.}

In our case, we started from the SED libraries as built up in \S \ref{sss:eps} to make four mock samples of galaxies.
We have taken the bc03 model with a \cite{Salpeter1955a} IMF as an example.
From the SED library, a total of 10000 model SEDs are randomly selected to make a mock sample with 10000 galaxies.
These model SEDs are shifted to a random  redshift $z$ ranging from 0 to 6, while also consider the effect of IGM.
We demand that the age of a galaxy must be smaller than the age of Universe at that redshift.
Then, the model SEDs are convolved with the transmission function of filters to obtain model fluxes.
Finally, some random noises with a Gaussian distribution are added to these model fluxes.
\hl{The Gaussian distribution has a zero mean and a dispersion equal to $10\%$ (S/N) of the model flux.}
The filters and corresponding errors are selected to mimic the Ks-selected sample of galaxies in COSMOS/UltraVISTA field that will be studied in the next section.

\hl{It should be mentioned that there are other methods to build more realistic mock sample of galaxies.}
For example, the distribution of luminosity and redshift \hl{could} be drawn from a luminosity function of galaxies, or the distributions of physical parameters of galaxies \hl{could} be that predicted by a model for the formation and evolution of galaxies.
However, these more realistic mock samples are not very necessary for a reasonable test of a SED-fitting code.
A good SED-fitting code should be able to properly recover the physical parameters of galaxies from their multi-wavelength observations, regardless of how these parameters are distributed.
Nevertheless, we should keep in mind that not all physical parameters  of galaxies can be recovered equally well, even though the best possible SED-fitting code has been used.
Besides the code itself, there are many other factors that can take part in determining the possibility of recovery.
For example, the number of available filters and corresponding S/N, the relative importance of a parameter for determining the shape of SED, the degeneracies between parameters, and so on.
\hl{So, the mock sample of galaxies is only used to check the internal consistency of BayeSED and the effects of intrinsic degeneracies between parameters of a SED model \citep[see e.g.][]{Walcher2008a}.}

\subsection{Interpreting and Results}
\label{ss:mock_results}
We have applied our BayeSED code to interpret the mock sample of galaxies that are built up in \S \ref{ss:mock_build}.
For the interpolation of model SEDs, both ANN and KNN methods have been used, and the results obtained will be compared here.
It is worth to mention that the results presented here represent an overall verification of everything that are involved in the BayeSED code.
Since the mock sample is built up with the original SED library, all potential errors that are hidden in the programing of code, or the PCA, ANN, KNN and MultiNest algorithms could propagate into the final results.

\begin{figure*}
  \centering 
  \includegraphics[scale=1.4]{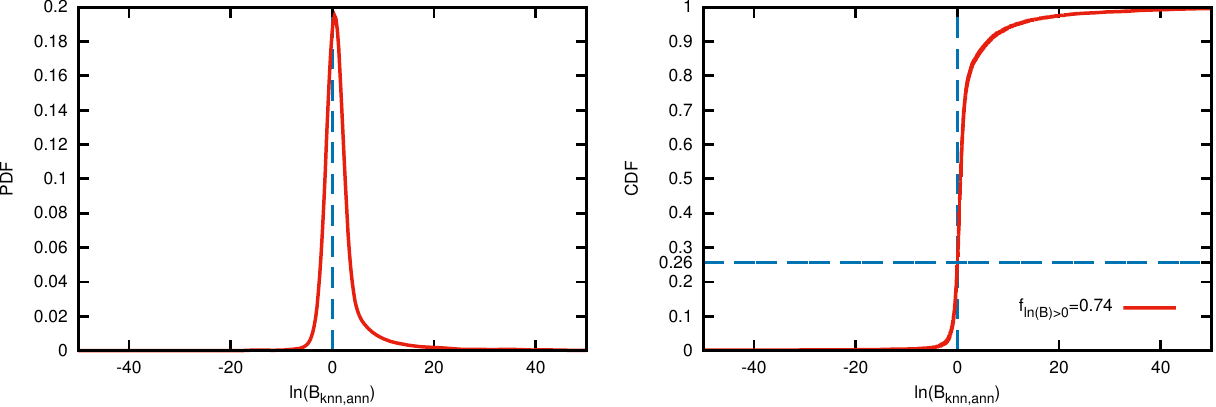} 
  \caption
  {
  \hl{  The probability density distribution function (PDF) and cumulative distribution function (CDF) of the Bayes factor $\rm ln(B_{knn,ann})$ for the mock sample of galaxies.}
  The bc03 model with a \cite{Salpeter1955a} IMF was used in this example.
  \hl{  The KNN method gives a much better approximation to the original bc03 model than the ANN method, since only $26\%$ of the mock galaxies favor the latter.}
  }
  \label{fig:ev_mock}
\end{figure*}

Since the ANN and KNN methods are used to approximate the original bc03 model, they can be considered as two special versions of that model.
So, it is meaningful to check the Bayesian evidences of them for the mock observations that are built up from the original bc03 model.
\hl{In Figure \ref{fig:ev_mock}, we show the probability density distribution function (PDF)\footnote{The PDFs hereafter in this paper are  obtained with the method of kernel density estimation.} and cumulative distribution function (CDF) of Bayes factor $\rm ln(B_{knn,ann})$ for the mock sample of galaxies.}
It is clear from the figure that the results obtained with KNN method has larger Bayesian evidence than that obtained with ANN method.
Except for the method of the interpolation, anything else is the same \hl{for the two approaches}.
So, this indicate that KNN method should be a better approximation to the original bc03 model.

\begin{figure}
  \centering 
  \includegraphics[scale=0.7]{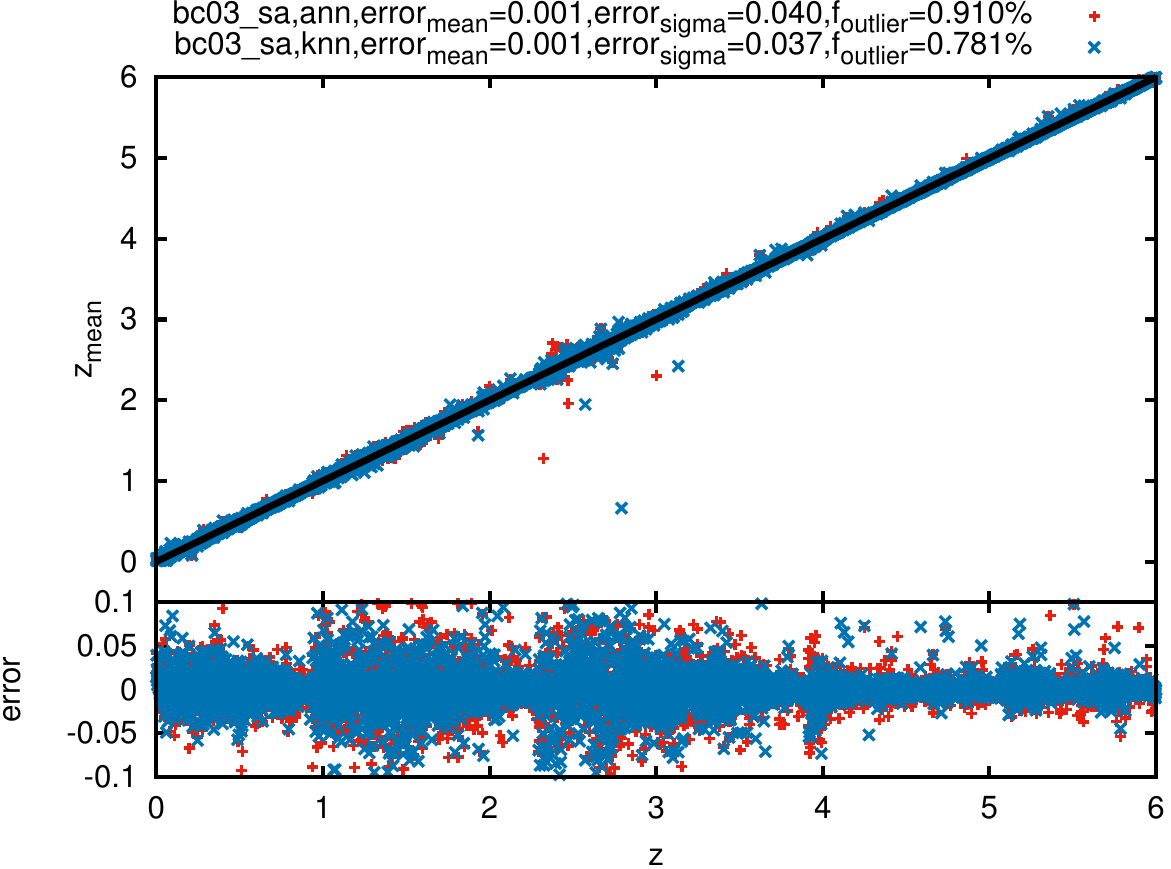} 
  \caption
  {
  The recovered values of redshift ($z_{\rm mean}$) compared with the true values ($z$) of the mock sample of galaxies.
  The \hl{redshifts} and corresponding \hl{errors} are estimated from the mean and standard deviation of the posterior distribution.
  The bc03 model with a \cite{Salpeter1955a} IMF was used in this example.
  \hl{  On the top of this figure, the mean, dispersion and the fraction of outliers with an error larger than $3\sigma$ of the distribution of errors are shown.}
  }
  \label{fig:mock_test_z}
\end{figure}

\begin{figure}
  \centering 
  \includegraphics[scale=0.7]{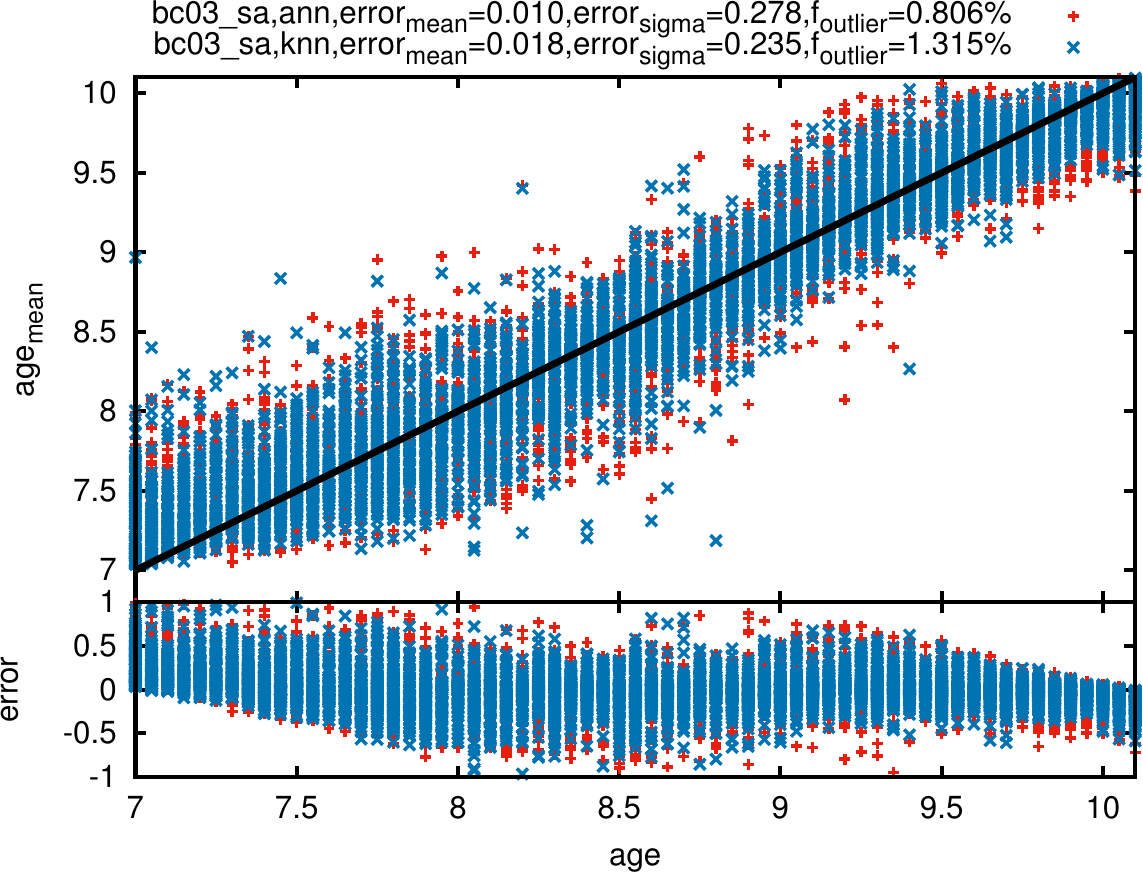} 
  \caption
  {
  Similar to Figure \ref{fig:mock_test_z}, but for the age of galaxies.
  \hl{  The age tends to be overestimated at the lower end while underestimated at the higher end.}
  \hl{  This is mainly caused by the limited parameter range.}
  }
  \label{fig:mock_test_age}
\end{figure}

\begin{figure}
  \centering 
  \includegraphics[scale=0.7]{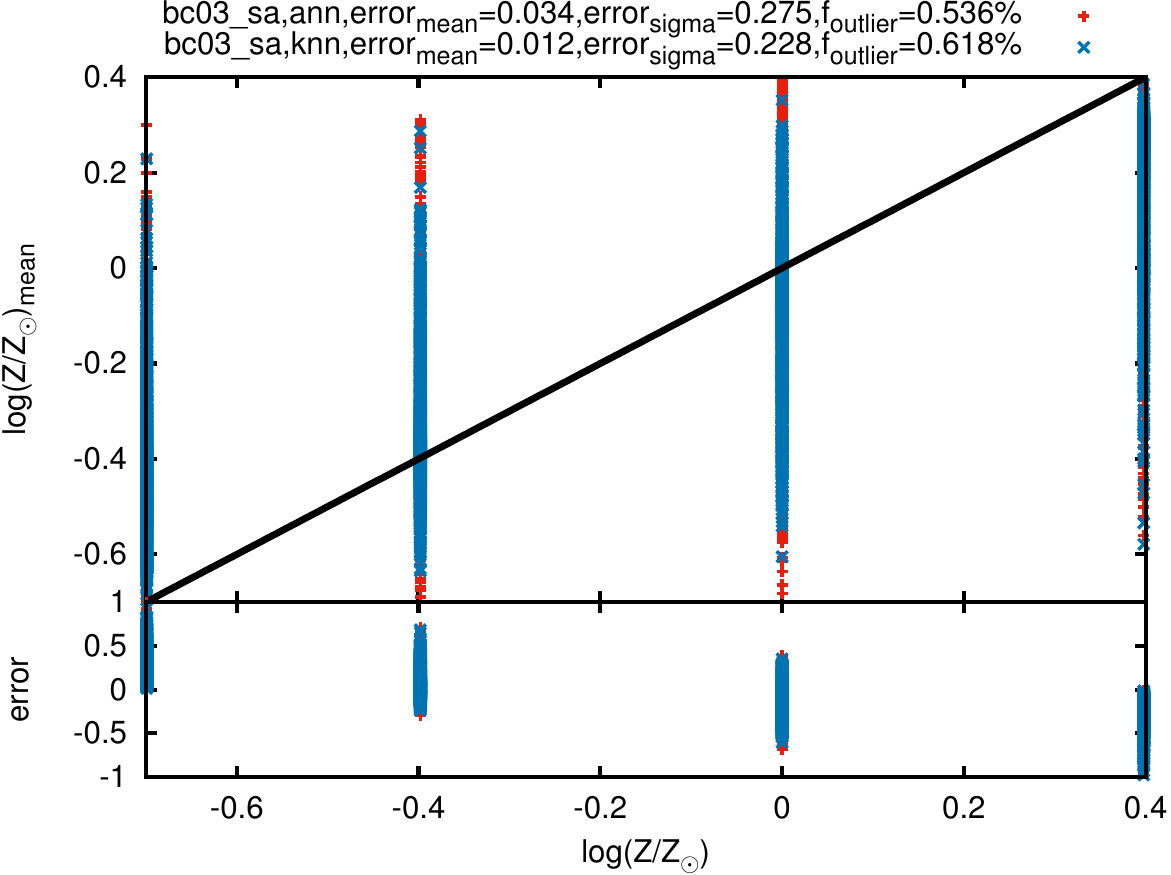} 
  \caption
  {
  Similar to Figure \ref{fig:mock_test_z}, but for the metallicity ($\rm Z$) of galaxies.
  \hl{  The anti-correlation of errors with the true values for this poorly constrained parameter is mainly caused by the limited parameter range (see text for more discussions about this).}
  }
  \label{fig:mock_test_Zm}
\end{figure}

\begin{figure}
  \centering 
  \includegraphics[scale=0.7]{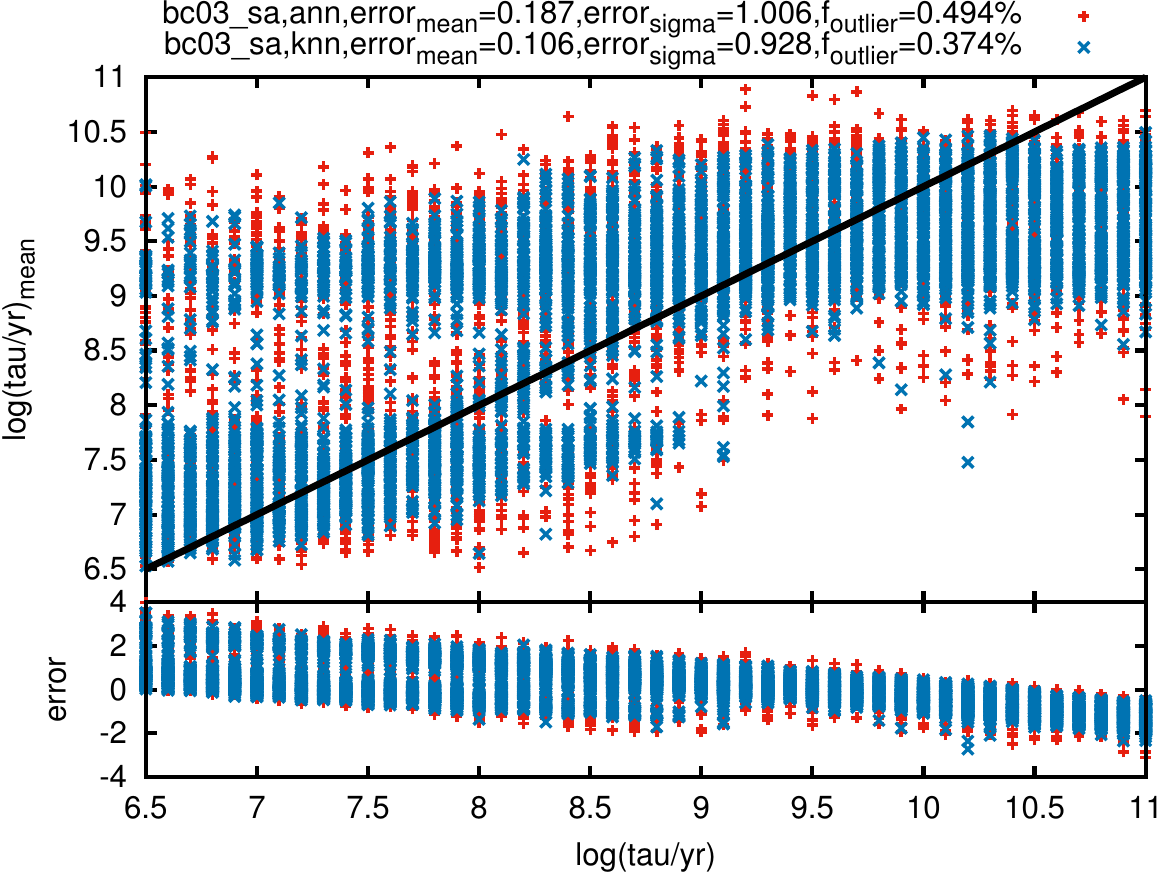} 
  \caption
  {
  Similar to Figure \ref{fig:mock_test_z}, but for the e-folding time ($\rm tau$) of the star formation history of galaxies.
  }
  \label{fig:mock_test_tau}
\end{figure}

\begin{figure}
  \centering 
  \includegraphics[scale=0.7]{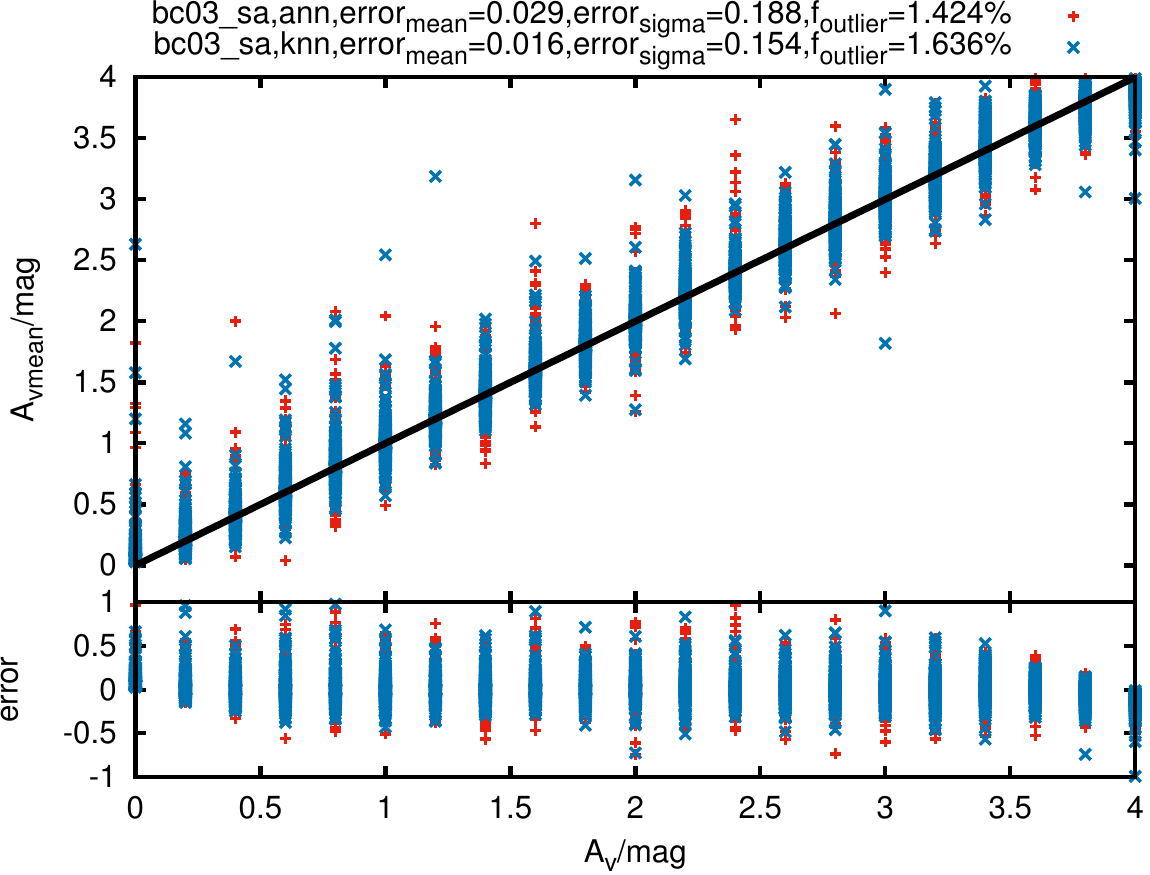} 
  \caption
  {
  Similar to Figure \ref{fig:mock_test_z}, but for the dust extinction ($A_{\rm v}$) of galaxies.
  }
  \label{fig:mock_test_Av}
\end{figure}

\begin{figure*}[htp]
  \centering
  \includegraphics[scale=0.90]{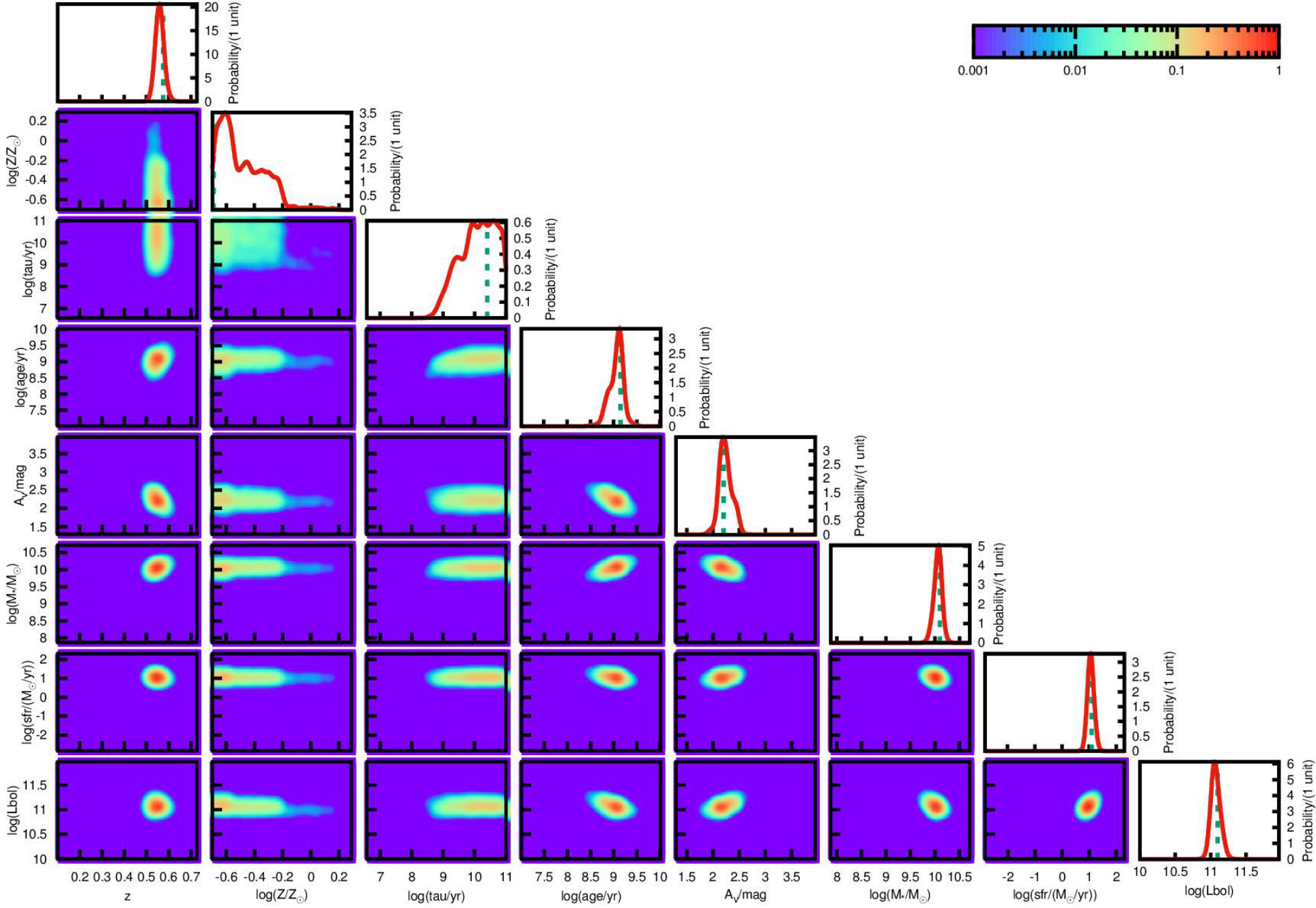}
  \caption
  {
  The posterior probability density functions (PDFs) of parameters for a mock galaxy with $z=0.58$, $\rm log(Z/Z_{\odot})=-0.70$, $\rm log(tau/yr)=10.4$, $\rm log(age/yr)=9.15$, $\rm A_{\rm v}/mag=2.2$, $\rm log(M_*/M_{\odot})=10.1$, $\rm log(sfr/(M_{\odot}/yr))=1.1$, and $\rm log(Lbol/L_{\odot})=11.1$.
  \hl{  Except for the metallicity (Z) and e-folding time ($\rm tau$), the PDFs of all other parameters show a sharp peak around the true values.}
  \hl{  The PDFs of $\rm log(age/yr)$ and $\rm A_{\rm v}/mag$ show a weak second peak, indicative of the degeneracy between them.}
  \hl{  Meanwhile, the PDFs of $\rm log(Z/Z_{\odot})$ and $\rm log(tau/yr)$ show even more peaks, indicative of serious degeneracies with other parameters.}
  }
  \label{fig:mock_dist}
\end{figure*}

\hl{In Figures \ref{fig:mock_test_z}, \ref{fig:mock_test_age}, \ref{fig:mock_test_Zm}, \ref{fig:mock_test_tau}, and \ref{fig:mock_test_Av}, we compare the recovered values of redshift, age, metallicity, e-folding time, and dust extinction with their true values.}
\hl{Among these, the redshift can be best recovered.}
\hl{For both the ANN and KNN methods, the mean for the distribution of errors are almost zero, while the dispersion and the fraction of outliers with an error larger than $3\sigma$ is also very small.}
\hl{The age and dust extinction of galaxies can be recovered moderately well.}
\hl{This can be better understood from the posterior probability density functions (PDFs) of them for a mock galaxy in Figure \ref{fig:mock_dist}.}
\hl{The PDFs of $\rm log(age/yr)$ and $\rm A_{\rm v}/mag$ show a weak second peak, indicating the degeneracy between them.}
\hl{Finally, the metallicity and e-folding time cannot be recovered properly.}
\hl{This seems because these two parameters are less important for shaping the SED, and so more easily be affected by errors in the observations and the degeneracies with other parameters.}
\hl{As can be noticed in Figure \ref{fig:mock_dist}, the PDFs of $\rm log(Z/Z_{\odot})$ and $\rm log(tau/yr)$ show many peaks, indicating serious degeneracies with other parameters.}
\hl{Meanwhile, it is clear from Figures \ref{fig:mock_test_Zm} and \ref{fig:mock_test_tau} that the distribution of errors for them show a clear anti-correlation with the true value.}
\hl{This kind of trends just imply that the two parameters can not be constrained effectively by the observations, but mainly constrained by the allowed range of them.}
\hl{Since a flat prior is assumed and the parameters are estimated with the mean of PDF, the parameters tend to be overestimate at the lower end and underestimated at the higher end.}
\hl{In practice, we found that if the redshift is fixed to the right value, the two parameters can be recovered much better.}
\hl{However, it seems still difficult to recovery them properly with photometric data only.}
\begin{figure}
  \centering 
  \includegraphics[scale=0.7]{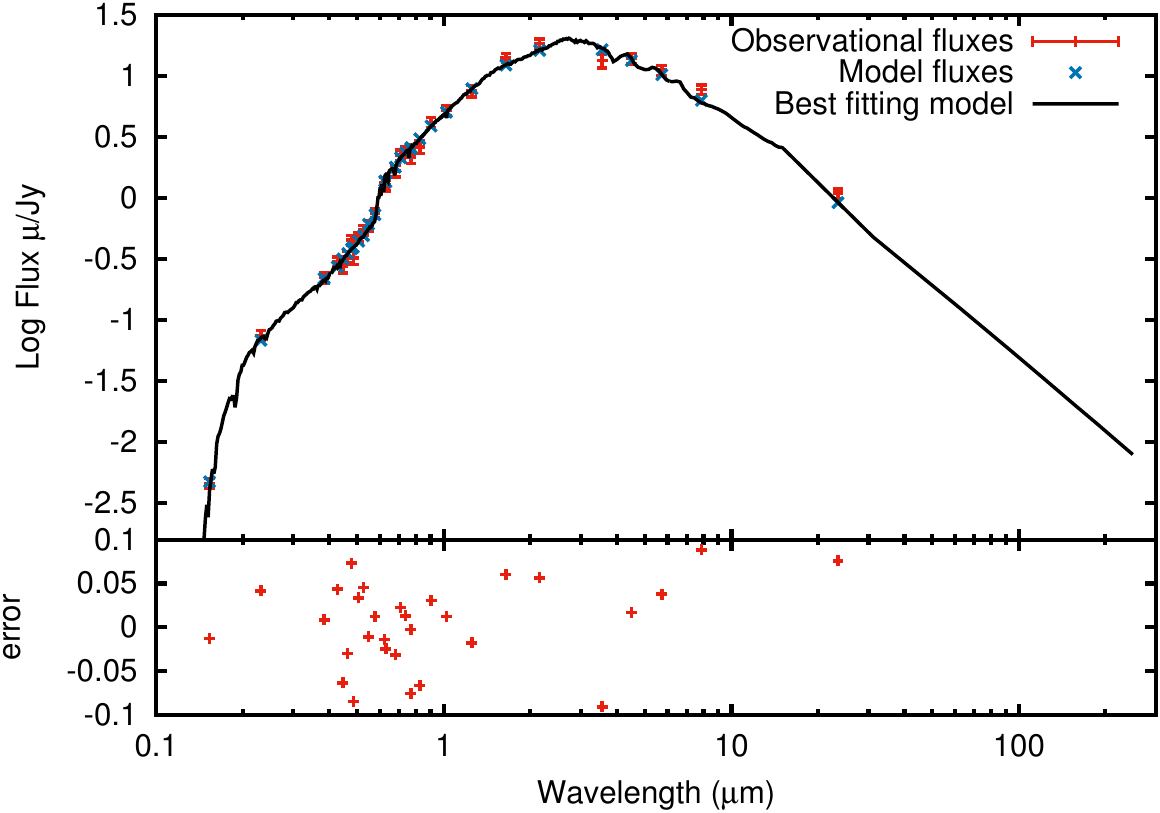} 
  \caption
  {
  \hl{  The best fitting results for the same mock galaxy as in Figure \ref{fig:mock_dist}.}
  }
  \label{fig:mock_bestfit}
\end{figure}

With the estimated value of free parameters of an evolutionary population model, we are able to derive other parameters, such as the stellar mass and star formation rate.
However, it would be more convenient to be able to estimate the free parameters and other derived parameters simultaneously.
This is allowed in our BayeSED code.
We achieved this by building another set of ANNs or KNNs to derive other parameters from the free parameters of the SED model.
This is similar to that for SED, except that the output of ANNs or KNNs are the derived parameters instead of SED.
With this method, we are able to simultaneously estimate any number of parameters that are derived from the free parameters of the SED model.

\begin{figure}
  \centering 
  \includegraphics[scale=0.7]{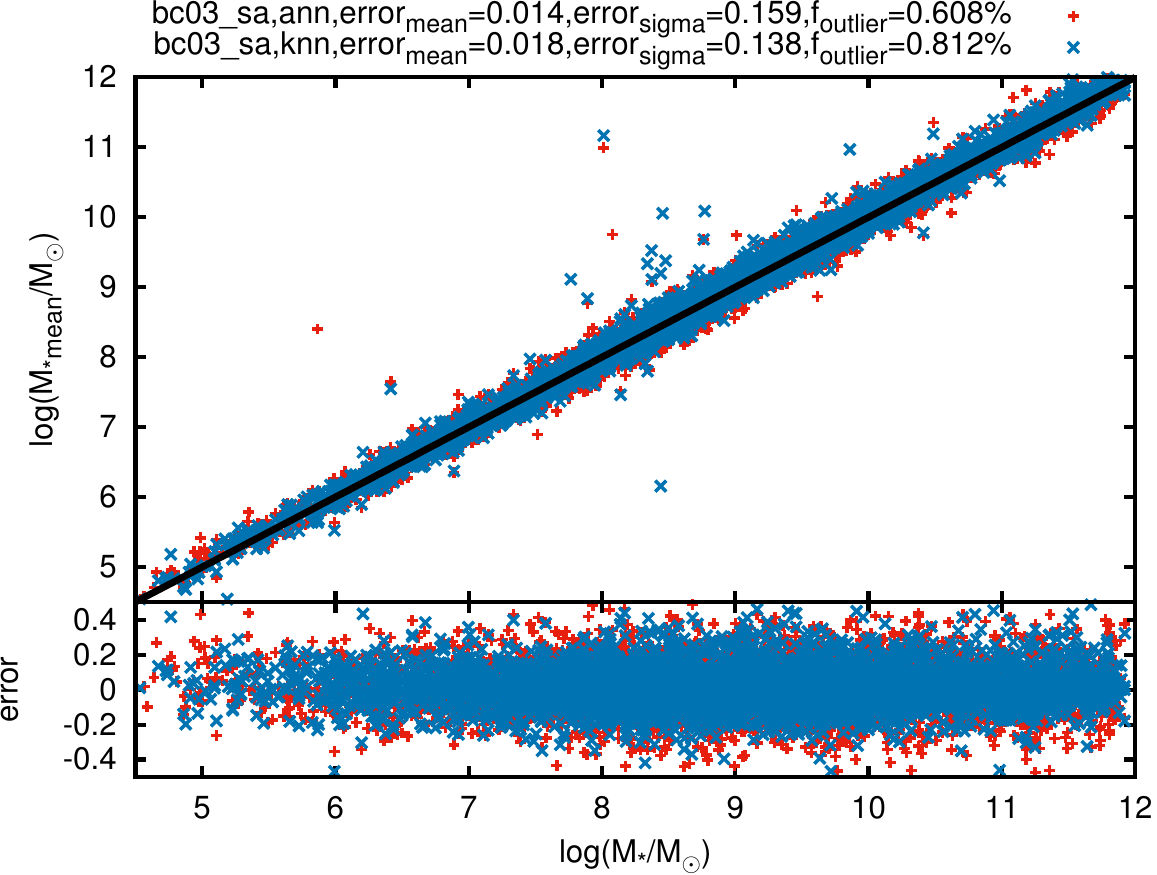} 
  \caption
  {
  The derived stellar mass ($\rm mass_{mean}$) compared with the true values ($\rm mass$) of the mock sample of galaxies.
  The \hl{stellar mass is estimated} with the mean of the posterior distribution of stellar mass.
  The bc03 model with a \cite{Salpeter1955a} IMF was used in this example.
  \hl{On} the top of this figure, the mean, dispersion and the fraction of outliers with an error larger than $3\sigma$ of the distribution of errors are shown.
  }
  \label{fig:mock_test_mass}
\end{figure}

\begin{figure}
  \centering 
  \includegraphics[scale=0.7]{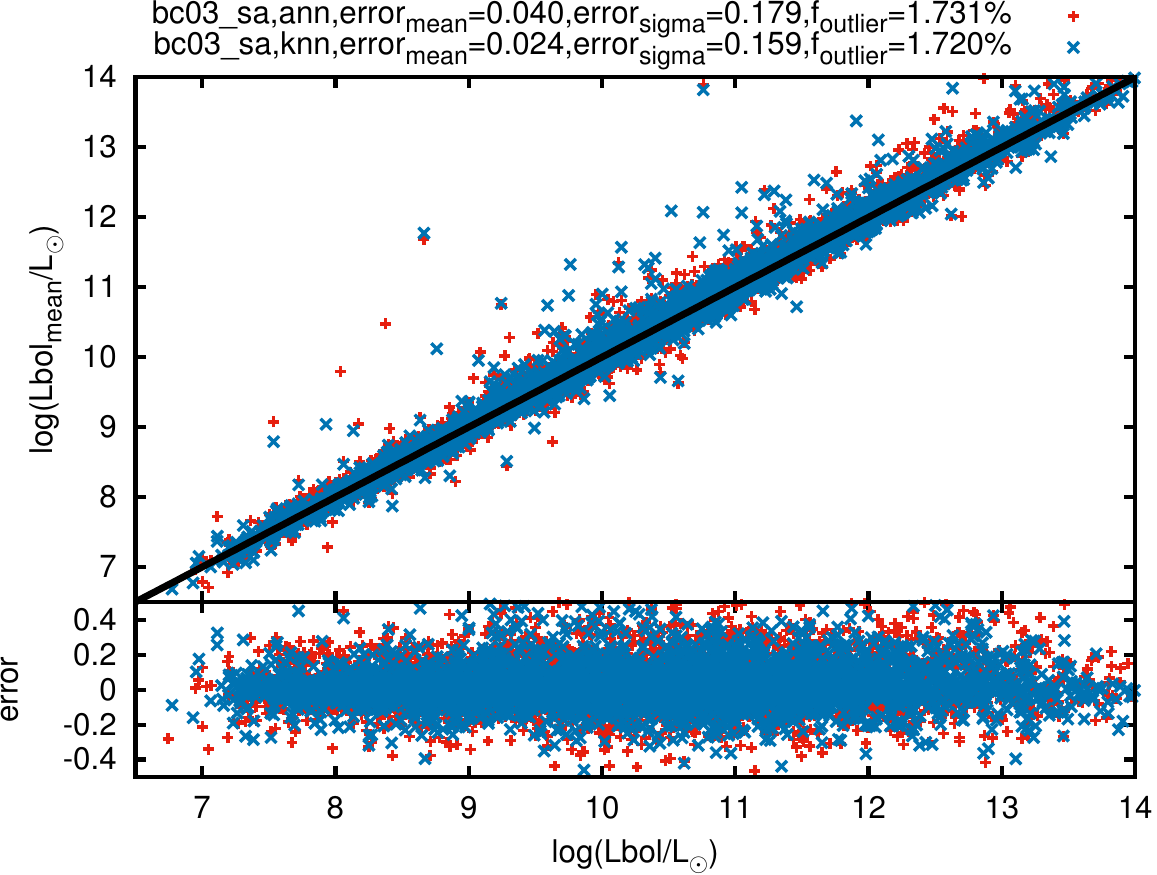} 
  \caption
  {
  Similar to Figure \ref{fig:mock_test_mass}, but for the bolometric luminosity ($\rm Lbol$).
  }
  \label{fig:mock_test_Lbol}
\end{figure}

\begin{figure}
  \centering 
  \includegraphics[scale=0.7]{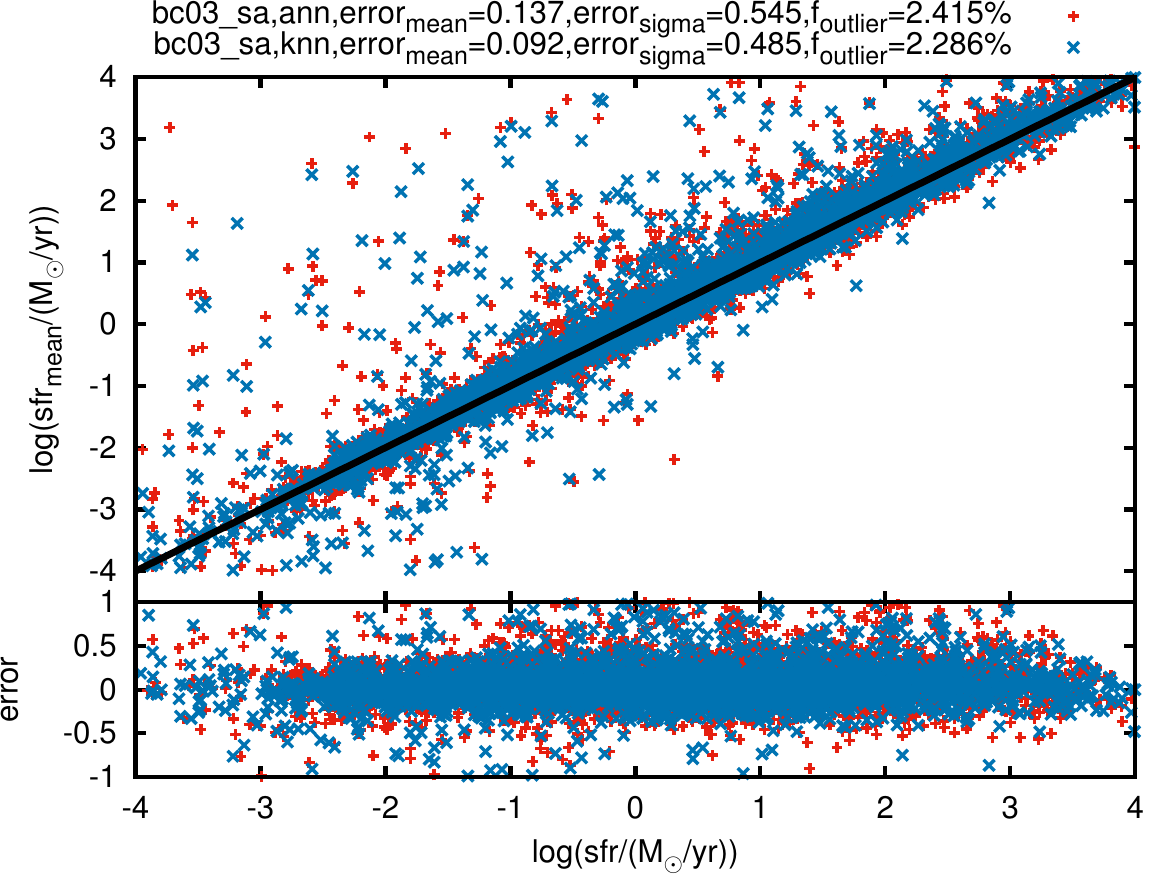} 
  \caption
  {
  Similar to Figure \ref{fig:mock_test_mass}, but for the star formation rate ($\rm sfr$).
  }
  \label{fig:mock_test_sfr}
\end{figure}

\begin{figure}
  \centering 
  \includegraphics[scale=0.7]{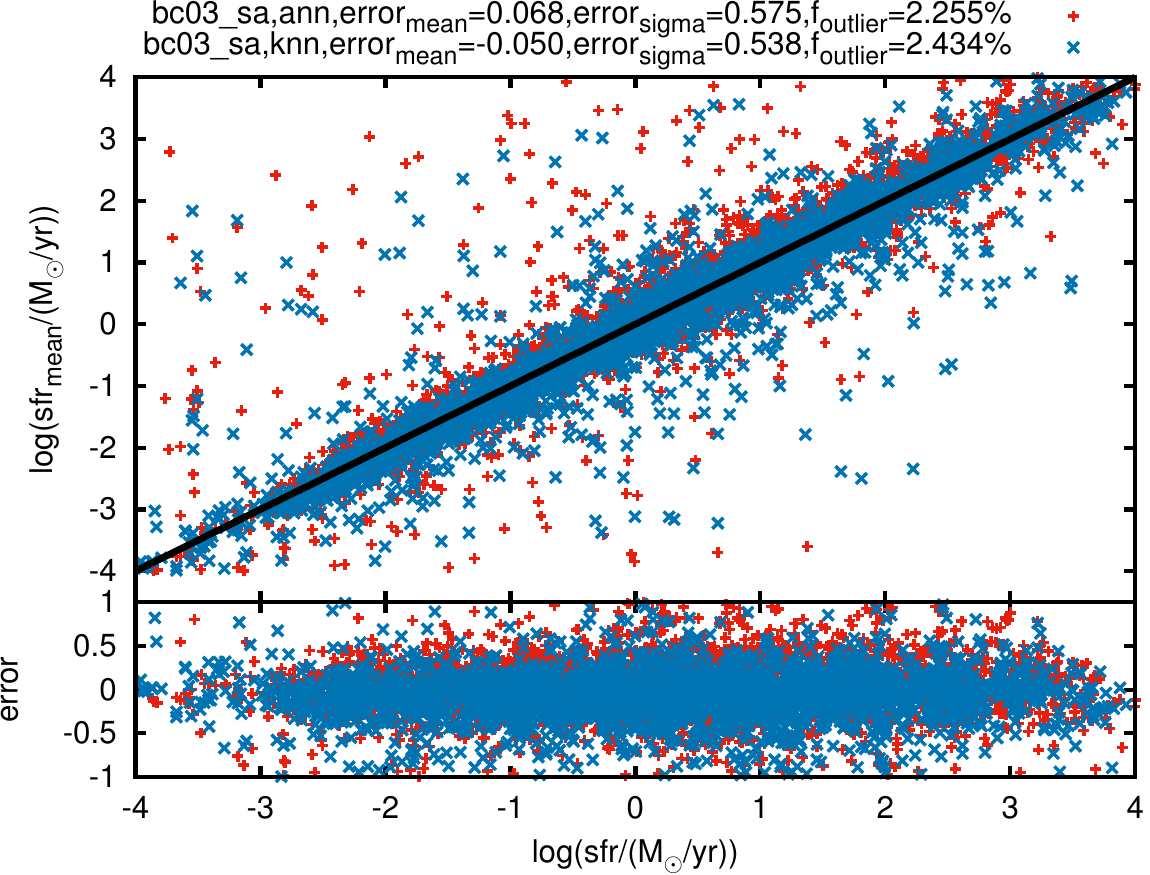} 
  \caption
  {
  Similar to Figure \ref{fig:mock_test_sfr}, but now the MAL \hl{estimations of the star formation rates} of galaxies have been used.
  }
  \label{fig:mock_test_sfr2}
\end{figure}

As examples, in Figures \ref{fig:mock_test_mass}, \ref{fig:mock_test_Lbol}, and \ref{fig:mock_test_sfr}, we compare the derived values of stellar mass, stellar bolometric luminosity, and star formation rate with their true values.
Generally, these parameters can be recovered properly except for a small fraction of extreme outliers.
Besides, the recovered value of these parameters seem to be biased for the outliers.
This is especially clear for the estimation of star formation rate.
We found that this bias is mainly depends on the method used for summary statistics.
In Figure \ref{fig:mock_test_sfr2}, we show the results obtained with MAL estimation.
We can see that the distribution of errors is more symmetric with MAL estimation, and the mean of errors is much closer to zero.
However, the dispersion and the fraction of outliers are not changed too much.
So, for a population of galaxies, the Mean estimation and MAL estimation  should be equally good.


In this section, we have systematically tested the reliability of BayeSED to recover the free and derived parameters  of a stellar population synthesis model from its observed multi-wavelength photometric SEDs by employ mock sample of galaxies.
Generally, we \hl{believe} that the results obtained with BayeSED code are acceptable.
Indeed, there are some extreme outliers in the recovered values, and they \hl{could} be partly resulted from the BayeSED code itself.
However, we \hl{believe} that the errors in the recovered value of parameters are dominated by the nature of these parameters themselves and the limited information about them in the photometric data.

\section{Application to a Ks-selected sample in the COSMOS/UltraVISTA field}
\label{s:application_real}
With the systematic tests of BayeSED code in \S \ref{s:application_mock}, we \hl{believe} it should be able to obtain reliable results for real galaxies.
In this section, we will apply BayeSED to interpret the SEDs of galaxies in a Ks-selected sample in the COSMOS/UltraVISTA field.
Since the tests in \S \ref{s:application_mock} showed that the results obtained with KNN method have larger Bayesian evidence than that obtained with ANN, we only show the results obtained with KNN method in this section.

\subsection{A Ks-selected Catalog in the COSMOS/UltraVISTA Field}
\label{ss:sample_selection}
The UltraVISTA survey \citep{McCracken2012a} is a NIR sky survey with a unique combination of area and depth.
When fully complete, the survey  will cover an area of 1.8 deg$^2$ down to K$_{s}$ $\sim$ 24.0.
Meanwhile, the survey field of UltraVISTA is the COSMOS field \citep{Scoville2007a}, which has the most extensive multi-wavelength coverage and is an attractive field for the studies of distant galaxies.
\cite{Muzzin2013a} presented a catalog covering 1.62 $\rm degree^2$ of the COSMOS/UltraVISTA field.
The catalog provides photometry in 30 photometric bands including the available GALEX, Subaru, Canada-France-Hawaii Telescope, VISTA, and Spitzer data. 
The sources in the catalog have been selected from the DR1 UltraVISTA K$_{s}$ band imaging that reaches a depth of K$_{s,tot}$ $=$ 23.4 AB with 90\% completeness.
 
In the study of this section, we have selected a subset of \cite{Muzzin2013a} catalog with \texttt{star} = 0, \texttt{contamination} = 0, \texttt{nan\_contam} $<$ 5, K$_{s}$ $<$ 23.9 (5$\sigma$ depth of the survey).
These objects are considered to be galaxies with good photometry.
Besides, we only selected objects with known spectroscopic redshifts for this illustrative study.
This results in a sample of 5467 galaxies with $0<z<2$.
A more comprehensive study of all galaxies in the \cite{Muzzin2013a} catalog with BayeSED will be presented in a future work.
In addition to the photometric catalog, \cite{Muzzin2013a} also provide a catalog of photometric redshifts computed with the EAZY code \citep{Brammer2008a} and a catalog of stellar masses and stellar population parameters determined using the FAST SED-fitting code \citep{Kriek2009a} for all galaxies in the survey.
We will compare our results obtained with BayeSED code with that of \cite{Muzzin2013a} obtained with FAST code.

\subsection{Interpreting and Results}
\label{ss:sample_results}
To be compared with the results of \cite{Muzzin2013a}, we have used the \cite{Bruzual2003a} or \cite{Maraston2005a} SED models with solar metallicity, \cite{Calzetti2000a} dust extinction law, and an exponentially declining star formation history to interpret the SEDs of galaxies in our sample.
\cite{Chabrier2003a} or \cite{Salpeter1955a} IMF has been used for the bc03 model, while \cite{Kroupa2001a} or \cite{Salpeter1955a} IMF has been used for ma05 model.
In total, four free parameters are involved, including $\rm{log}(\tau/yr)$ in the range of $[6.5,11]$, $\rm{log}(age/yr)$ in the range of $[7.0,10.1]$, and visual attenuation $A_{\rm v}$ in the range of $[0,4]$.
During the sampling with MultiNest, the four parameters are allowed to be uniformly and continuously selected from the allowed parameter space.
As another prior, the age of a galaxy is forced to be less than the age of the universe at the redshift of the galaxy. 
In BayeSED, the scale factors of model SEDs are not considered as free parameters during the sampling of parameter space with MultiNest.
Instead, they are uniquely determined using the efficient iterative algorithm of \cite{Sha2007a} as in the EAZY code.
With this algorithm, a linear combination of multiple SED models can be used to interpret the observed SEDs of galaxies.
\hl{In this paper, only the stellar population synthesis model of bc03 or ma05 model is employed to interpret the observed SEDs.}
So, to be consistent with this selection of SED model, the Spitzer bands with wavelengths longer than $4.5\micron$ have not been used during the fitting of SEDs, since no model for dust emission has been used.

\subsubsection{Comparison with the results of \cite{Muzzin2013a}}
\label{sss:comp_Muzzin2013}
In \S \ref{s:application_mock}, we have verified the internal consistency of BayeSED by using a mock sample of galaxies.
\hl{Here, we instead check the external consistency of BayeSED with the widely used FAST code as employed by \cite{Muzzin2013a}.}
\hl{Due to the very different methodologies employed by the two codes, we expect some differences for the results obtained by them.}
\hl{However, the results obtained by the two codes should be generally consistent with each other.}

\begin{figure}
  \centering 
  \includegraphics[scale=0.65]{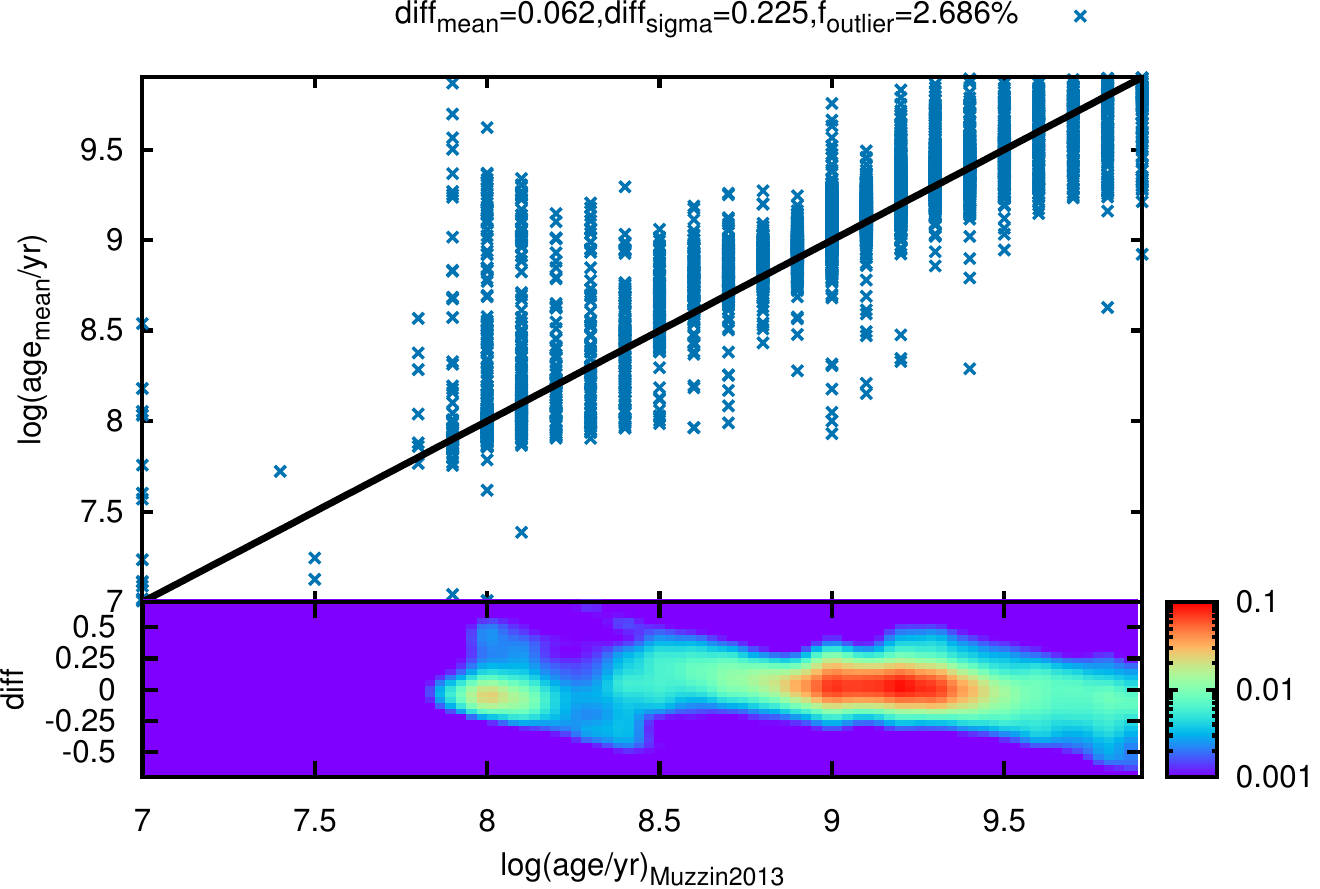}
  \caption
  {
  The age of galaxies obtained with our BayeSED code compared with that obtained with the FAST code by \cite{Muzzin2013a}.
  \hl{  The parameter is estimated with the mean of the posterior distribution.}
  \hl{  The probability density distribution of the difference between the results of the two codes is shown in the lower panel.}
  \hl{  On the top of the figure, the mean, dispersion and the fraction of outliers with a difference larger than $3\sigma$ is shown.}
  }
  \label{fig:ULTRAVISTA0_Z0_0bc03_ch_age}
\end{figure}

\begin{figure}
  \centering 
  \includegraphics[scale=0.65]{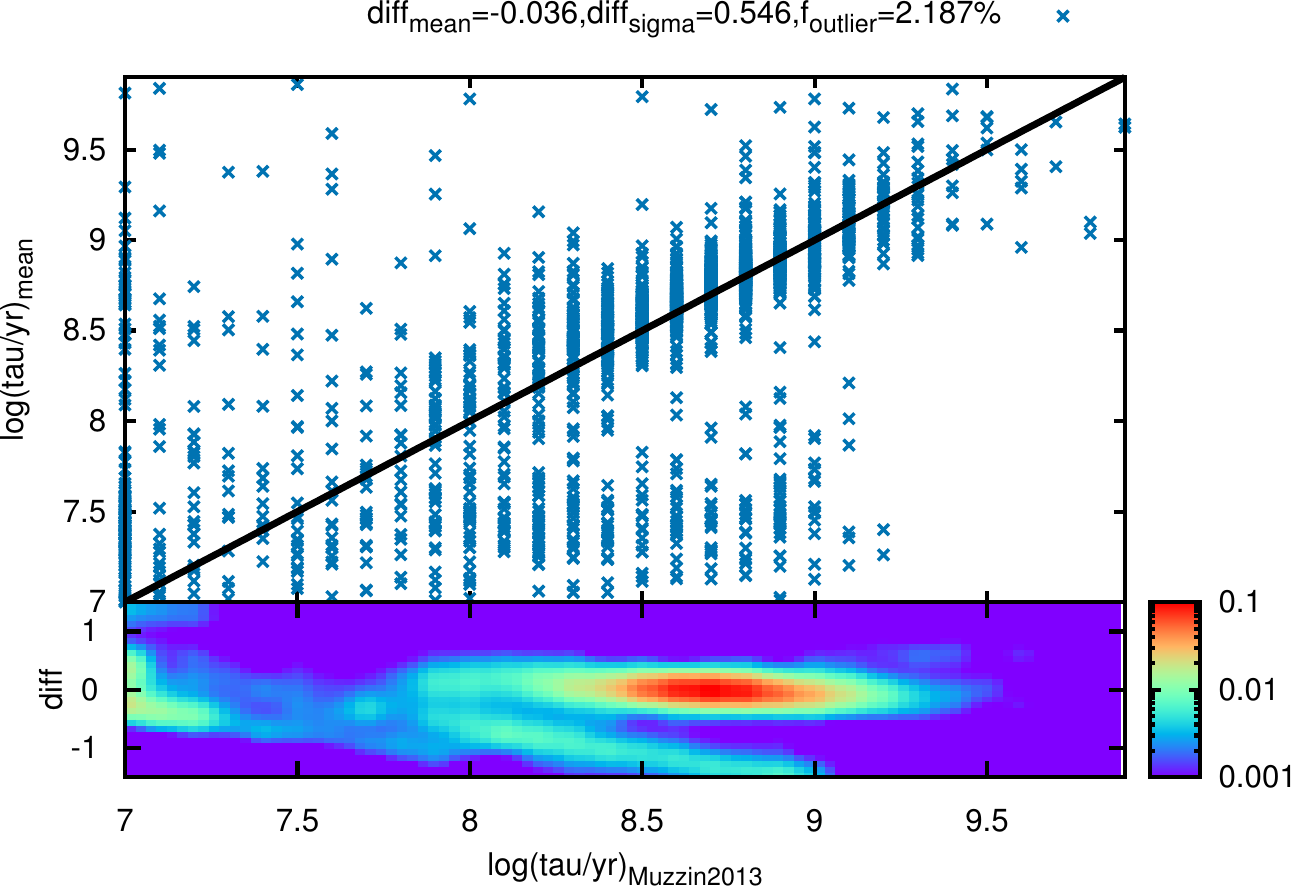}
  \caption
  {
  Similar to Figure \ref{fig:ULTRAVISTA0_Z0_0bc03_ch_age}, but for the e-folding time $\rm tau$.
  }
  \label{fig:ULTRAVISTA0_Z0_0bc03_ch_tau}
\end{figure}

\begin{figure}
  \centering 
  \includegraphics[scale=0.65]{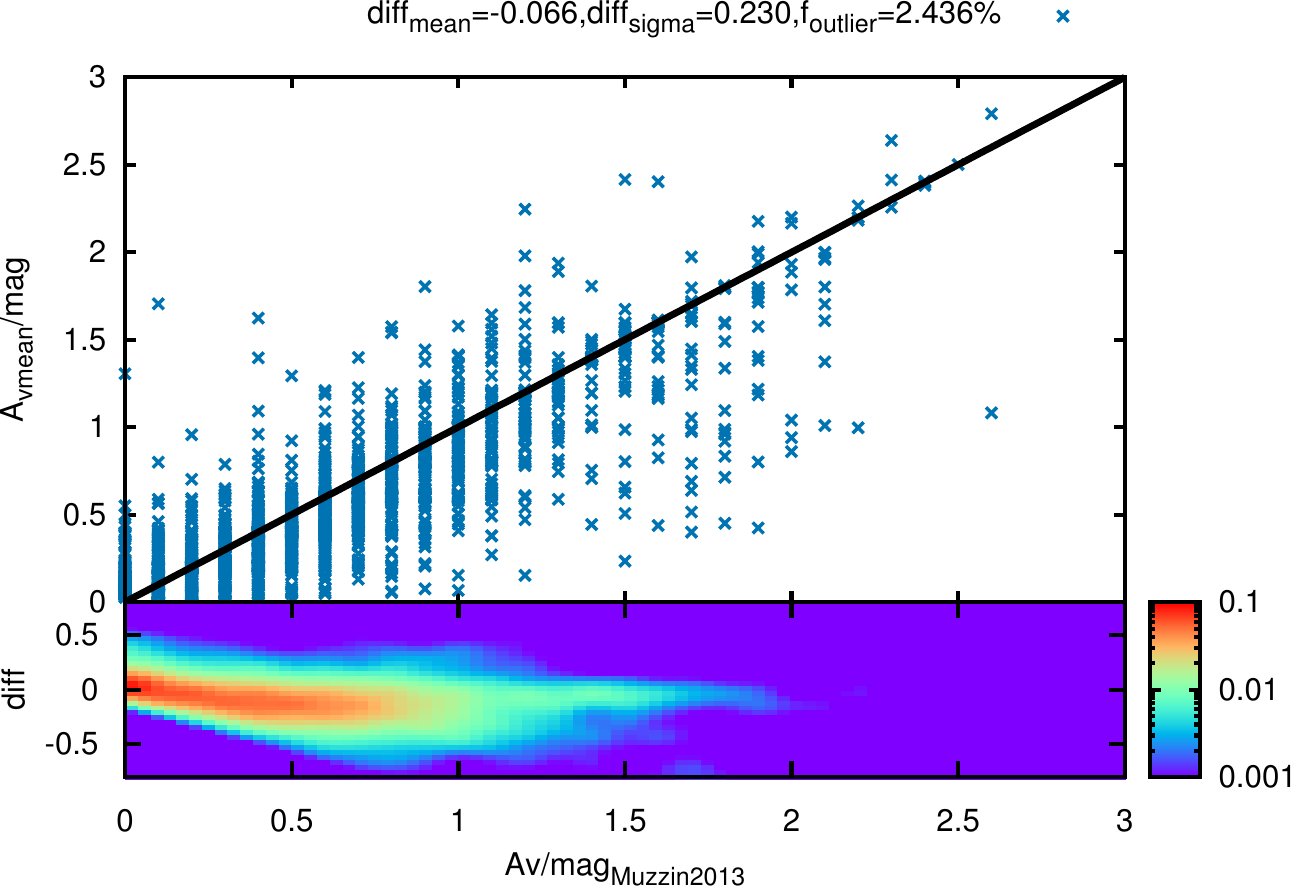}
  \caption
  {
  Similar to Figure \ref{fig:ULTRAVISTA0_Z0_0bc03_ch_age}, but for the dust extinction $A_{\rm v}$.
  }
  \label{fig:ULTRAVISTA0_Z0_0bc03_ch_Av}
\end{figure}

\begin{figure}
  \centering 
  \includegraphics[scale=0.65]{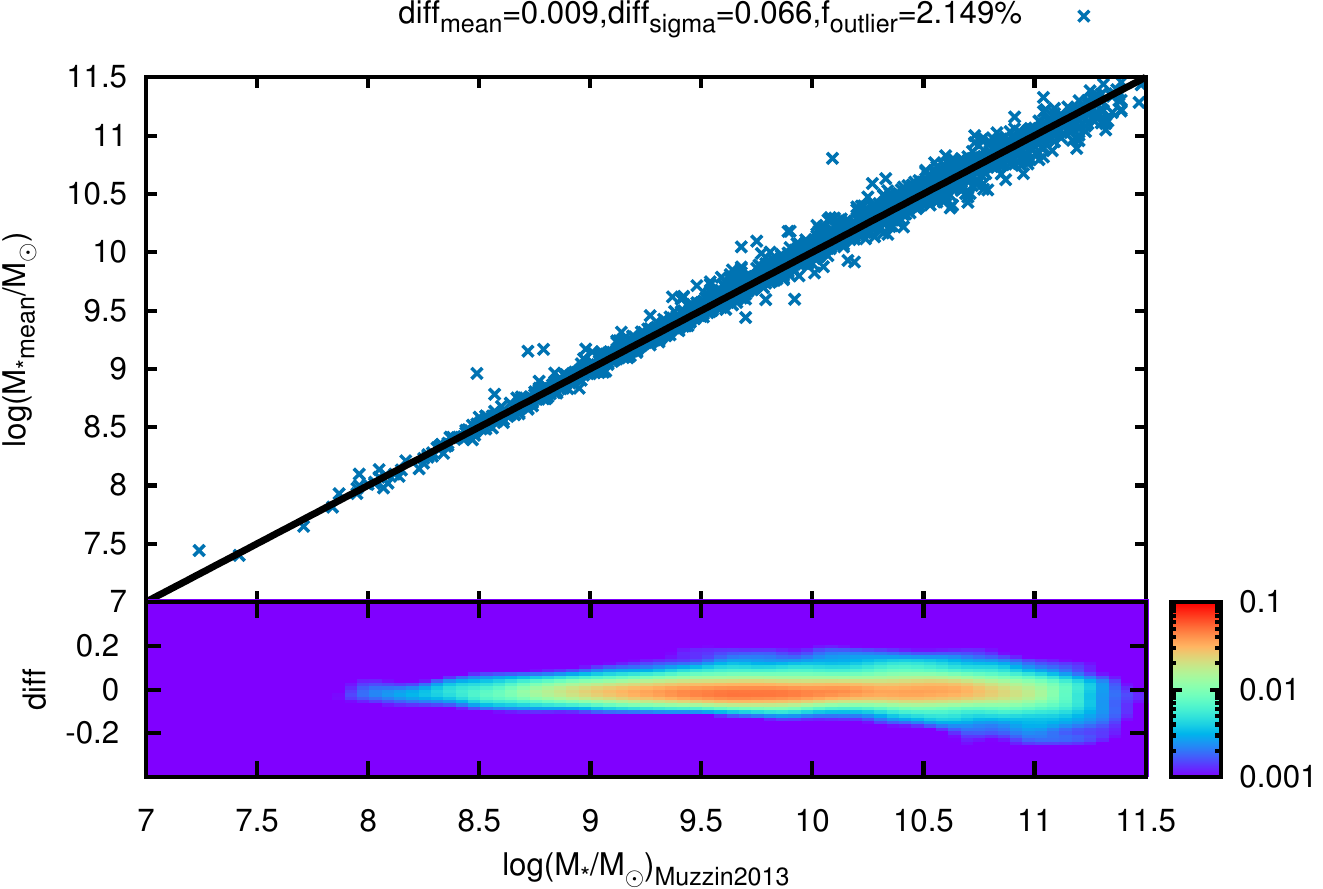}
  \caption
  {
  Similar to Figure \ref{fig:ULTRAVISTA0_Z0_0bc03_ch_age}, but for the stellar mass.
  }
  \label{fig:ULTRAVISTA0_Z0_0bc03_ch_mass}
\end{figure}

\begin{figure}
  \includegraphics[scale=0.65]{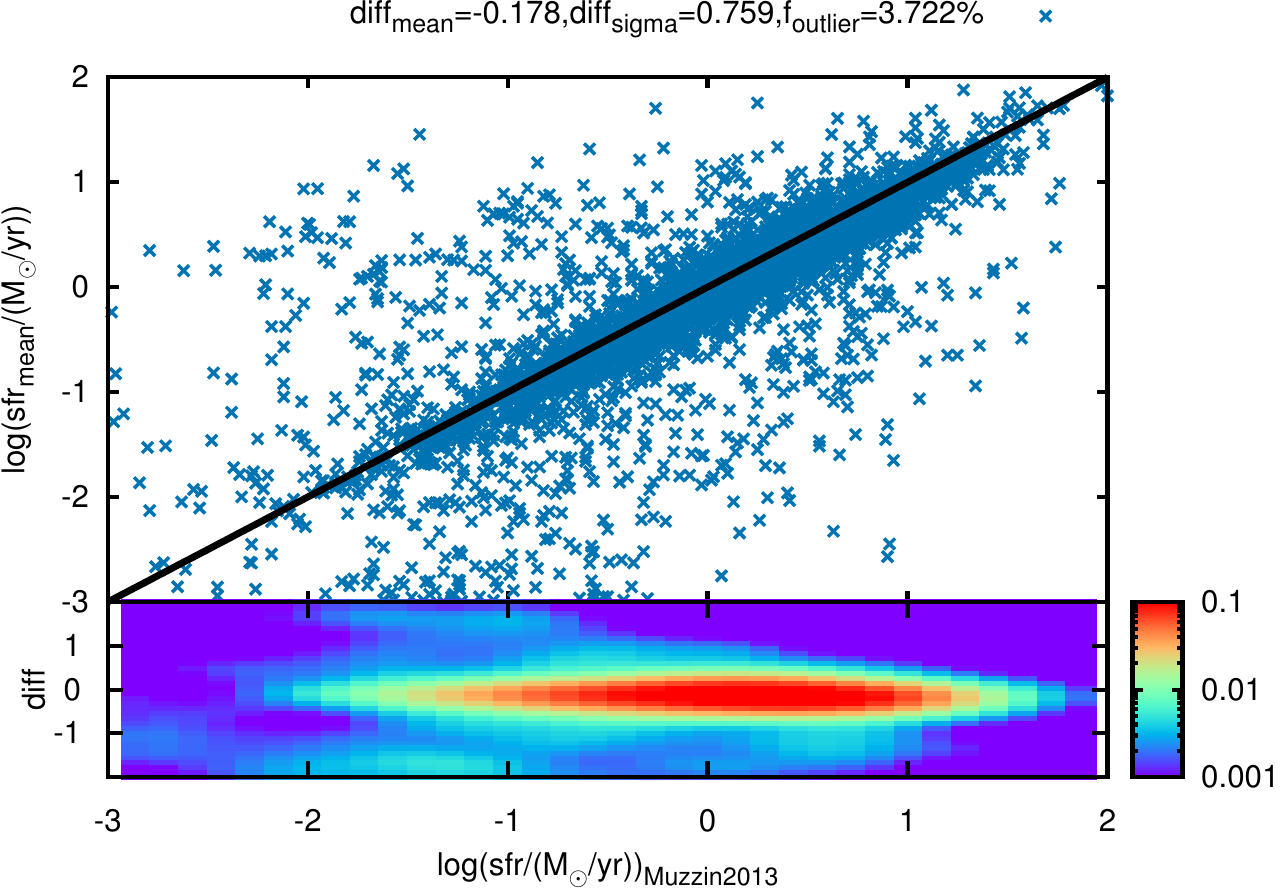}
  \caption
  {
  Similar to Figure \ref{fig:ULTRAVISTA0_Z0_0bc03_ch_age}, but for the star formation rate.
  }
  \label{fig:ULTRAVISTA0_Z0_0bc03_ch_sfr}
\end{figure}

\hl{In Figures \ref{fig:ULTRAVISTA0_Z0_0bc03_ch_age}, \ref{fig:ULTRAVISTA0_Z0_0bc03_ch_tau}, \ref{fig:ULTRAVISTA0_Z0_0bc03_ch_Av}, \ref{fig:ULTRAVISTA0_Z0_0bc03_ch_mass}, and \ref{fig:ULTRAVISTA0_Z0_0bc03_ch_sfr}, we compare the values of age, e-folding time, dust extinction, stellar mass, and star formation rate obtained using BayeSED with that obtained using FAST by \cite{Muzzin2013a}.}
Only the results obtained with \cite{Bruzual2003a} model and \cite{Chabrier2003a} IMF have been shown.
\hl{As shown in the figures, except for a small fraction of extreme outliers, our results are generally consistent with that of \cite{Muzzin2013a}.}
\hl{This is more clear for those parameters that are more likely to be well constrained, such as stellar mass.}
\hl{Besides, the estimation of e-folding time ($\rm tau$) seems very similar for the two codes.}
\hl{It is much better than the case for a mock sample of galaxies as shown in Figure \ref{fig:mock_test_tau}.}
\hl{This is mainly because the redshift of a galaxy is fixed to the spectroscopic redshift, instead of as an additional free parameter that needs to be estimated from the observations.}


\begin{figure}
  \includegraphics[scale=0.7]{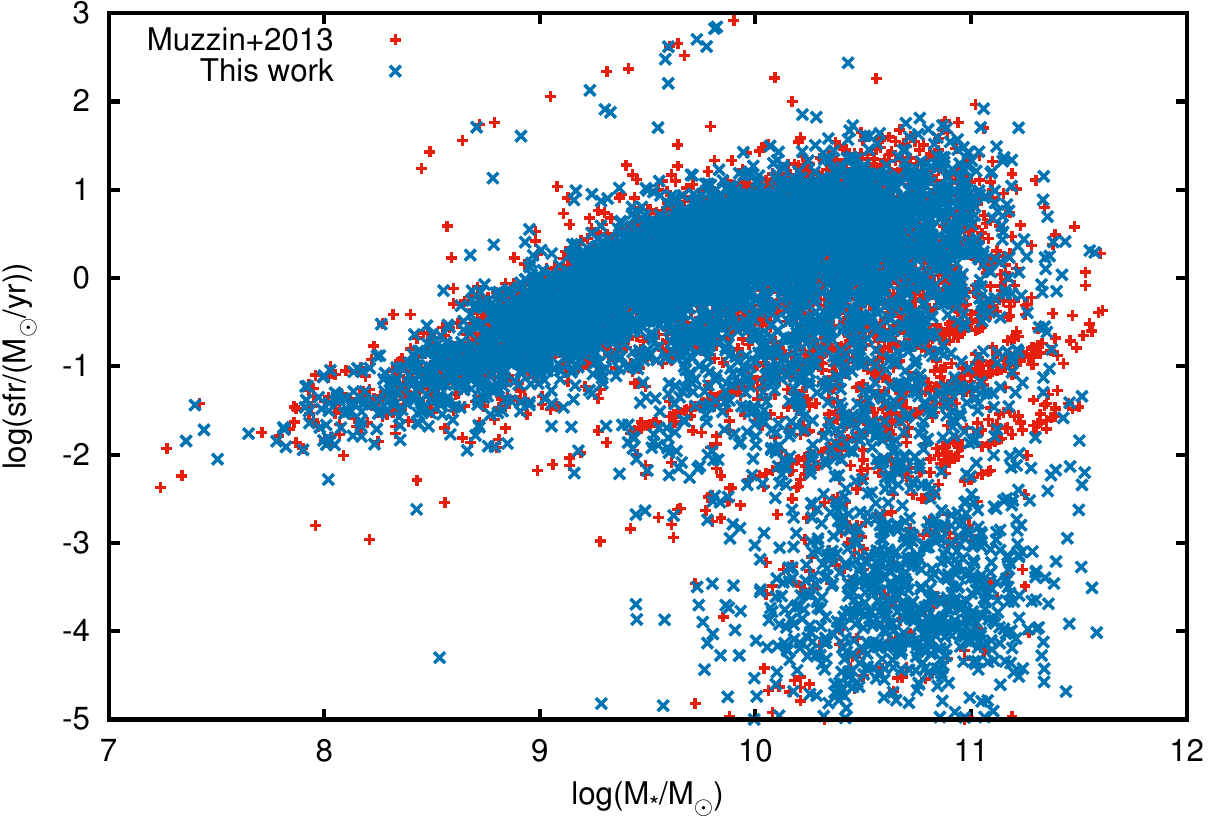}
  \caption
  {
  The distribution of galaxies in the $M_*$-SFR diagram.
  The results obtained with BayeSED in \hl{this work are generally consistent with those} of \cite{Muzzin2013a} who employed the FAST code, while the former show a more natural distribution.
  }
  \label{fig:ULTRAVISTA0_Z0_0bc03_ch_sfr_mass}
\end{figure}

An important advantage of BayeSED over traditional grid-based SED-fitting methods is that the parameter space of SED models can be sampled extensively and continuously.
With the grid-base methods, the estimation of parameters are only allowed to be within the precomputed set of grid points.
This is a very unnatural restriction when interpreting the SEDs of real galaxies, and can result in biased results.
In Figure \ref{fig:ULTRAVISTA0_Z0_0bc03_ch_sfr_mass}, we have shown the distribution of stellar mass vs.  star formation rate with results obtained with BayeSED and that obtained with grid-base FAST SED-fitting code.
Generally, the two set of results are consistent with each other, especially for the star-forming main sequence.
However, the results obtained with grid-base code show some unnatural parallel groups, which does not exist in results obtained with BayeSED.
Actually, this kind of issues are even more clear for the free parameters of a SED model.

\begin{figure*}
  \centering 
  \includegraphics[scale=1.4]{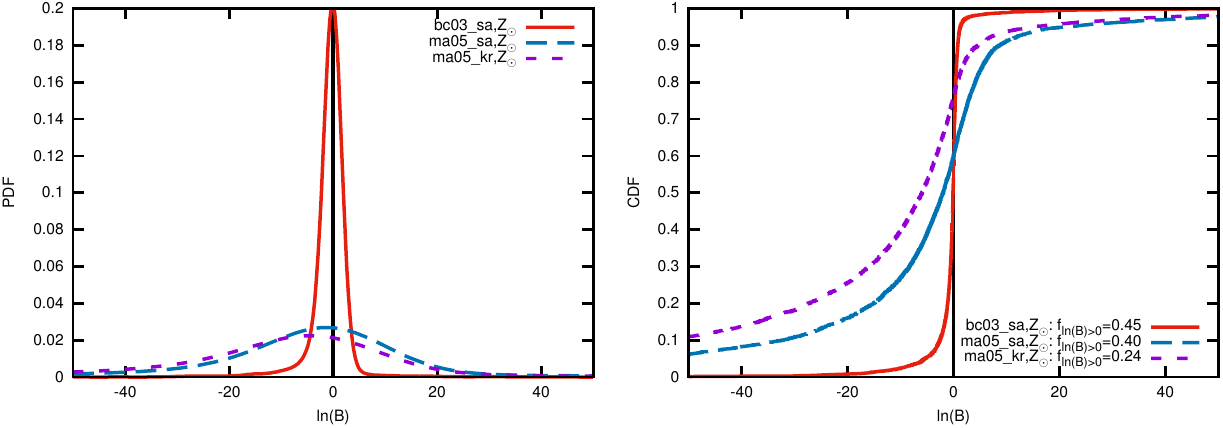}
  \caption
  {
  \hl{  The probability density distribution function (PDF) and cumulative distribution function (CDF) of Bayes factors $\rm ln(B)$ for 5467 galaxies in the Ks-selected sample with spectroscopic redshift.}
  \hl{  The four combinations of SED model and IMF are considered as four different models for Bayesian model comparison.}
  \hl{  The \cite{Bruzual2003a} model with a \cite{Chabrier2003a} IMF and solar metallicity is used as the base model for the computation of all Bayes factors hereafter in this paper.}
  \hl{ The distributions show that the base model statistically has larger Bayesian evidence than the other three models, since only $45\%$, $40\%$, and $24\%$ of galaxies support them, respectively.}
  \hl{  In general, the \cite{Bruzual2003a} model statistically has larger Bayesian evidence than the \cite{Maraston2005a} model, regardless of which IMF has been used.}
  }
  \label{fig:ULTRAVISTA0_Z0_ev}
\end{figure*}

\hl{As mentioned in \S \ref{s:intro}, there are many uncertainties in the population synthesis modeling of galaxy SEDs.}
\hl{So, with the ability of BayeSED code for an efficient computation of Bayesian evidence in hand, it would be very interesting to do a Bayesian model comparison for different population synthesis modeling.}
\hl{In Figure \ref{fig:ULTRAVISTA0_Z0_ev}, we show the probability density distribution function (PDF) and cumulative distribution function (CDF) of Bayes factor $\rm ln(B)$ for 5467 galaxies in the Ks-selected sample with spectroscopic redshift.}
\hl{Four different combinations of population synthesis model and IMF are considered as four different models for Bayesian model comparison.}
\hl{The \cite{Bruzual2003a} model with \cite{Chabrier2003a} IMF and solar metallicity, the one used in the work of \cite{Muzzin2013a}, is used as the base model for the computation of all Bayes factors hereafter in this paper.}
\hl{The distributions show that the base model statistically has larger Bayesian evidence than the other three models, since only $45\%$, $40\%$, and $24\%$ of galaxies support them, respectively.}
\hl{In general, the \cite{Bruzual2003a} model statistically has larger Bayesian evidence than \cite{Maraston2005a} model, no matter which IMF has been used.}
\hl{The \cite{Maraston2005a} includes a more advanced treatment of the thermally pulsating AGB stars.}
\hl{However, whether it is  a better treatment of this phase remains an open issue \citep{Kriek2010a,Zibetti2013a}.}
\hl{The results in Figure \ref{fig:ULTRAVISTA0_Z0_ev} show that \cite{Maraston2005a} model is better than \cite{Bruzual2003a} model only for less than $10\%$ of galaxies in sample.}

\subsubsection{Metallicity and redshift as additional free parameters}
\label{sss:free_Z_z}
\begin{figure*}
  \centering 
  \includegraphics[scale=1.4]{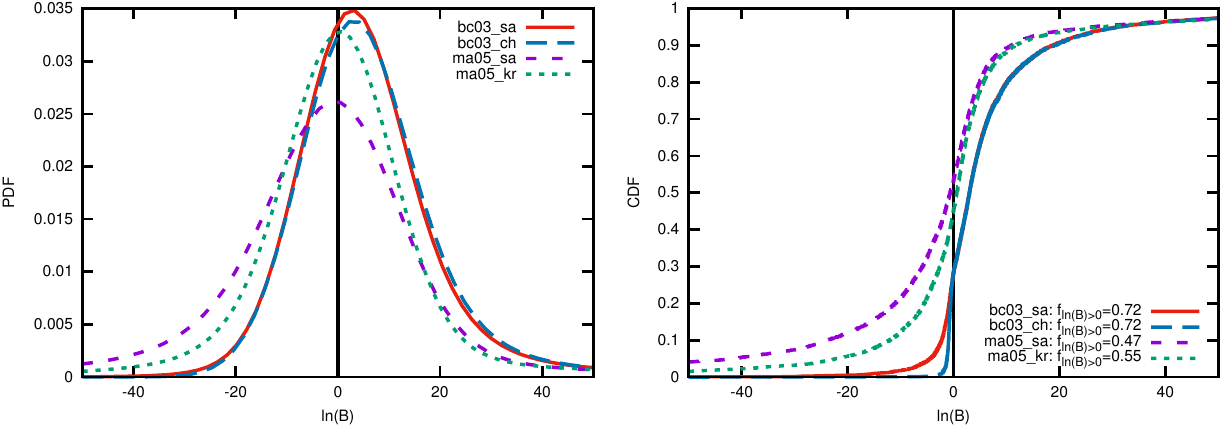}
  \caption
  {
  Similar to Figure \ref{fig:ULTRAVISTA0_Z0_ev}, but now the metallicity is set to be an additional free parameter ranging from $\rm 0.2Z_{\odot}$ to $\rm 2Z_{\odot}$.
  It is clear that the Bayesian evidence of all models have increased significantly.
  \hl{  Now, the two bc03 models have a much higher support rate than the base model.}
  \hl{  Meanwhile, for the two ma05 models, only the one with a \cite{Kroupa2001a} IMF has a slightly higher support rate than the base model.}
  \hl{  So, it becomes even clearer that the \cite{Bruzual2003a} model statistically has a larger Bayesian evidence than the \cite{Maraston2005a} model.}
  \hl{  Meanwhile, it seems that the latter model is more sensitive to the different choice of IMF than the former model.}
  }
  \label{fig:ULTRAVISTA0_ev}
\end{figure*}

For many works of SED-fitting of galaxies, the solar metallicity of stellar population is commonly assumed, even for high redshift galaxies.
Apparently, this is not a very reasonable assumption for galaxies in the real universe.
An excuse for this assumption is that it is usually very hard to determine the stellar population metallicity of galaxies with photometric data only.
On the other hand, the SED-fitting of galaxies would be much more time-consuming, especially for grid-base methods.
However, biased results could be obtained if solar metallicity is assumed for all galaxies.
To test the importance of the assumption about metallicity, we have employed BayeSED code to interpret the SEDs of the same sample of galaxies, but with metallicity as an additional free parameter ranging from $\rm 0.2Z_{\odot}$ to $\rm 2Z_{\odot}$.
In practice, we found that this will not obviously increase the time of computation for our BayeSED code.
\hl{In Figure \ref{fig:ULTRAVISTA0_ev}, we show the PDF and CDF of Bayes factor $\rm ln(B)$ for the same sample of galaxies in this case.}
\hl{Compared with the results in Figure \ref{fig:ULTRAVISTA0_Z0_ev}, the Bayesian evidences of all models are clearly increased.}
\hl{Now, the two bc03 models statistically have larger Bayesian evidence than the base model.}
\hl{Meanwhile, for the two ma05 models, only the one with \cite{Kroupa2001a} IMF has slightly larger support rate than the base model.}
\hl{Besides, it is worth to notice that \cite{Maraston2005a} model seems more sensitive to the different selection of IMF than the \cite{Bruzual2003a} model.}
\hl{Generally, with metallicity as an additional free parameter, it becomes more clear that the \cite{Bruzual2003a} model has statistically higher Bayesian evidence than \cite{Maraston2005a} model.}

\begin{figure*}
  \centering 
  \includegraphics[scale=1.4]{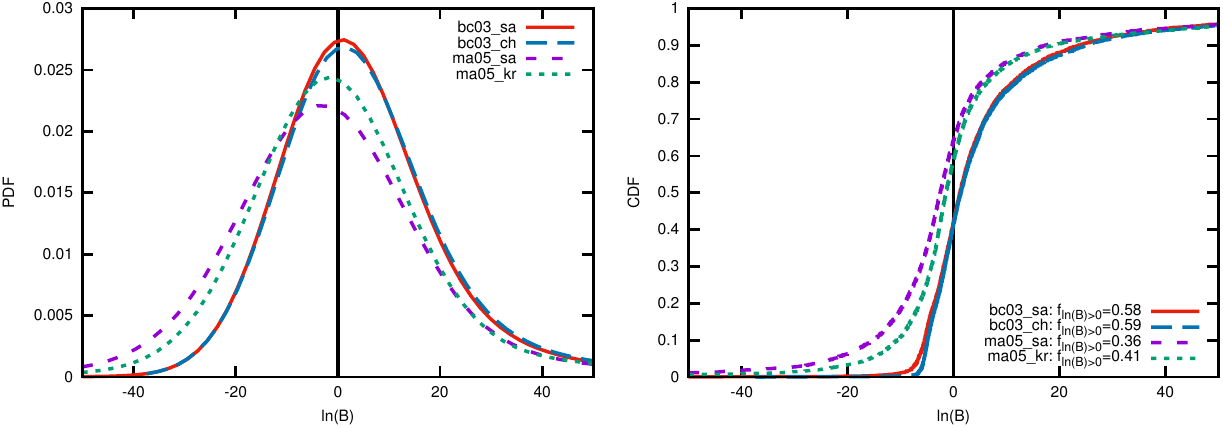}
  \caption
  {
  Similar to Figure \ref{fig:ULTRAVISTA0_ev}, but now the redshift is also set to be an additional free parameter.
  \hl{  In this case, the Bayesian evidence of all models have decreased a little.}
  \hl{  This demonstrates that an additional free parameter not necessarily increases the Bayesian evidence of a model if the increased complexity of the model is not rewarded with a much better fitting to the observations.}
  }
  \label{fig:ULTRAVISTA1_ev}
\end{figure*}

\begin{figure}
  \centering 
  \includegraphics[scale=0.7]{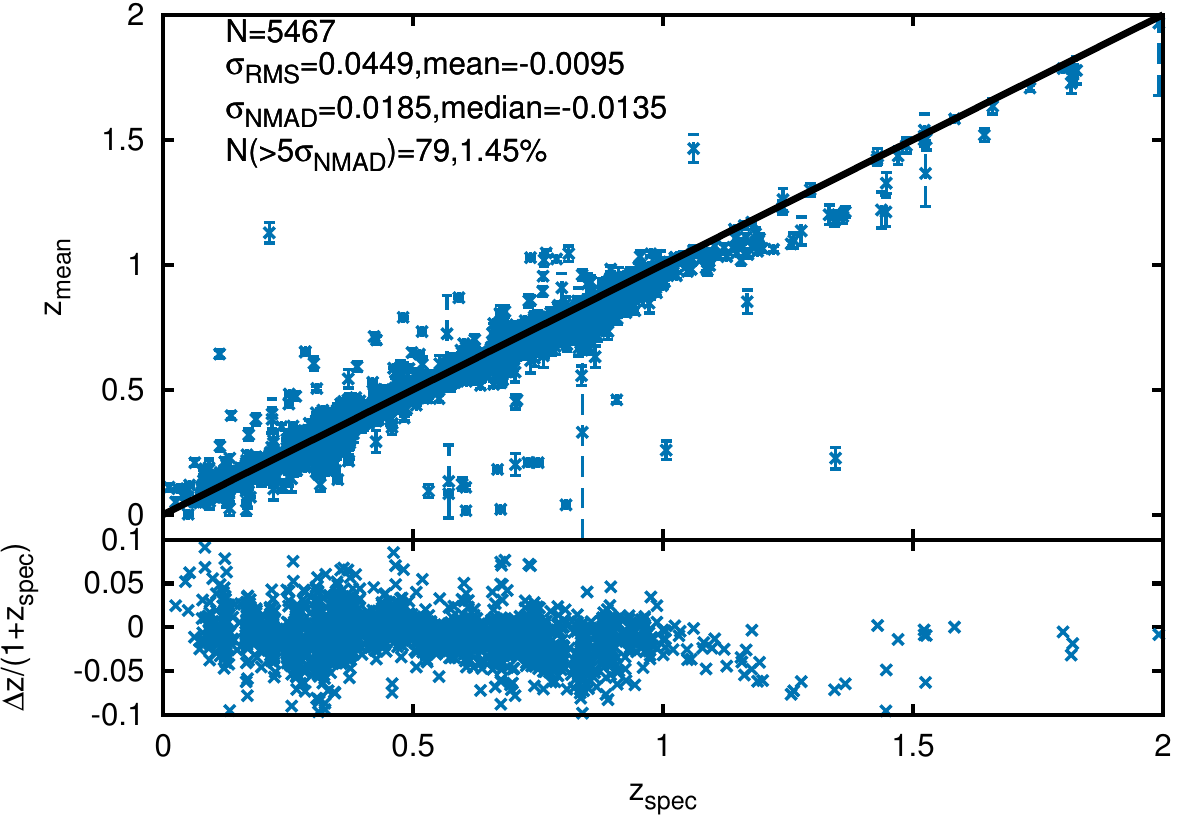}
  \caption
  {
  Photometric redshifts vs. Spectroscopic redshifts for galaxies in our sample.
  By using BayeSED, it is now possible to obtain photometric redshift and the stellar population parameters all simultaneously and self-consistently.
  \hl{  The performance of BayeSED for the estimation of photometric redshift is shown in the figure (see text for more details).}
  }
  \label{fig:ULTRAVISTA1_0bc03_ch_z}
\end{figure}

As mentioned in \S \ref{ss:bayesed}, the redshift of galaxy can be set as a free parameter as well in BayeSED code.
So, it is possible to simultaneously obtain photometric redshift and stellar population parameters of a galaxy with BayeSED, while using the same set of SED models and so more self-consistently.
Here, we test the reliability of BayeSED for the determination of photometric redshift of galaxies.
The distribution of Bayesian evidences for this case are shown in Figure \ref{fig:ULTRAVISTA1_ev}.
\hl{Compared with the results shown in Figure \ref{fig:ULTRAVISTA0_ev}, where the spectroscopic redshifts have been used, the Bayesian evidence of all models decrease a little in this case.}
In Figure \ref{fig:ULTRAVISTA1_0bc03_ch_z}, we compared the estimated photometric redshifts with the spectroscopic redshifts for galaxies in our sample.
\hl{The performance of a code for photometric redshift estimation is usually judged by the root mean square (RMS) of $(z_p-z_s)/(1+z_s)$.}
\hl{In our case, $\rm \sigma_{RMS}=0.0449$.}
\hl{When outliers with error larger than $3\sigma$ (48/5467=0.88\%) are removed, $\rm \sigma_{RMS}=0.0254$.}
\hl{This outperforms the result of \cite{Benitez2000a} (0.06) who have employed a Bayesian approach as well and employed more informative physical priors.}
\hl{Our results is not as good as that of \cite{Muzzin2013a} ($\rm \sigma_{RMS}=0.013$), who have employed the EAZY code and considered the effect of zero-point offset from an iterative procedure.}
\hl{However, they only considered 5119 galaxies with high-quality spectroscopic redshifts and uncontaminated photometry.}
\hl{A better judgement for the performance of photometric redshift estimation can be achieved by using the normalized median absolute deviation of $\Delta z = \zp-\zs$, which is defined \citep{Brammer2008a} as:}
\begin{equation}
\sigma_\textsc{nmad} = 1.48 \times \mathrm{median}\left( \left| \frac{\Delta z-\mathrm{median}(\Delta z)}{1+\zp} \right| \right).
\label{eq:nmad}
\end{equation}
\hl{$\sigma_{\rm NMAD}$ is less sensitive than $\rm \sigma_{RMS}$ to the outliers.}
\hl{In our case, $\sigma_{\rm NMAD}=0.0185$, while the median of the error is $-0.0135$ and the fraction of outliers is $1.45\%$.}
\hl{Except for the larger value of median, this is better than the result of \cite{Brammer2008a} with EAZY code, which is applied to a smaller sample with 1989 galaxies but spans a wider range of redshift.}
\hl{Given the difficulty of using a population synthesis model instead of using carefully selected templates or employing some kind of empirical training procedure to estimate the photometric redshift of galaxies, the degree of accuracy achieved by BayeSED is acceptable.}
\hl{In the future version of BayeSED, we plan to add more optimizations that have been adopted by many codes for a better estimation of photometric redshift.}

\subsubsection{The distributions of galaxy color and stellar population parameters}
\label{sss:distributions}
The observed distributions of galaxy properties provide important clues for understanding the formation and evolution of galaxies, and benchmarks for discriminations between different semi-analytic modelings and hydrodynamical simulations of galaxy formation.
\hl{However, a detailed mathematical characterization of the distributions of these properties and a full theoretical understanding of them in the context of galaxy formation and evolution \citep{Somerville2008a,Schaye2010a,Buzzoni2011a} is beyond the scope of this paper.}
Here, we show some well known features for the distribution of galaxy properties that are obtained with BayeSED code.
For the results presented in this subsection, the spectroscopic redshifts of galaxies have been used, and the metallicity is set to be a free parameter ranging from $\rm 0.2Z_{\odot}$ to $\rm 2Z_{\odot}$.
Besides, only the results obtained with \cite{Bruzual2003a} model and \cite{Chabrier2003a} IMF have been presented, since it has the largest Bayesian evidence as shown in Figure \ref{fig:ULTRAVISTA0_ev}.

\begin{figure}
  \includegraphics[scale=0.7]{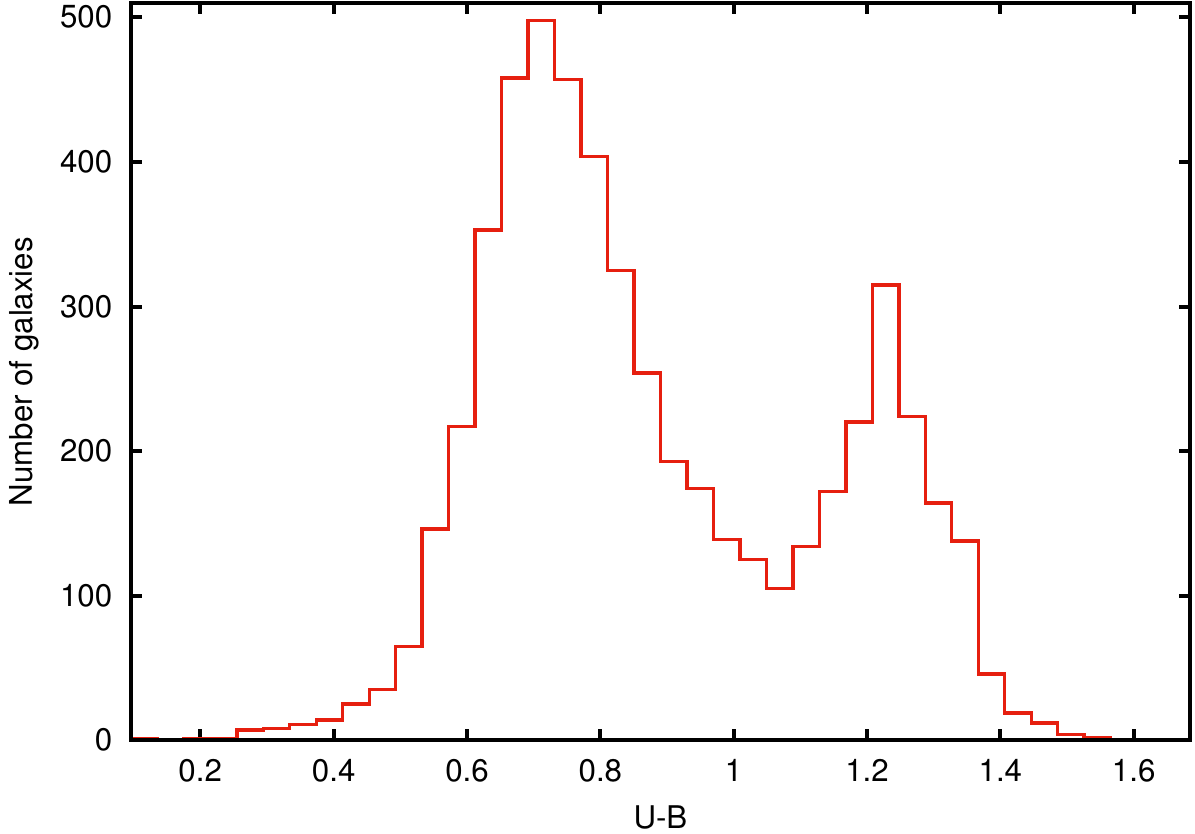}
  \caption
  {
  The bimodal distribution of rest-frame $\rm U-B$ \hl{colors} for galaxies in our sample.
  }
  \label{fig:ULTRAVISTA0_0bc03_ch_dist_UB}
\end{figure}

\begin{figure}
  \includegraphics[scale=0.7]{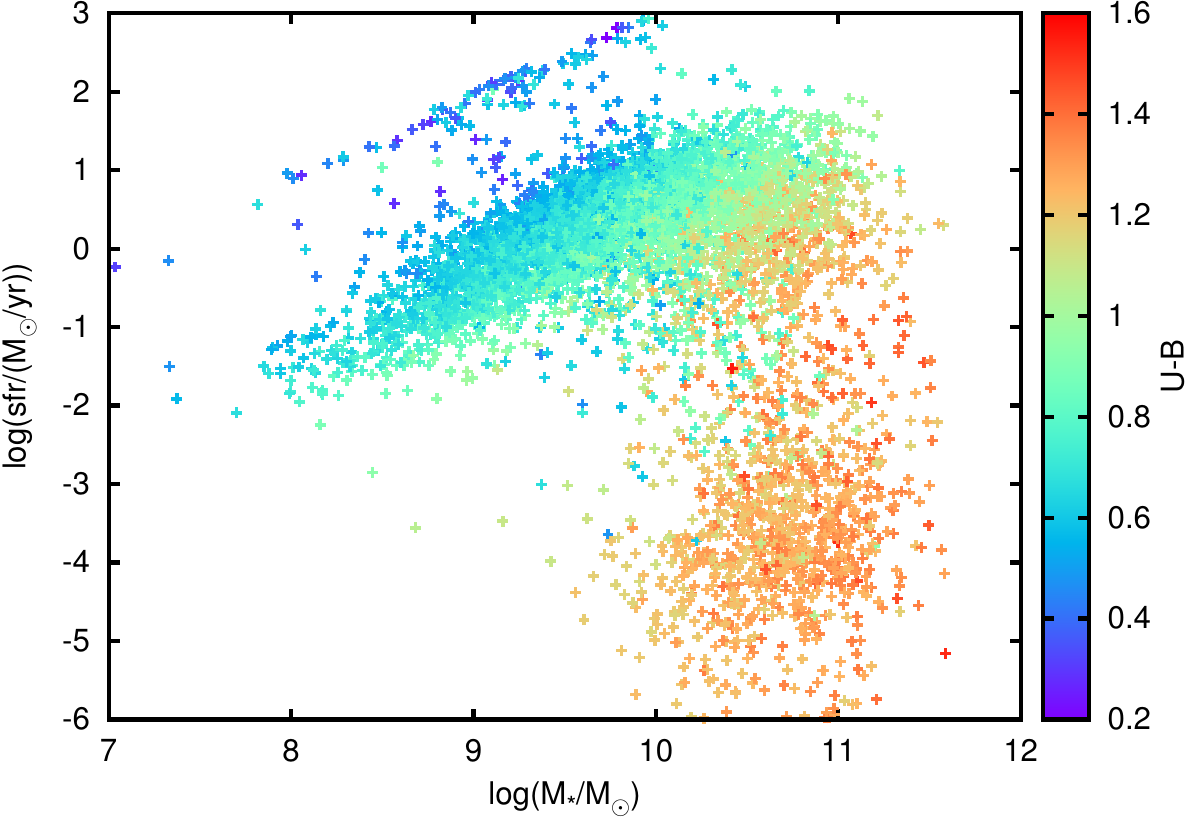}
  \caption
  {
  The distribution of rest-frame $\rm U-B$ \hl{colors} in the  $M_*$-SFR diagram.
  }
  \label{fig:ULTRAVISTA0_0bc03_ch_sfr_mass_UB}
\end{figure}

In Figure \ref{fig:ULTRAVISTA0_0bc03_ch_dist_UB}, we show the distribution of the rest-frame $\rm U-B$ color of galaxies in our sample.
The well known bimodal distribution of the color of galaxies is clearly shown in the figure.
These galaxies can be divided to the red ones with $\rm U-B\gtrsim1.1$, and the blue ones with $\rm U-B\lesssim1.1$.
The red galaxies and blue are thought to be fundamentally different populations of galaxies.
This can be more clearly noticed in the $M_*$-SFR diagram, as shown in Figure \ref{fig:ULTRAVISTA0_0bc03_ch_sfr_mass_UB}.
Generally, blue galaxies are mainly the star-forming galaxies in the ``main sequence'', while red galaxies are mainly the quiescent galaxies with negligible, if any, star formation.
Meanwhile, there are a few galaxies with $\rm SFR>10$, above the ``main sequence''.
These galaxies have the bluest color in our sample, and should be the starburst galaxies.
On the other hand, a portion of red galaxies are actually star-forming galaxies with $M_* \gtrsim 3*10^{10}\rm M_\odot$.

It is well known that many physical parameters, such as metallicity, dust attenuation, and age, can affect the color of galaxies.
These parameters are degenerated with each other to a certain extent, preventing them to be determined accurately all simultaneously.
However, they should be accurate enough do some qualitative analysis.
Moreover, we can check if the distributions of these parameters are reasonable in the context of galaxy evolution.

\begin{figure*}
  \centering 
  \subfigure[]
  {
  \includegraphics[scale=0.7]{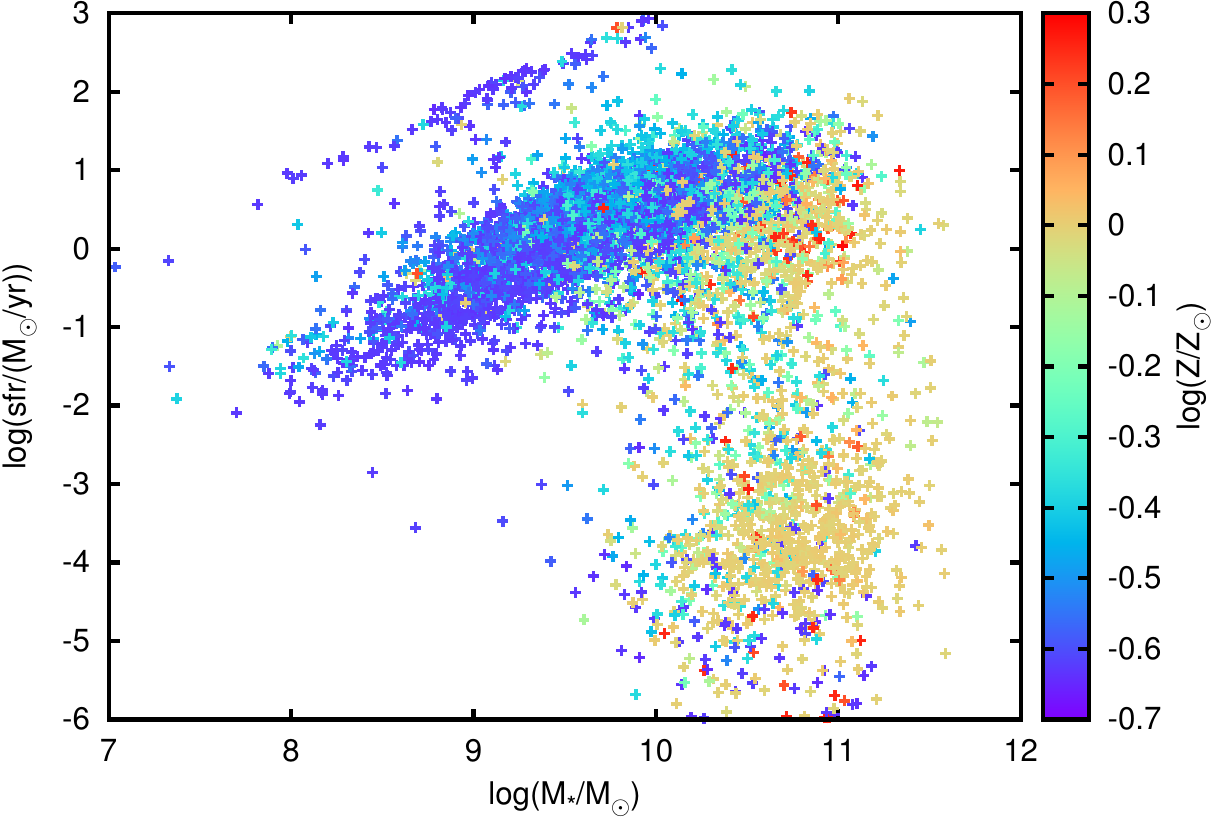}
  \label{fig:ULTRAVISTA0_0bc03_ch_sfr_mass_Zm}
  }
  \subfigure[]
  {
  \includegraphics[scale=0.7]{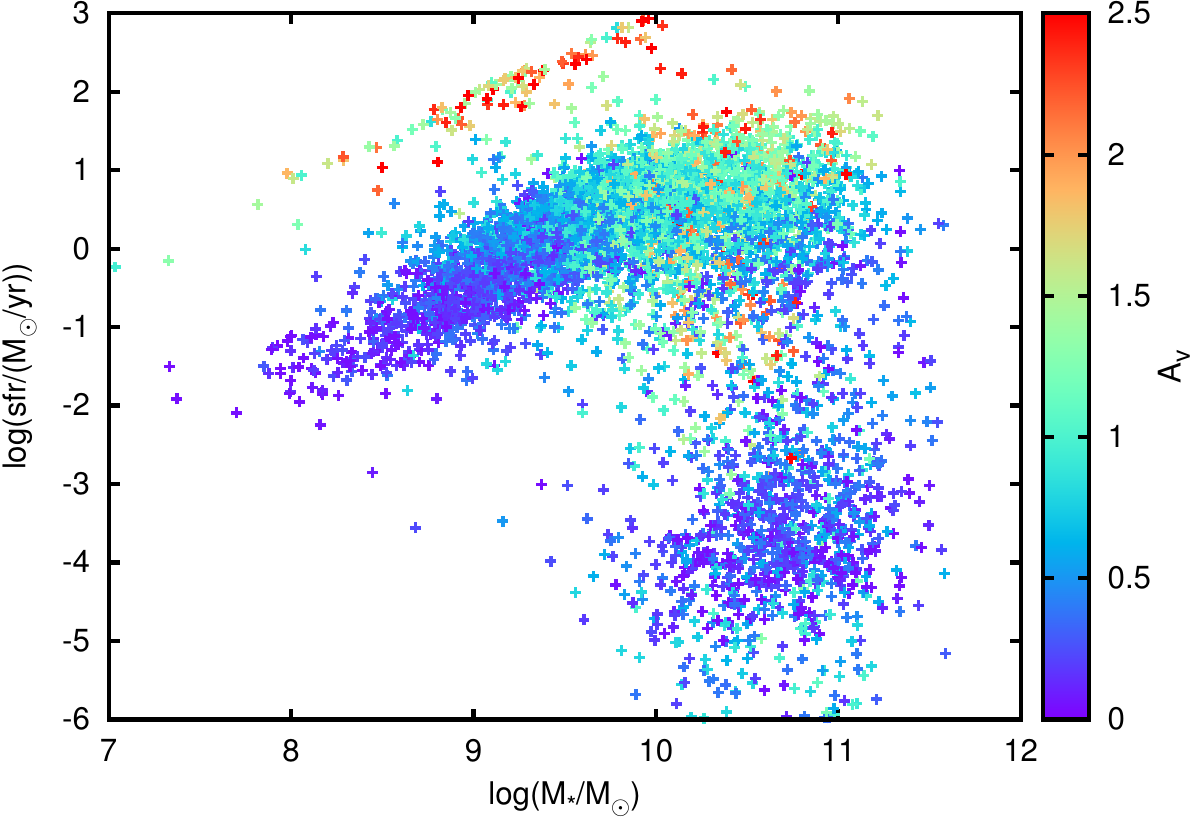}
  \label{fig:ULTRAVISTA0_0bc03_ch_sfr_mass_Av}
  }
  \\
  \subfigure[]
  {
  \includegraphics[scale=0.7]{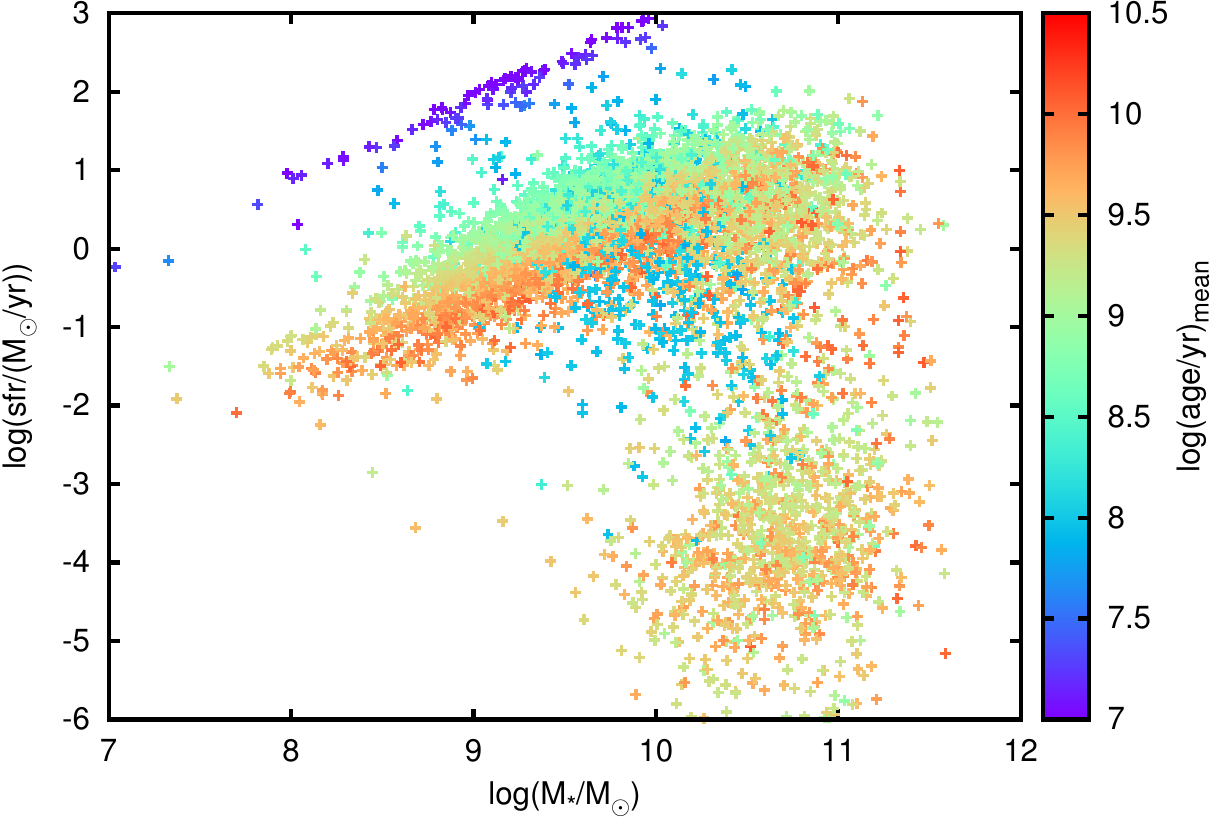}
  \label{fig:ULTRAVISTA0_0bc03_ch_sfr_mass_age}
  }
  \subfigure[]
  {
  \includegraphics[scale=0.7]{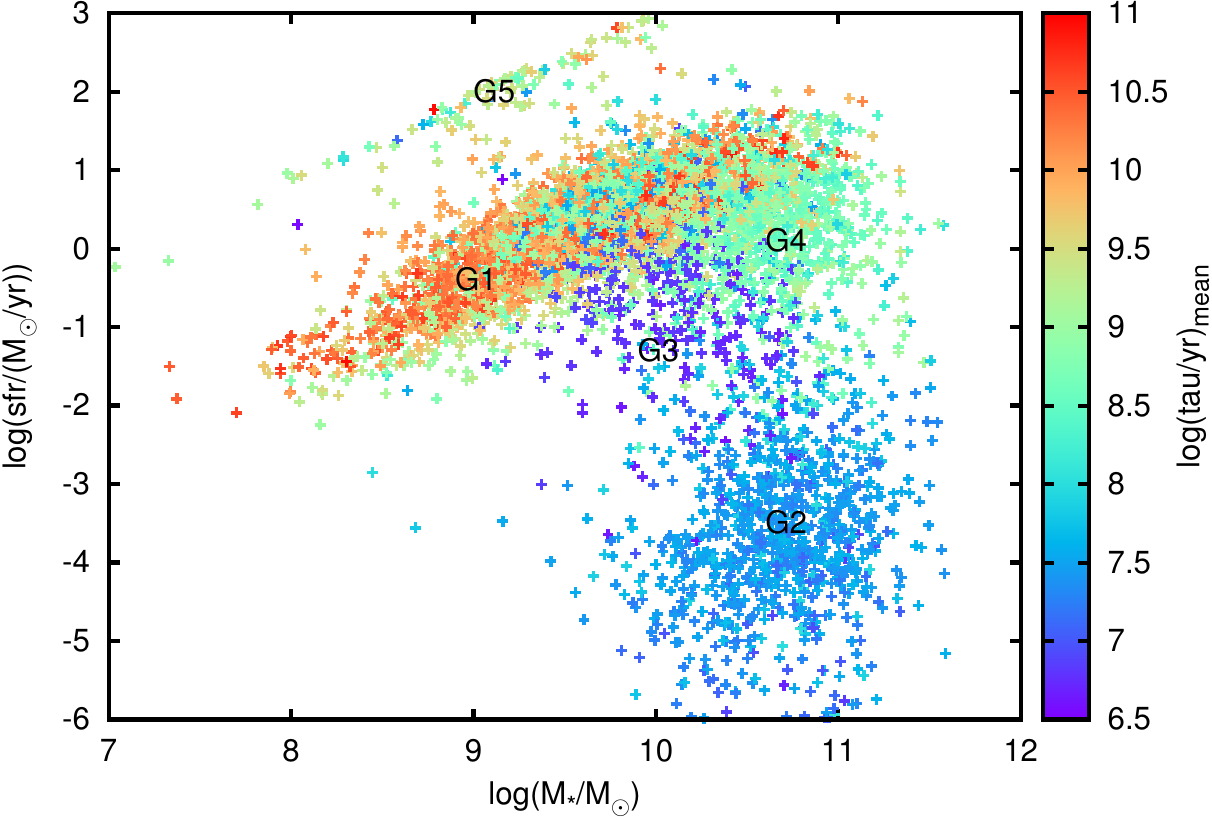}
  \label{fig:ULTRAVISTA0_0bc03_ch_sfr_mass_tau}
  }
  \\
  \subfigure[]
  {
  \includegraphics[scale=0.7]{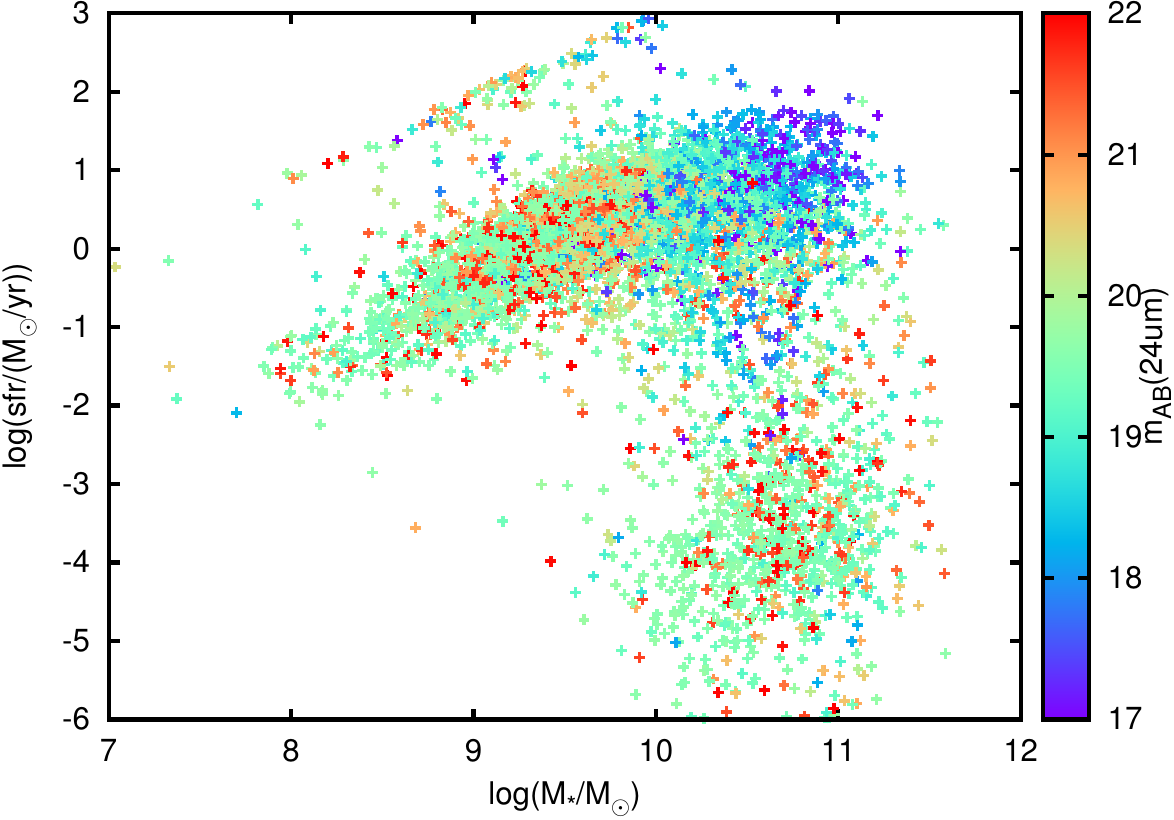}
  \label{fig:ULTRAVISTA0_0bc03_ch_sfr_mass_24um}
  }
  \subfigure[]
  {
  \includegraphics[scale=0.7]{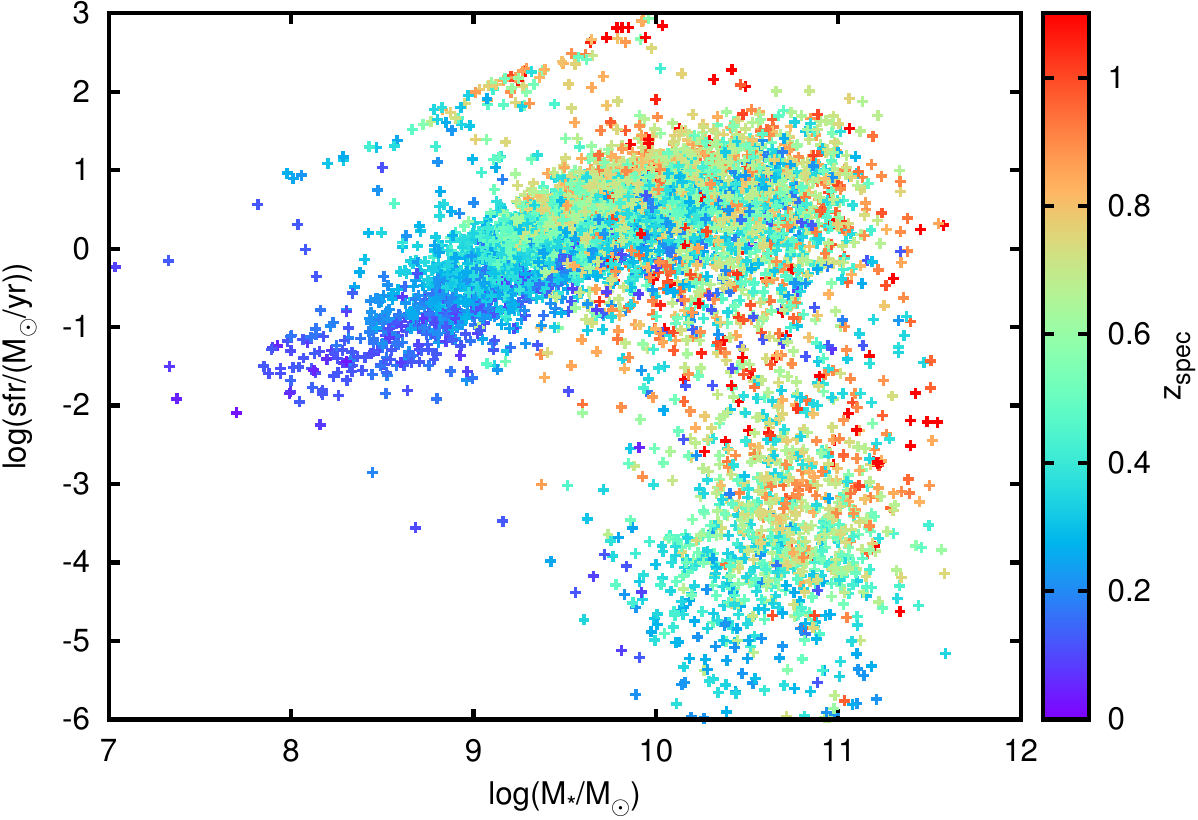}
  \label{fig:ULTRAVISTA0_0bc03_ch_sfr_mass_z}
  }
  \caption
  {
  The distribution of age (a), metallicity (b), dust extinction (c), e-folding time of SFH (d), apparent magnitude at $24\micron$ (e), and redshift (f) in the  $M_*$-SFR diagram.
  With these distributions, five populations of galaxies can be easily recognized, which are more \hl{obvious} in (d).
  }
  \label{fig:relations}
\end{figure*}

As shown in Figure \ref{fig:mock_test_Zm}, it is usually hard to determine the stellar metallicity of galaxies with photometric data only.
However, since the spectroscopic redshift of galaxies have been used, we expect that the estimated metallicity of galaxies here should be more accurate.
In Figure \ref{fig:ULTRAVISTA0_0bc03_ch_sfr_mass_Zm}, we show the distribution of stellar metallicity of galaxies in the  $M_*$-SFR diagram.
As clearly shown in the figure, the blue galaxies all have low metallicity, while the red galaxies have much higher, near \hl{solar}, metallicity.
This is reasonable in the context of galaxy evolution.
Since the blue galaxies are mainly star-forming and starburst galaxies, the stars in these galaxies should be formed recently from low-metallicity gas.
On the other hand, red galaxies are fully evolved galaxies, where many stars are formed from recycled-gas enriched by the last generations of star formation.
Meanwhile, in Figure \ref{fig:ULTRAVISTA0_0bc03_ch_sfr_mass_Av}, we show the distribution of dust extinction of the same galaxies in the  $M_*$-SFR diagram.
The starburst galaxies have the highest dust extinction, while the more massive star-forming galaxies show less but still important dust extinction.
In contrary, there are very little dust extinction in both the quiescent and low mass star-forming galaxies.

Usually, the stellar population age of galaxies is thought to be the main parameter for the color of galaxies, since galaxies become redder when they aging.
However, as shown in Figure \ref{fig:ULTRAVISTA0_0bc03_ch_sfr_mass_age}, it seems not easy to separate galaxies into red and blue with age alone.
The starburst galaxies, which are the youngest in our sample, are indeed blue galaxies.
However, some blue low mass star-forming galaxies, instead of the reddest quiescent galaxies, are the oldest galaxies in our sample.
As related to age, in Figure \ref{fig:ULTRAVISTA0_0bc03_ch_sfr_mass_tau}, we show the distribution of the e-folding time of SFH in the  $M_*$-SFR diagram.
Interestingly, the galaxies in our sample can be clearly divided into at least five groups with this distribution.
The blue star-forming galaxies in the main sequence (G1) have the longest e-folding time of SFH.
On the other hand, the red quiescent galaxies (G2) have much shorter e-folding time of SFH.
Most galaxies in our sample belong to this two groups.
The galaxies with the shortest e-folding time of SFH (G3) are located between the last two groups.
These galaxies are also unique in the distribution of age, as shown in Figure \ref{fig:ULTRAVISTA0_0bc03_ch_sfr_mass_age}.
Meanwhile, the galaxies at the high-mass end of the main sequence (G4) constitute the four group.
Finally, those starburst galaxies with \hl{very younge age, highest dust extinction and much shorter e-folding time} than normal star-forming galaxies are considered to be the fifth group of galaxies (G5).

The galaxies in the G3 and G4 groups are likely in the transition stages between red and blue galaxies.
To check if they are related to AGN activities, we show the distribution of the apparent magnitude at $24\micron$ in Figure \ref{fig:ULTRAVISTA0_0bc03_ch_sfr_mass_24um}.
Meanwhile, we show the distribution \hl{of} redshifts of these galaxies in  Figure \ref{fig:ULTRAVISTA0_0bc03_ch_sfr_mass_z}.
\hl{For most of these galaxies, the fluxes at $24\micron$ correspond to the rest-frame mid-IR fluxes, which are thought to be responsible by dust heated by AGN \citep{Nenkova02,Elitzur2006a,Fritz2006a,Nenkova2008a,Nenkova2008b,Han2012b}.}
As shown clearly in  Figure \ref{fig:ULTRAVISTA0_0bc03_ch_sfr_mass_24um}, galaxies in the high-mass end of the star-forming main sequence have the strongest mid-IR emissions.
These galaxies should be the composite galaxies.
Most galaxies in the G4 group show important mid-IR emissions, while most galaxies in the G3 group do not.
The galaxies in the G3 group could be galaxies at an earlier evolution stage, when the AGN activities are still weak or not launched yet.

\section{SUMMARY AND CONCLUSIONS}
\label{s:summary}
In this paper, we have described an updated version of BayeSED.
On the basis of a previous work \citep{Han2012a,Han2013a}, we have presented a major update to this code in this paper.
Firstly, the most up-to-date version of the MultiNest Bayesian inference (\S \ref{ss:bayes}) tool has been employed.
The MultiNest algorithm (\S \ref{ss:sampling}) has recently been improved by the importance nested sampling to allow a more efficient sampling of high-dimensional parameter space and more accurate calculation of Bayesian evidence.
Secondly, besides the ANN method, we have added the KNN method as another method for the interpolation of model SED libraries in BayeSED (\S \ref{ss:interpolation}).
For both ANN and KNN, we have defined a file format to store all necessary \hl{information} about a SED model such that almost any SED model can be easily used by BayeSED to interpret the observed SEDs of galaxies.
Thirdly, the redshift of a galaxy can be set as a free parameter, and the effects of IGM and Galactic extinction have been considered.
So, it is now possible to obtain the redshift and other physical parameters of a galaxy, all simultaneously and self-consistently.
Fourthly, the main body of BayeSED has been completely rewritten in C++ in an object-oriented programming fashion, and  parallelized with MPI.
So, with BayeSED, it is now possible to \hl{analyze} the SEDs of a large sample of galaxies with a detailed Bayesian approach that is based on an intensive sampling of the parameter space of SED models.

We have systematically tested the reliability of the BayeSED code to recover the physical parameters of galaxies from their multi-wavelength photometric SEDs with a mock sample of galaxies (\S \ref{s:application_mock}).
For both the ANN and KNN methods, the tests show that BayeSED can recover the redshift and stellar population parameters reasonably well.
We found that different parameters can be recovered with a varying degree of accuracy.
While the redshift, age, dust extinction, stellar mass, bolometric luminosity and star formation rate can be properly recovered, it is usually hard to recover metallicity and e-folding time of SFH with photometric data, especially when the spectroscopic redshift is not available.
We \hl{believe} that this is due to the nature of these parameters themselves, and the limited information about them, in the photometric data.
Meanwhile, the tests showed that the KNN interpolation method, although more \hl{memory- and time-consuming} than the ANN method, may lead to more accurate results.

\hl{We systematically applied BayeSED to interpret the observed SEDs of a large sample of galaxies in the COSMOS/UltraVISTA field}, with different evolutionary population synthesis models (\S \ref{s:application_real}).
\hl{A Bayesian model comparison of evolutionary population synthesis models has been accomplished for the first time, }
\hl{We found that the \cite{Bruzual2003a} model statistically has larger Bayesian evidence than the \cite{Maraston2005a} model for the Ks-selected sample of galaxies with spectroscopic redshifts and mostly less than one.}
\hl{Besides, we found that the \cite{Maraston2005a} model is more sensitive to the different choice of IMF than the \cite{Bruzual2003a} model.}
\hl{However, the conclusion is drawn in a statistical sense and could be different for samples of galaxies at higher redshift \citep{MarastonC2006a,Maraston2010a,Henriques2011}, which is worthy of further investigation.}
\hl{Meanwhile, we found that, by using stellar metallicity as an additional free parameter, the Bayesian evidences of stellar population synthesis models can be increased significantly.}
\hl{Therefore, we conclude that it is important to set metallicity as a free parameter to obtain unbiased results, even if this parameter cannot be estimated very accurately with photometric data only.}

\hl{We have compared our results obtained using our BayeSED code with that obtained using the widely used FAST code, and found a good agreement.}
\hl{Nevertheless, we found that the parameters estimated with BayeSED show more natural distributions than more conventional grid-based SED-fitting codes such as FAST.}
\hl{Besides, based on the rest-frame color and stellar population  parameters obtained with BayeSED, we recognized five distinct populations of galaxies in the Ks-selected sample of galaxies.}
\hl{They may represent galaxies at different evolution stages or in different environments.}

With the systematic tests \hl{using} a mock sample of galaxies and the comparison with a popular grid-based SED-fitting code for a real sample of galaxies, we conclude that the BayeSED code can be reliably applied to interpret the SEDs of large samples of galaxies.
Based on the MultiNest algorithm to allow intensive sampling of parameter space and efficient computation of Bayesian evidence of the SED models, BayeSED could be a powerful tool for investigating the formation and evolution of galaxies from the rich multi-wavelength observations currently available.
\hl{Particularly}, with the efficient computation of Bayesian evidence for SED models, BayeSED could be useful for the further development of evolutionary population synthesis models and other SED models for galaxies.
Besides, while we have only applied BayeSED to the photometric data \hl{so far}, it should be straightforward to apply similar methods to the spectroscopic data in the future.

\acknowledgments
\hl{We thank an anonymous referee for his/her valuable comments which help to improve the paper.}
The authors gratefully acknowledge the computing time granted by the Yunnan Observatories, and provided on the facilities at the Yunnan Observatories Supercomputing Platform.
\hl{We thank Prof. Houjun Mo for helpful discussion about Bayesian evidence and Lulu Fan for helpful discussion about Bayesian SED-fitting of galaxies.}
This work is supported by the National Natural Science Foundation of China (Grant Nos. 11303084, U1331117, 11390374), the Science and Technology Innovation Talent Programme of the Yunnan Province (Grant No. 2013HA005), and the Chinese Academy of Sciences (Grant No. XDB09010202).

\bibliography{ms.bbl}

\clearpage

\end{document}

%% file: libs.tab
\begin{tabular}{ccccccccc}
\hline
\hline
name&resolution&SFH&IMF&dust extinction law&log(tau/yr)&log(age/yr)&Av/mag&Z\\
\hline
bc03\_sa&pr(460)\tablenotemark{a}&exp&{\cite{Salpeter1955a}}&{\cite{Calzetti2000a}}&$[6.5,0.1,11]\tablenotemark{b}$&$[7.0,0.05,10.1]$&$[0,0.2,4]$&$0.004,0.008,0.02,0.05$\\
bc03\_ch&pr(460)&exp&{\cite{Chabrier2003a}}&{\cite{Calzetti2000a}}&$[6.5,0.1,11]$&$[7.0,0.05,10.1]$&$[0,0.2,4]$&$0.004,0.008,0.02,0.05$\\
ma05\_sa&pr(460)&exp&{\cite{Salpeter1955a}}&{\cite{Calzetti2000a}}&$[6.5,0.1,11]$&$[7.0,0.05,10.1]$&$[0,0.2,4]$&$0.001,0.01,0.02,0.04$\\
ma05\_kr&pr(460)&exp&{\cite{Kroupa2001a}}&{\cite{Calzetti2000a}}&$[6.5,0.1,11]$&$[7.0,0.05,10.1]$&$[0,0.2,4]$&$0.001,0.01,0.02,0.04$\\
\hline
\end{tabular}